# Exploring Rulial Space: The Case of Turing Machines

Stephen Wolfram

*As an example of the concept of rulial space, we explore the case of simple Turing machines. We construct the rulial multiway graph which represents the behavior of all possible Turing machines with a certain class of rules. This graph (which is a Cayley graph of a "Turing machine group") gives a map of the space of non-deterministic Turing machines. We investigate the subgraph formed by deterministic machines, and explore the relationship to the P vs. NP problem. We also consider the implications of features of rulial space for physics, including estimating the maximum speed ρ in rulial space, relations between rulial black holes and computational reducibility, and interpretations of hypercomputation.*

## Generalized Physics and the Theory of Computation

Let's say we find a rule that reproduces physics. A big question would then be: "Why this rule, and not another?" I think there's a very elegant potential answer to this question, that uses what we're calling rule space relativity—and that essentially says that there isn't just one rule: actually all possible rules are being used, but we're basically picking a reference frame that makes us attribute what we see to some particular rule. In other words, our description of the universe is a sense of our making, and there can be many other—potentially utterly incoherent—descriptions, etc.

But so how does this work at a more formal level? This bulletin is going to explore one very simple case. And in doing so we'll discover that what we're exploring is potentially relevant not only for questions of "generalized physics", but also for fundamental questions in the theory of computation. In essence, what we'll be doing is to study the structure of spaces created by applying all possible rules, potentially, for example, allowing us to "geometrize" spaces of possible algorithms and their applications.

A version of this document with immediately executable code is available at
https://www.wolframphysics.org/bulletins/2020/06/exploring-rulial-space-the-case-of-turing-machines/

In our models of physics, we begin by considering spatial hypergraphs that describe relations between "atoms of space". Then, looking at all possible ways a given rule can update these hypergraphs, we form what we call a multiway graph. The transversals of this graph define what we call branchial space, in which we can see the pattern of entanglements between quantum states.

But there's also a third level we can consider. Instead of just forming a multiway graph in which we do all possible updates with a given rule, we form a rulial multiway (or "ultramultiway") graph in which we follow not only all possible updates, but also all possible rules. The transversals to this rulial multiway graph define what we call rulial space. Causal invariance in the rulial multiway graph then implies "rule space relativity" which is what allows us to use different possible reference frames to describe the universe.

In the end, the Principle of Computational Equivalence implies a certain invariance to the limiting structure of the rulial multiway graph, independent of the particular parametrization of the set of possible rules. But for the sake of understanding rulial space and the rulial multiway graph—and getting intuition about these—this bulletin is going to look at a specific set of possible rules, defined by simple Turing machines.

The rules we'll use aren't a good fit for describing our universe. But the comparative simplicity of their structure will help us in trying to elucidate some of the complexities of rulial space. In addition, using Turing machines will make it easier for us to make contact with the theory of computation, where Turing machines are standard models.

## Turing Machines

Here's a representation of the rule for a particular 2-state ($s = 2$), 2-color ($k = 2$) Turing machine:

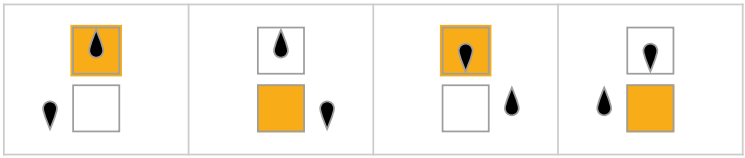

The pointer represents the "head" of the Turing machine and its orientation represents the state of the head. Here's what this particular Turing machine does over the course of a few steps starting from a "blank tape" (i.e. all squares white):

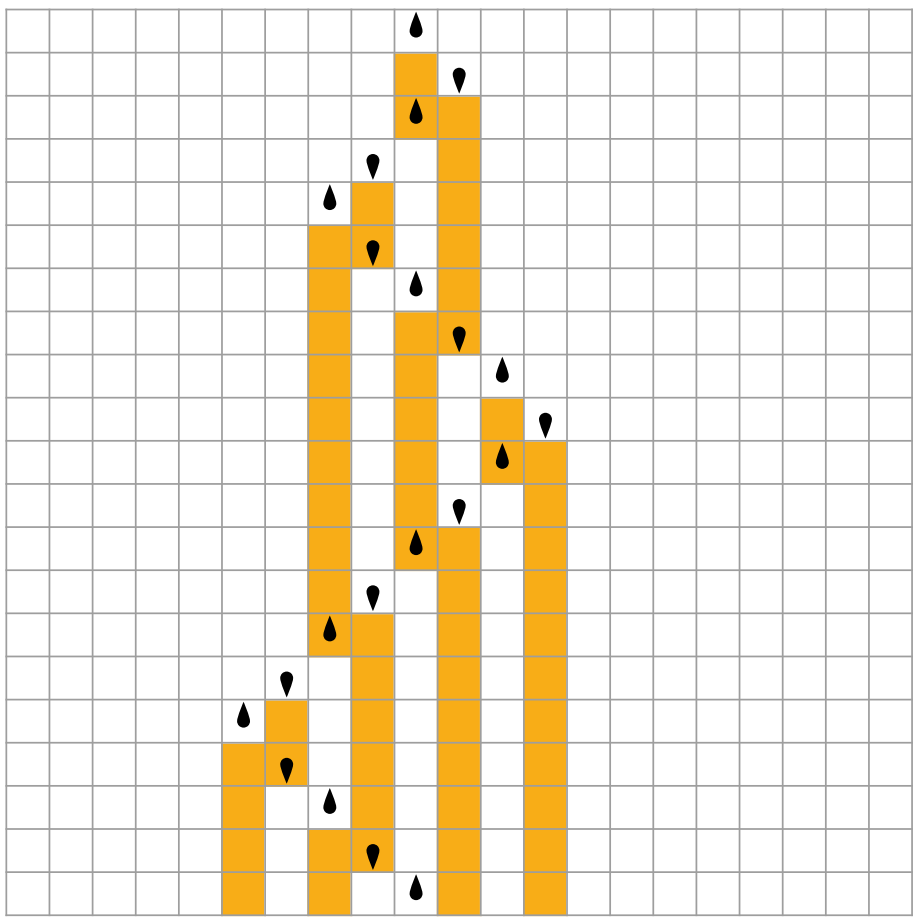

There are $(2\, s\, k)^{s\, k}$ possible $s$-state $k$-color Turing machines, or 4096 $s = 2$, $k = 2$ machines. Here are all the distinct behaviors that occur (up to left-right reflection) in these 4096 machines, starting from a blank tape:

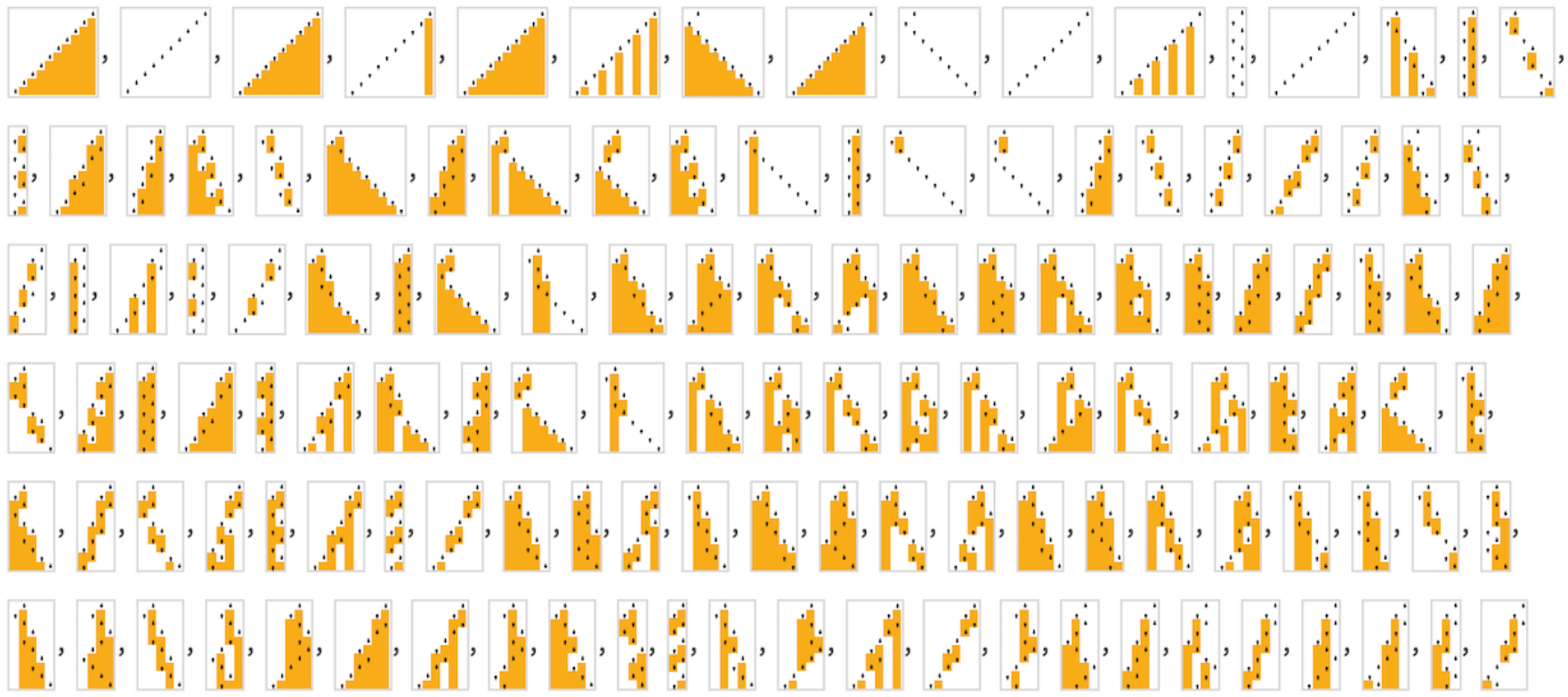

But note that all these are deterministic Turing machines, in the sense that, for a given machine, it is always the same rule that is applied at every step. But what about non-deterministic Turing machines? The idea here is to allow several different rules at any given step.

As a simple example, consider the pair of rules:

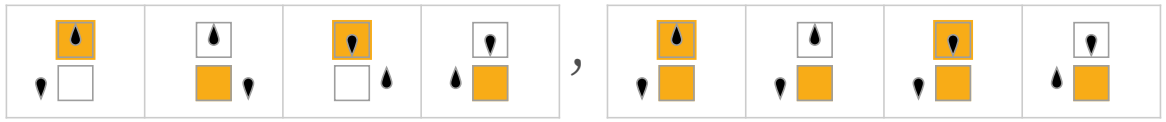

To show the possible evolution histories, we can construct a multiway system:

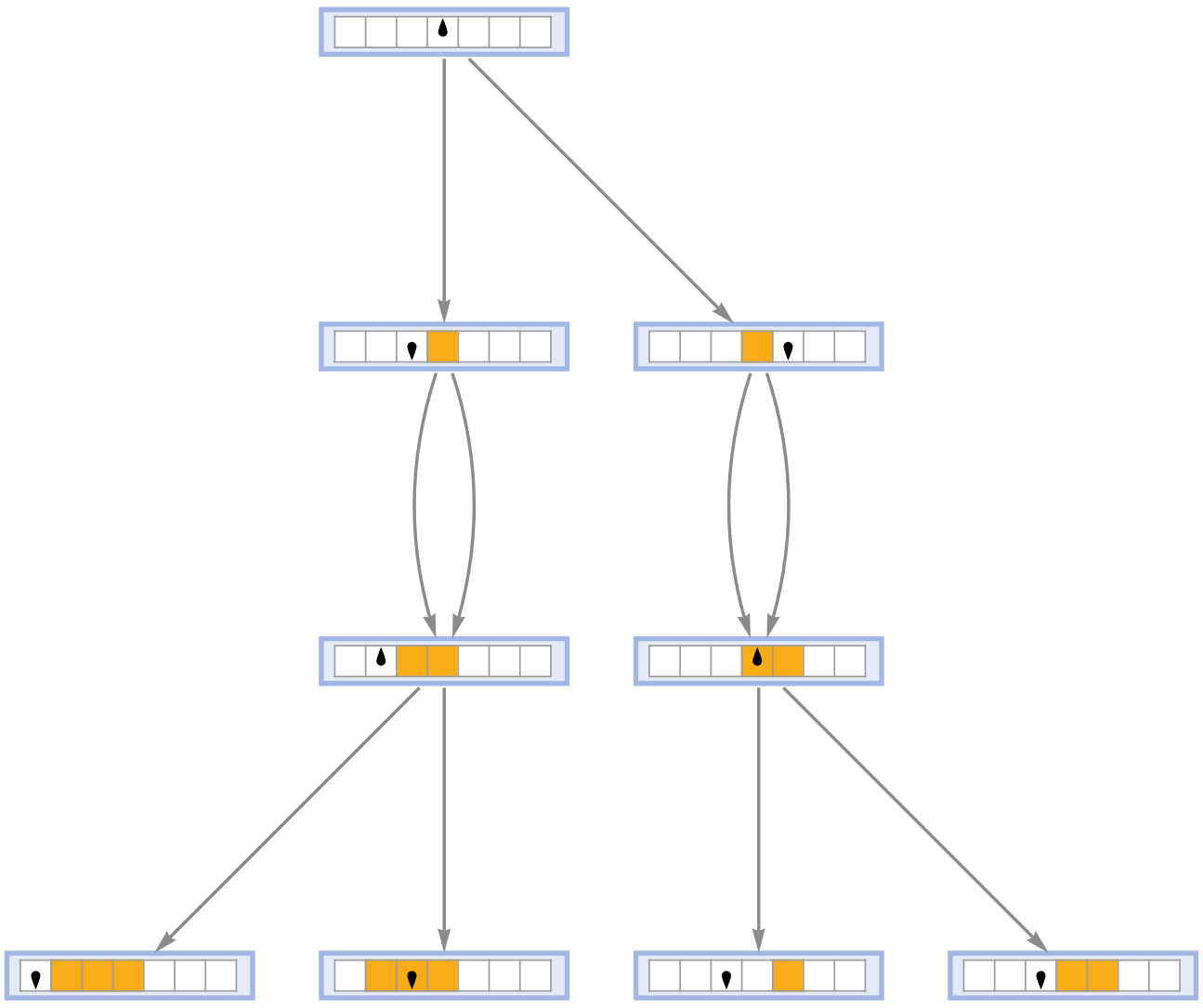

Each arrow represents an application of one of the two possible rules. A non-deterministic Turing machine is usually considered to follow a particular path, corresponding to a particular evolution history. Then a typical question is whether there exists some path that leads to a particular final state (which might be viewed as the solution to some particular problem). (In a quantum Turing machine, one considers collections of multiple states, viewed as being in a quantum superposition.)

In what we're going to be doing here, we want to study the "extreme case" of non-determinism—where at every step, every possible Turing machine rule (at least with a given *s*, *k*) is applied. Then we'll be interested in the full rulial multiway graph that's created.

## The Rulial Multiway Graph

Start with a blank tape. Then apply all possible 2,2 Turing machine rules. This is the rulial multiway graph that's formed after 1 step:

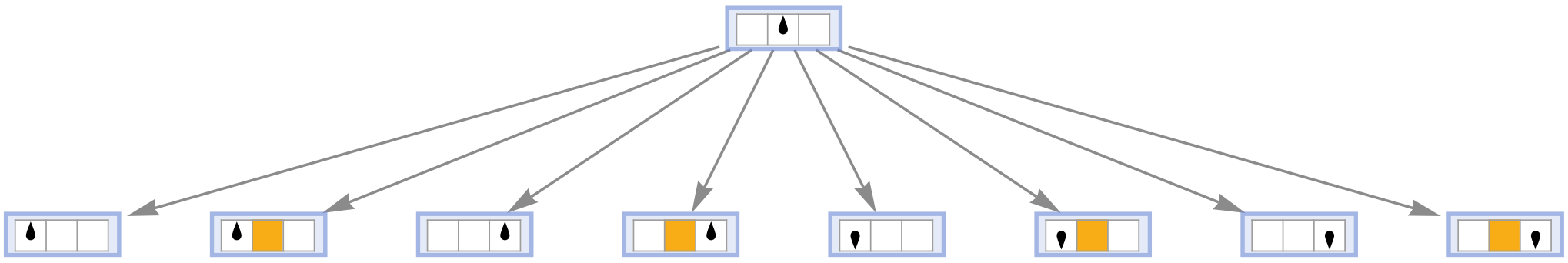

Each individual arrow here represents many possible rules from all the 4096 discussed above. Here's what happens after 2 steps:

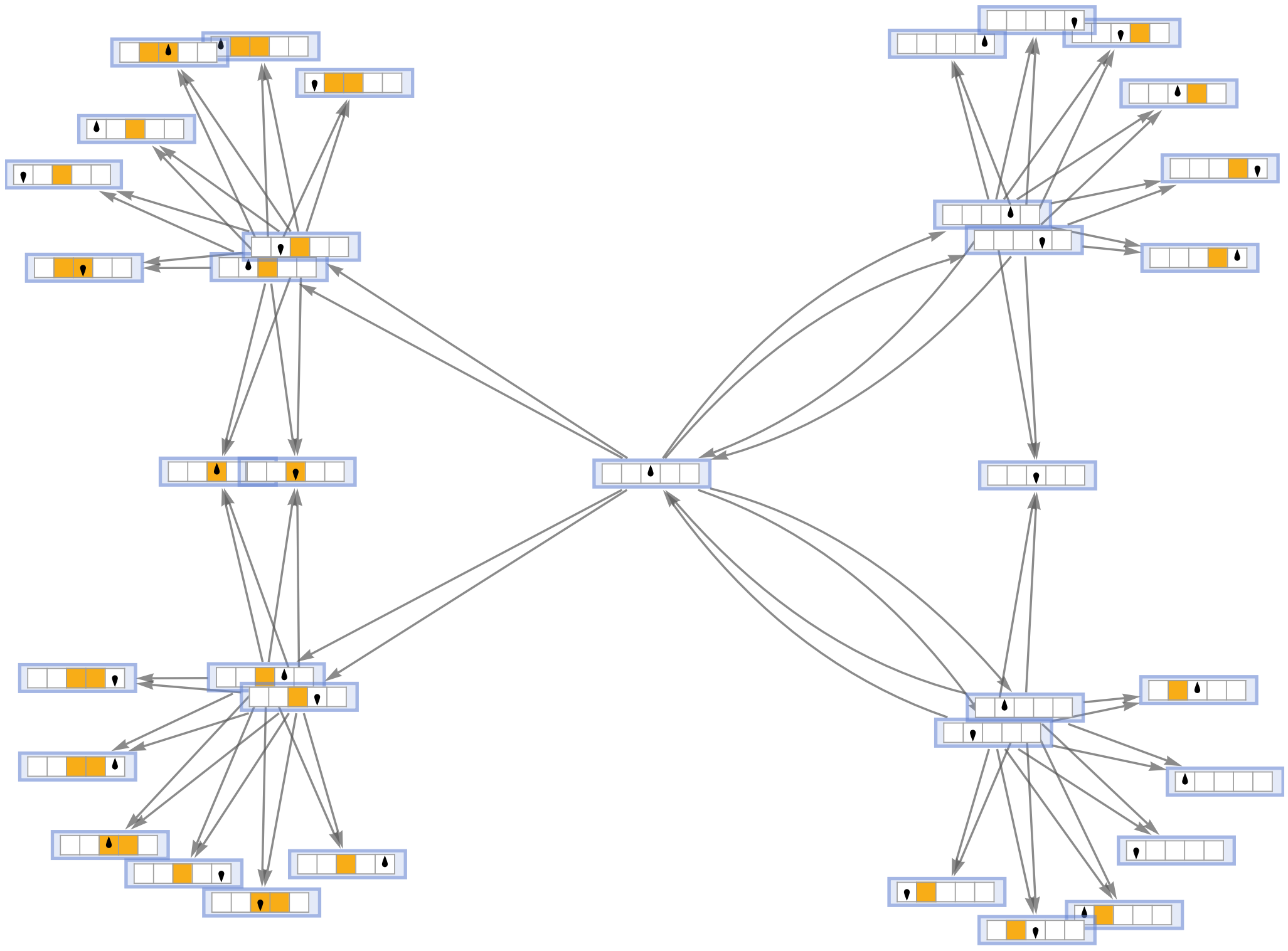

In layered form, this becomes:

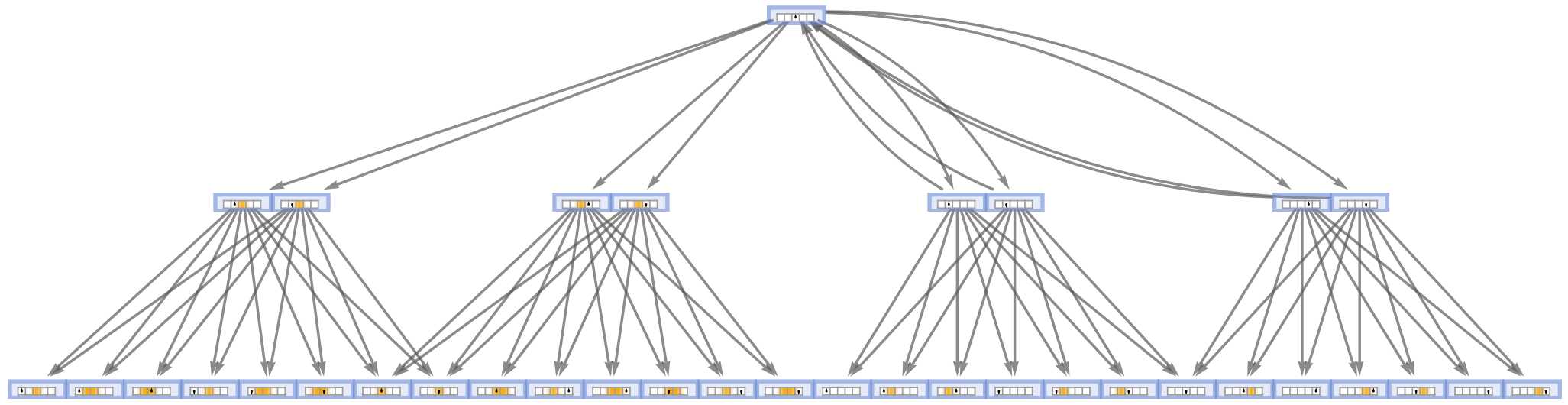

In creating these rulial multiway graphs, there's an important simplification we're able to make. We don't need to separately apply all 4096 2,2 Turing machine rules to each state at each step—because all that ever matters is the one case of the rule that's relevant to the particular state one has. There are 4 possible individual cases—and for each of these there are 8 possible outcomes, leading to 32 "micro-rules":

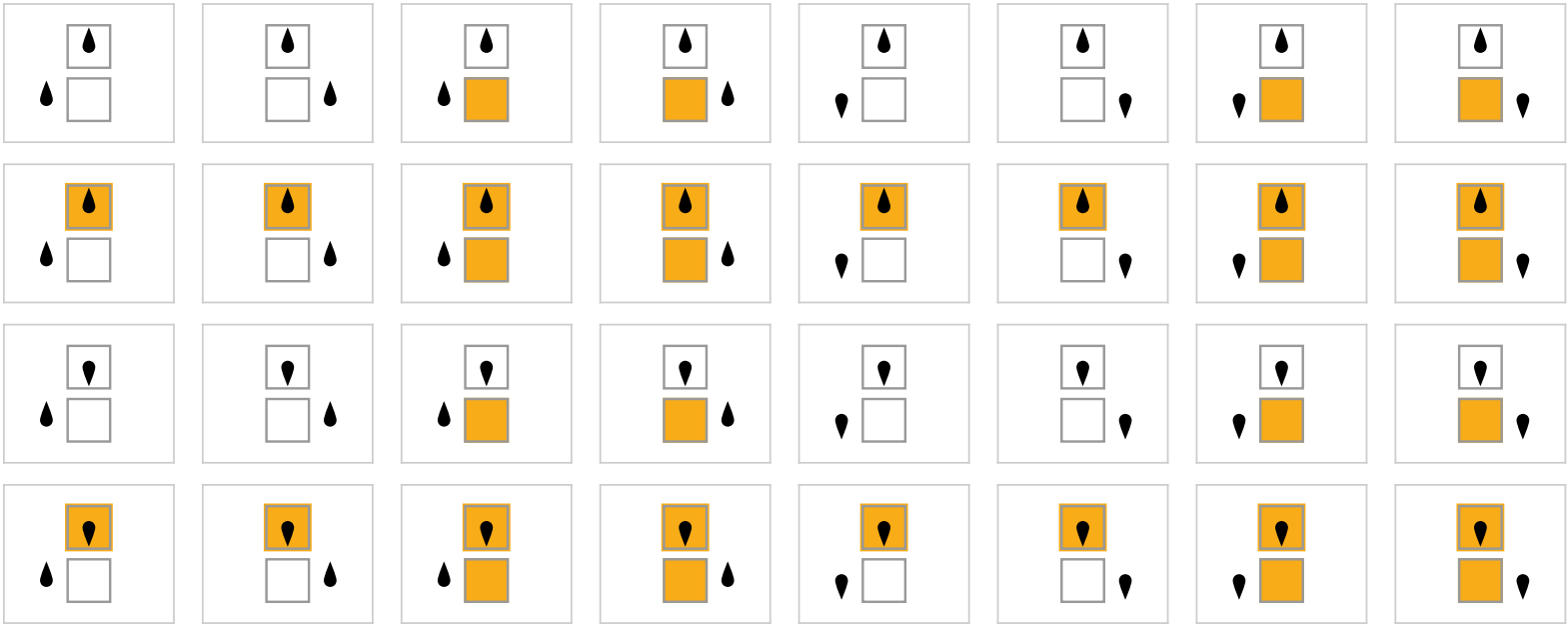

After 3 steps, the rulial multiway graph has the form:

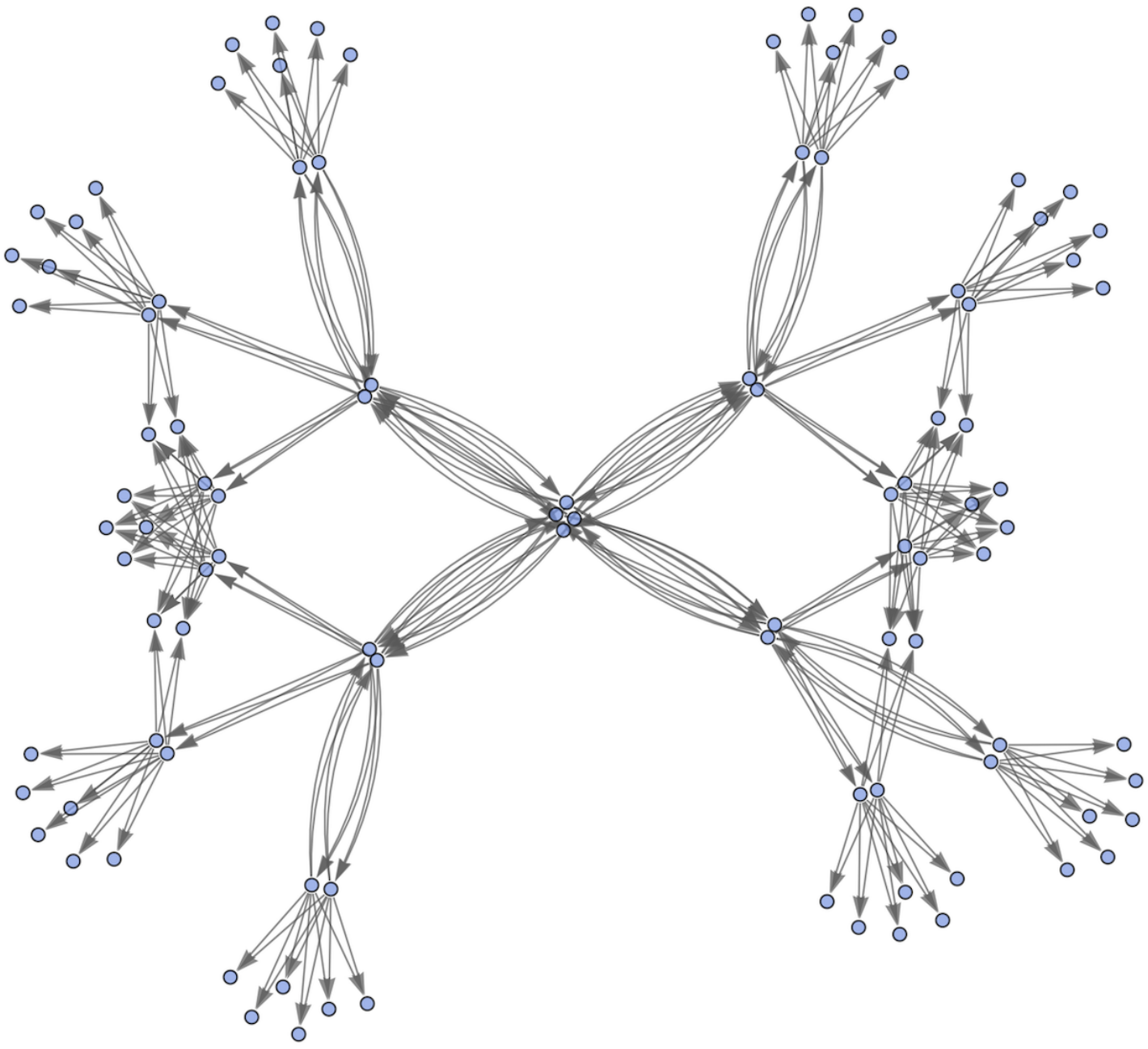

In 3D this becomes:

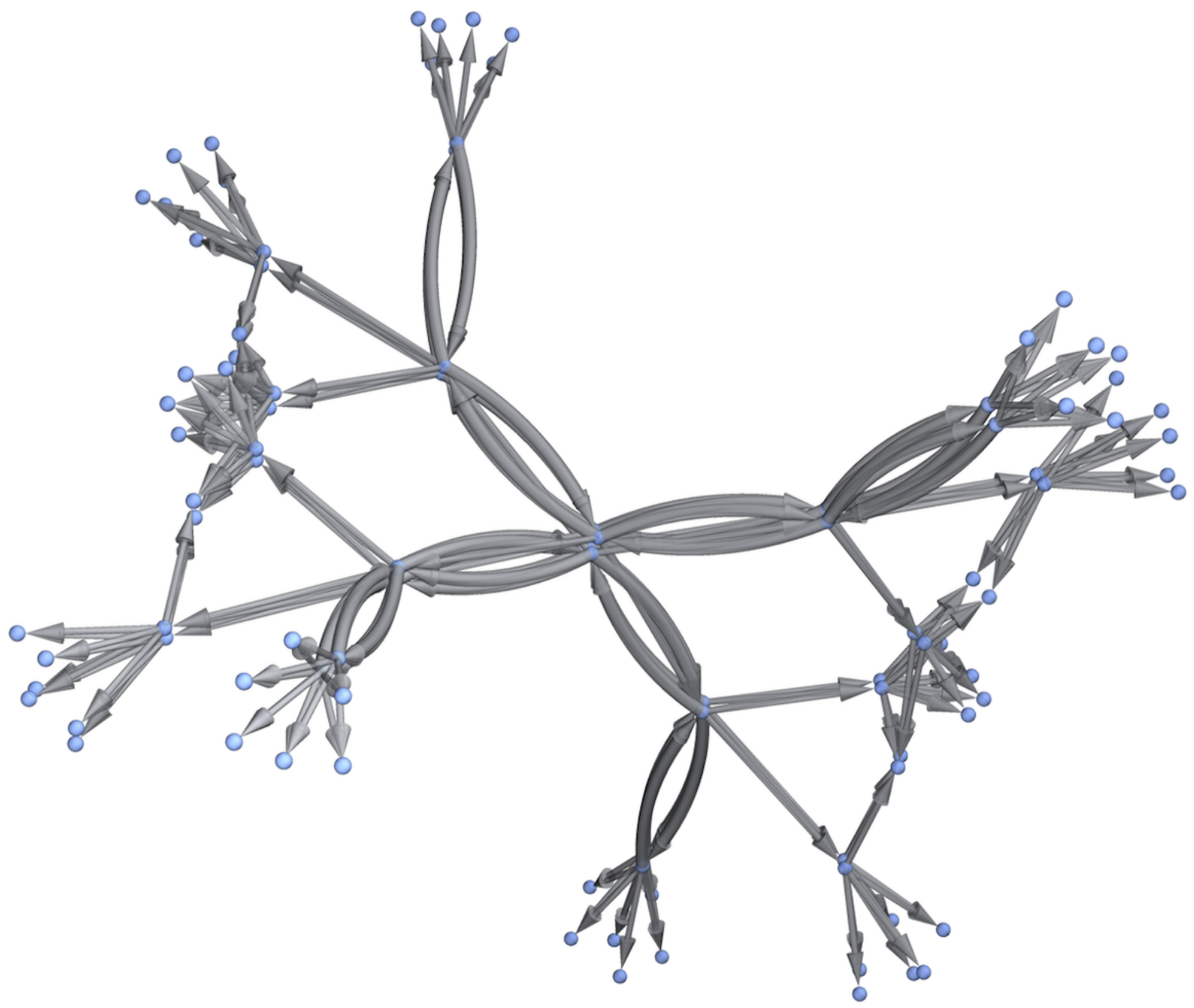

In layered form it is:

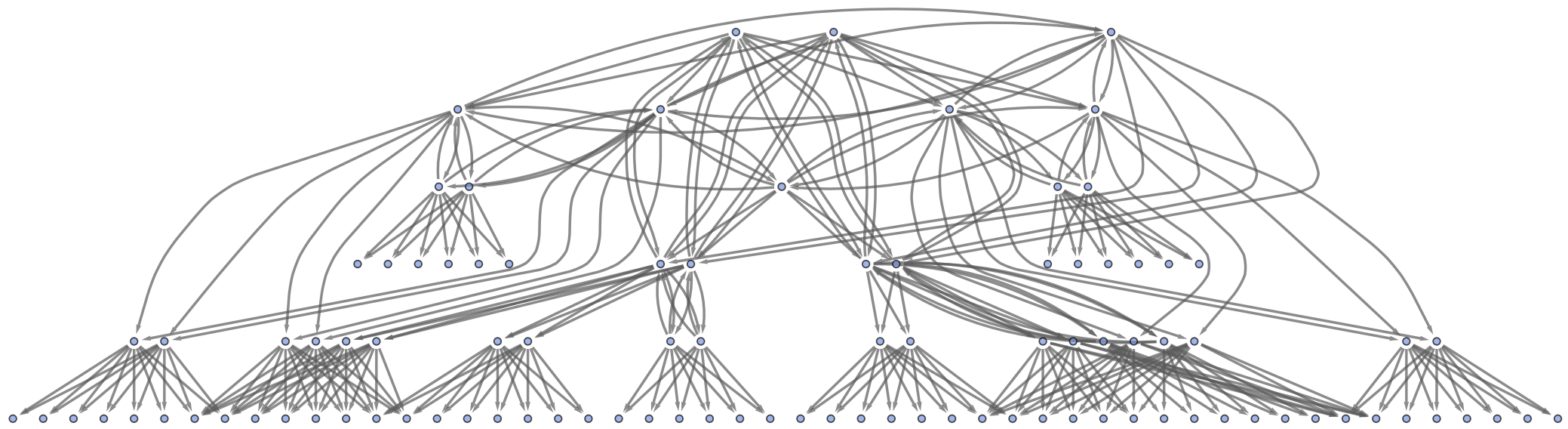

After 5 steps, the rulial multiway graph is:

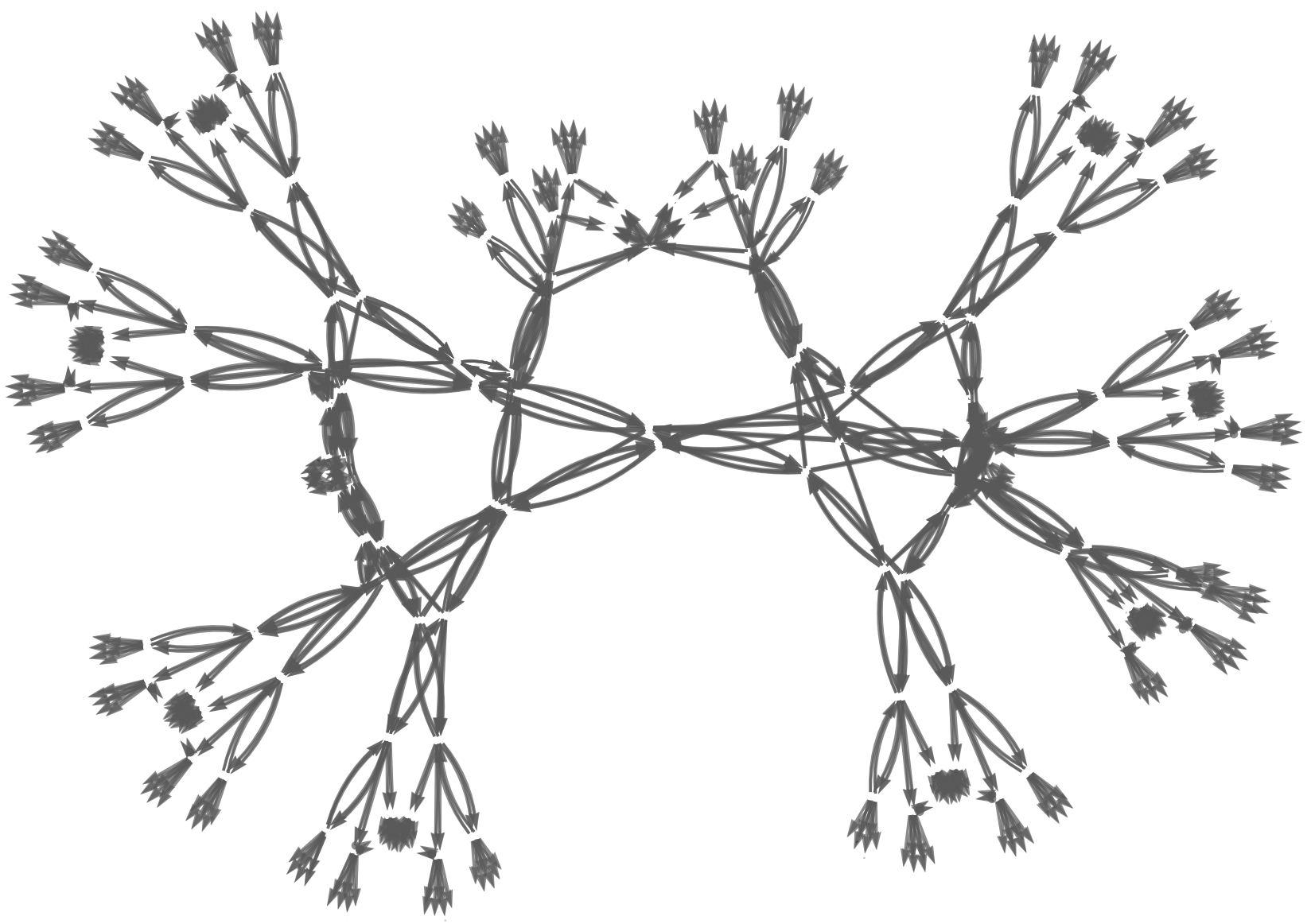

What about other numbers of states and colors? Here are the rulial multiway graphs after 3 steps for Turing machines with various numbers of states and colors $\{s,k\}$:

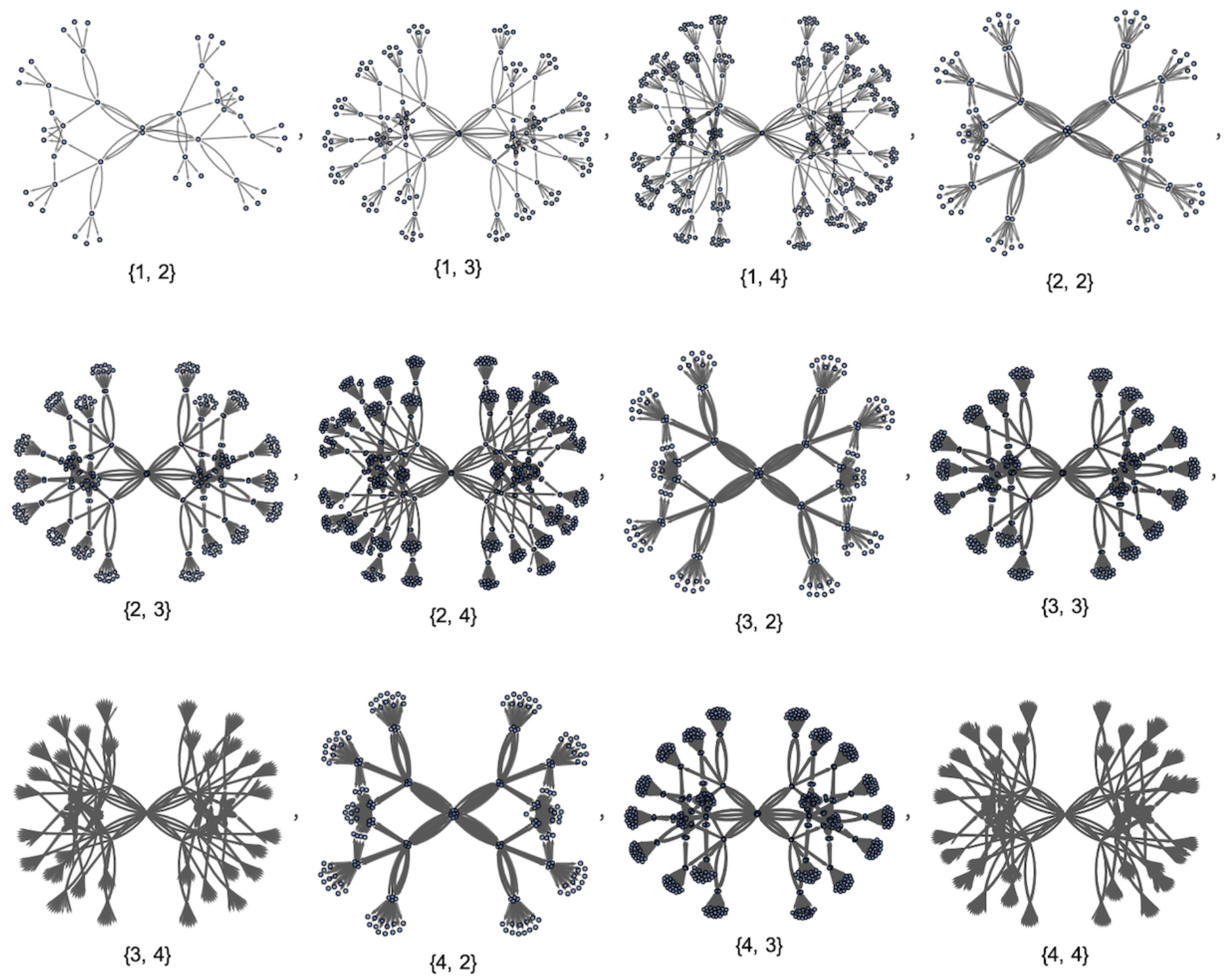

Notice the presence of $s = 1$ examples. Even with a single possible state, it is already possible to form a nontrivial rulial multiway system. Here are the results for $k = 2$ after 1 step:

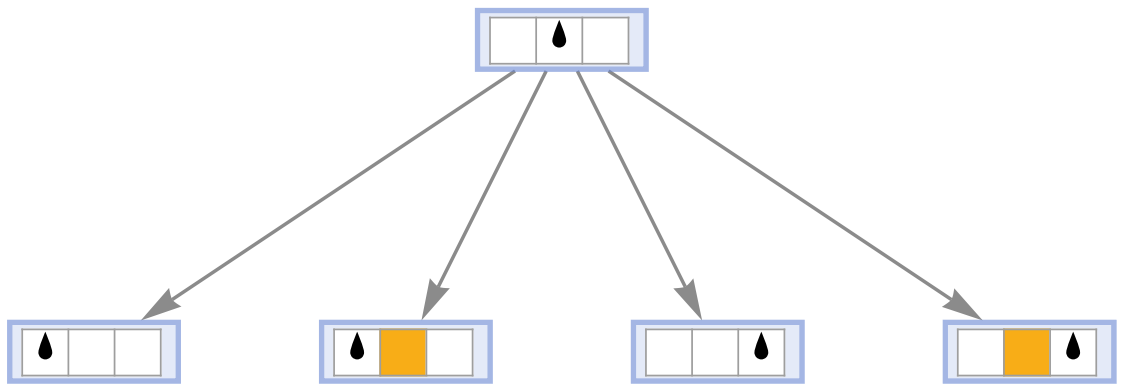

2 steps:

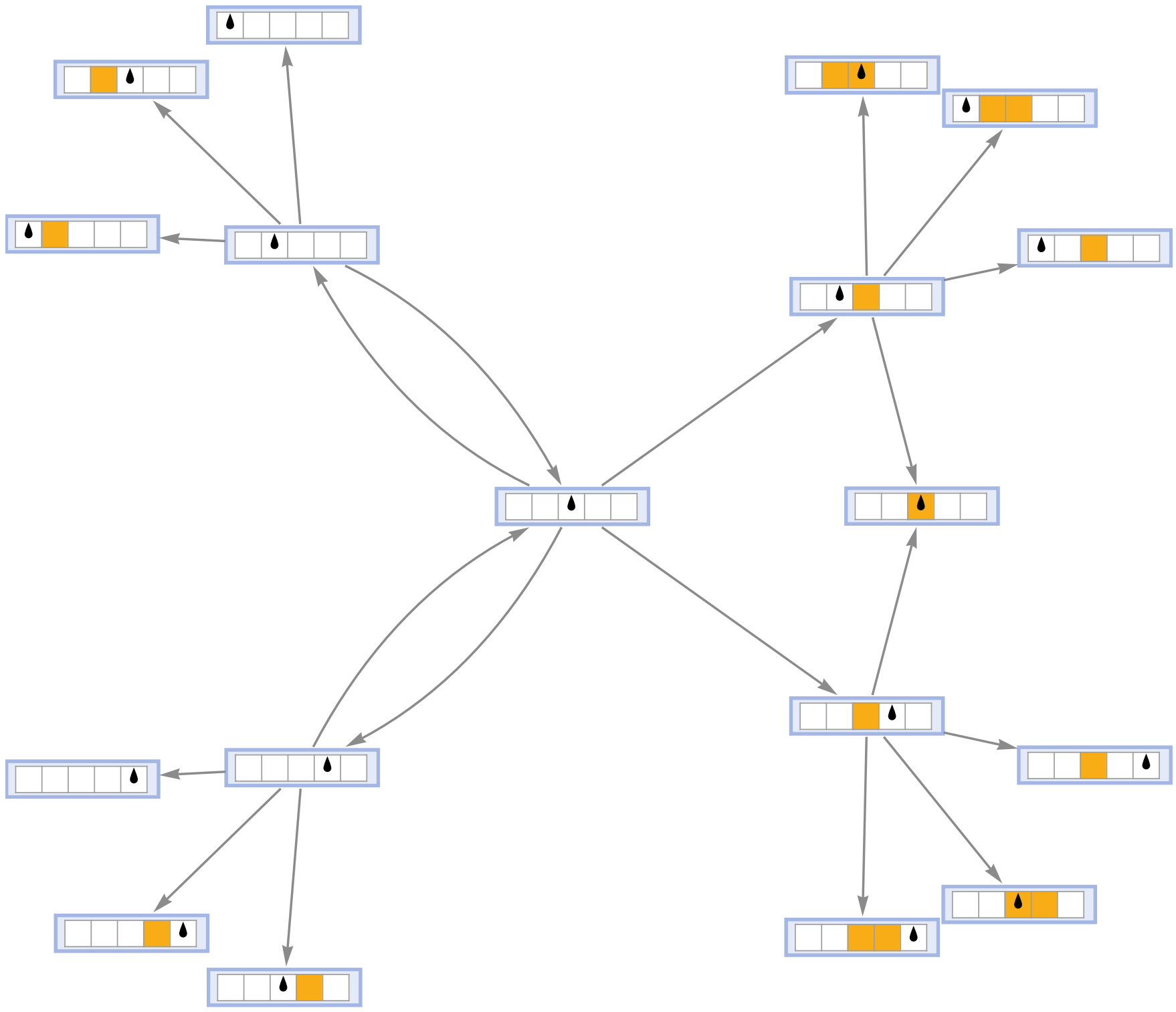

6 steps:

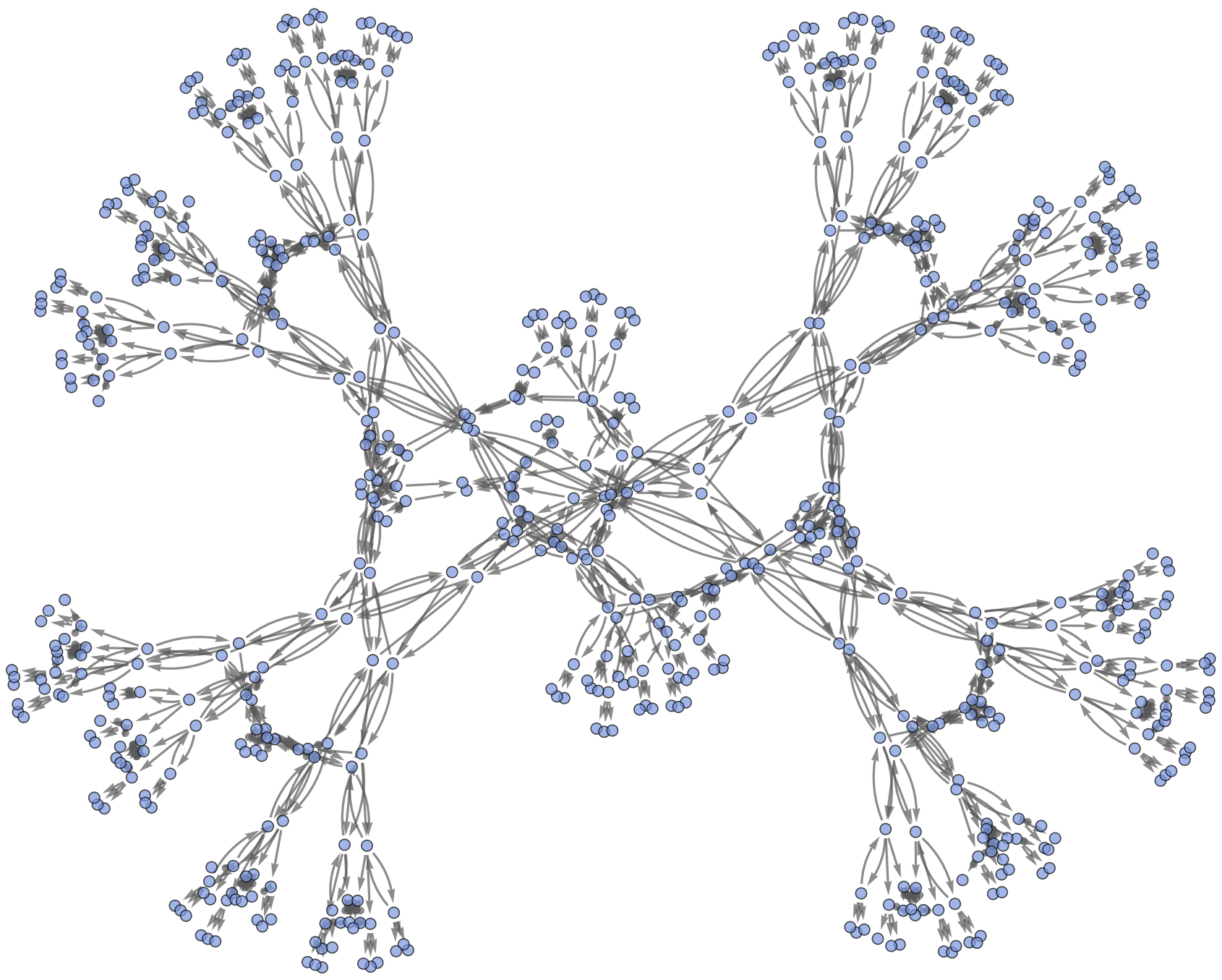

In principle one can also consider the even simpler case of $s = k = 1$. After 1 step one gets:

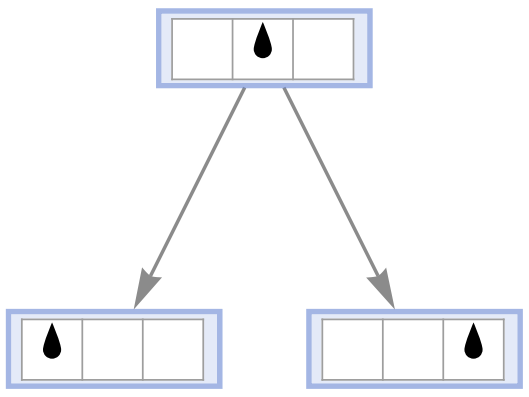

After 2 steps this becomes:

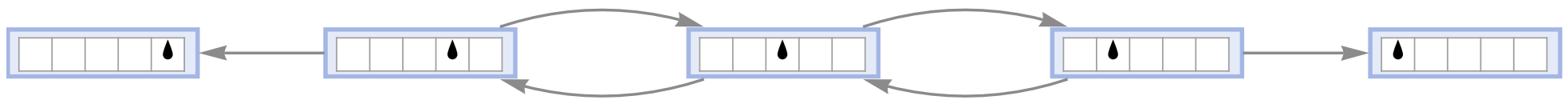

And after 3 steps:

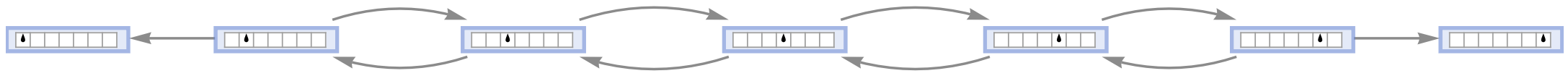

As a layered graph rendering, this becomes:

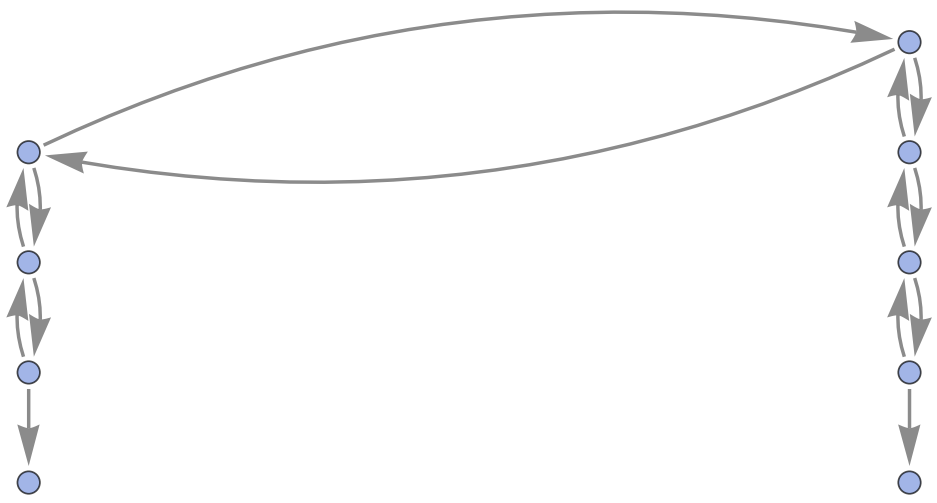

With $s = 2$, $k = 1$ one gets after 1 step:

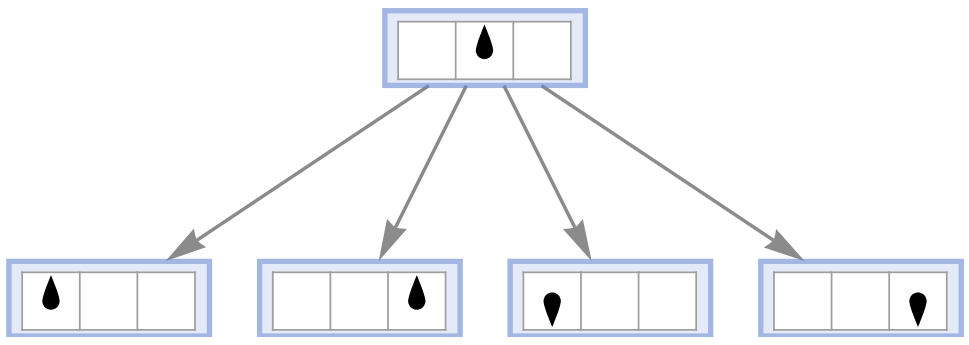

After 2 steps this becomes:

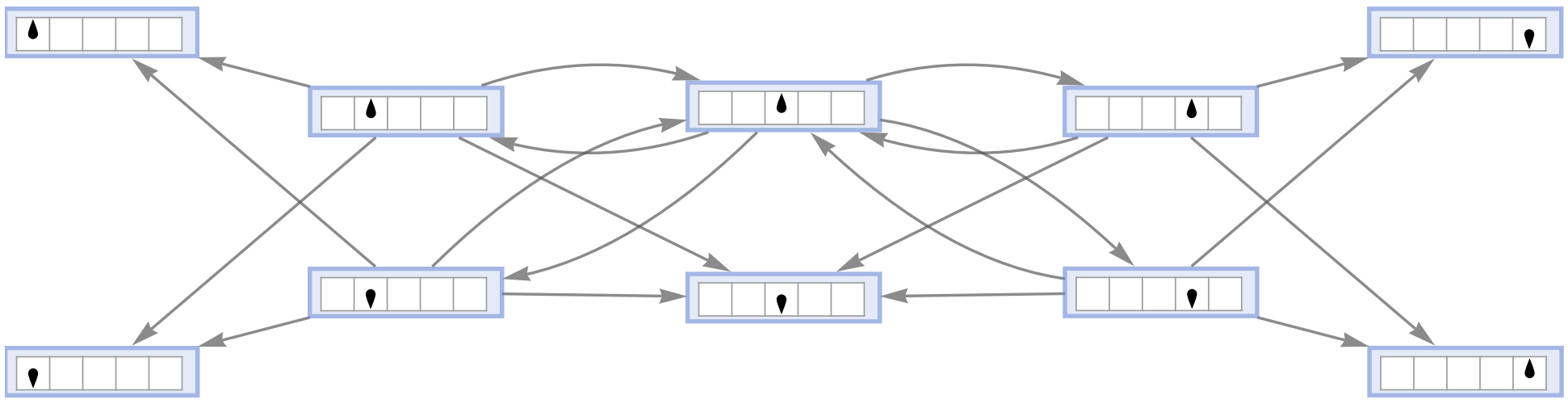

And after 3 steps:

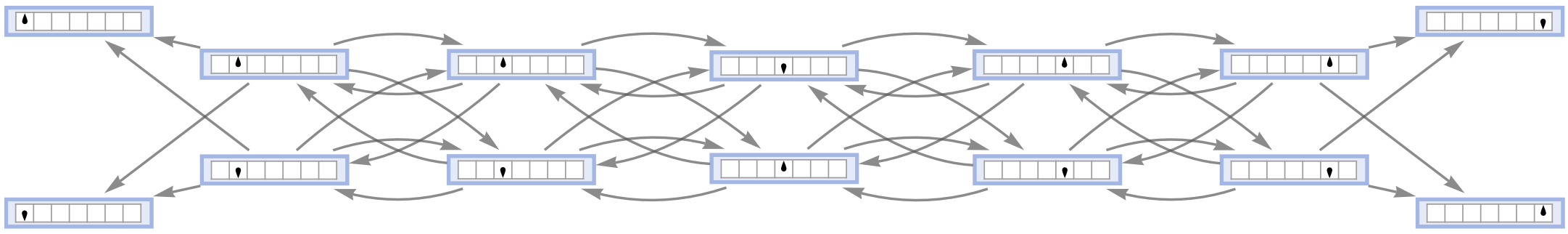

After 6 steps this becomes:

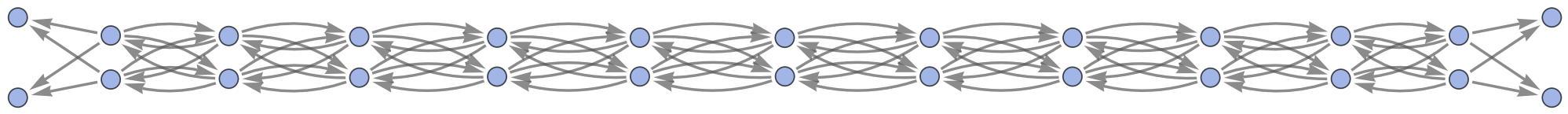

At least in the central part of this graph, there are edges going in both directions. Combining these, and treating the graph as undirected, one gets:

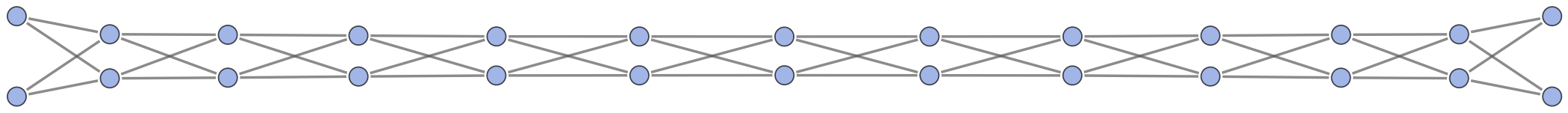

With 3 states ($s = 3$) there are 3 tracks:

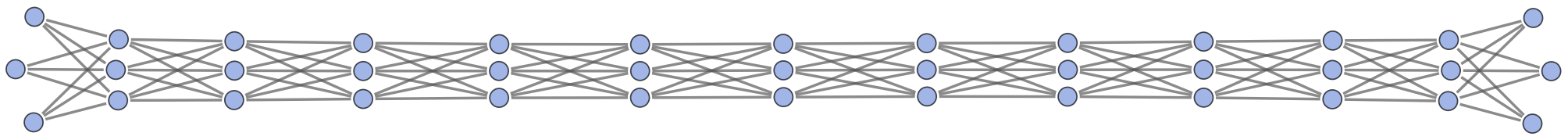

One can also consider Turing machines in which the head can not only move left or right, but can also stay still. In this case, with $s = 2$, $k = 1$ one gets after 1 step:

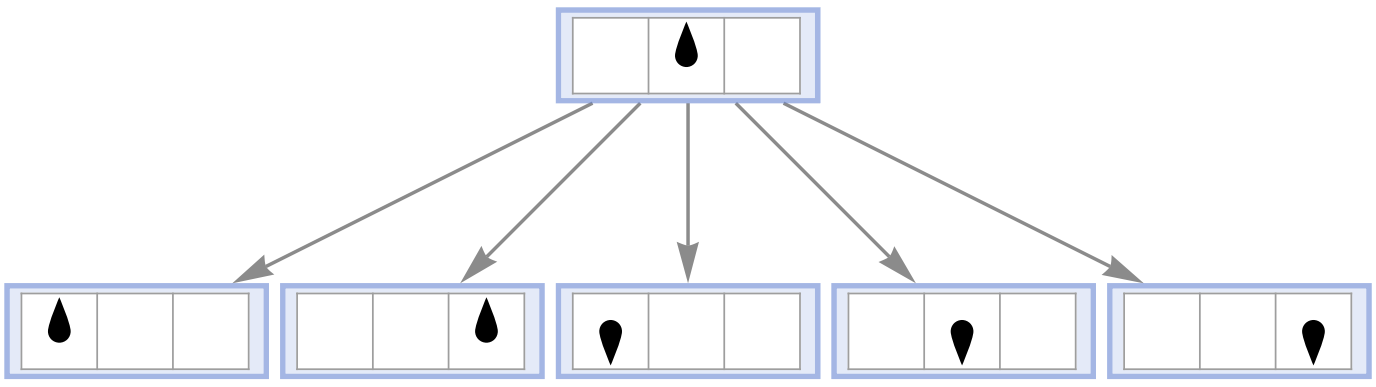

After 2 steps this becomes:

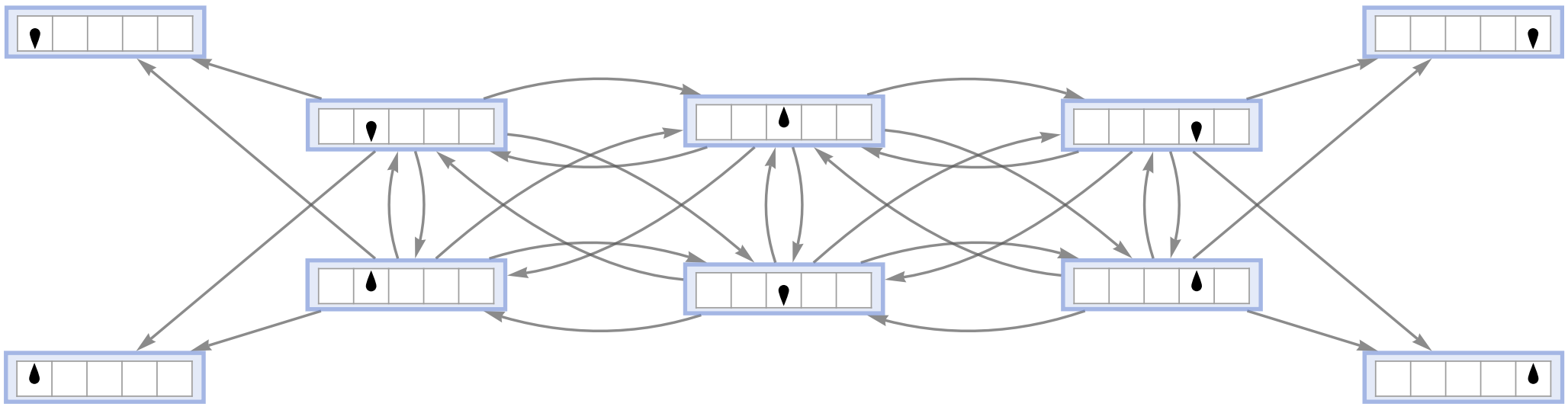

After 4 steps, removing repeated edges, etc. this gives:

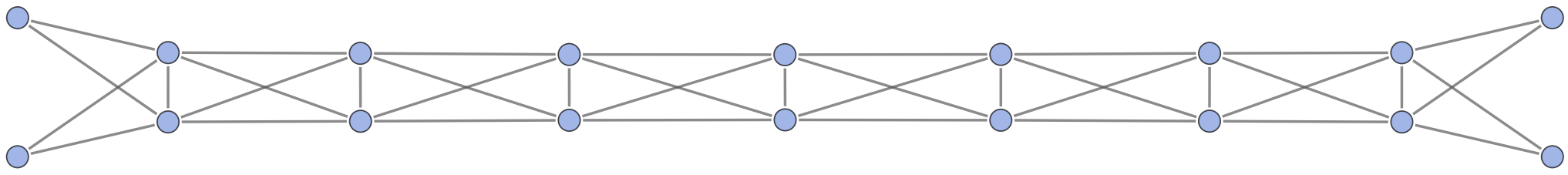

As another generalization, one can consider Turing machines that can move not just one, but also two squares:

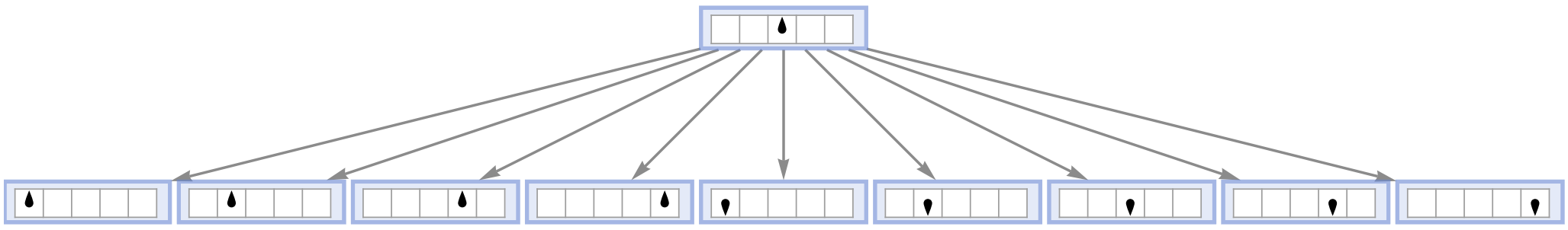

After 2 steps one gets:

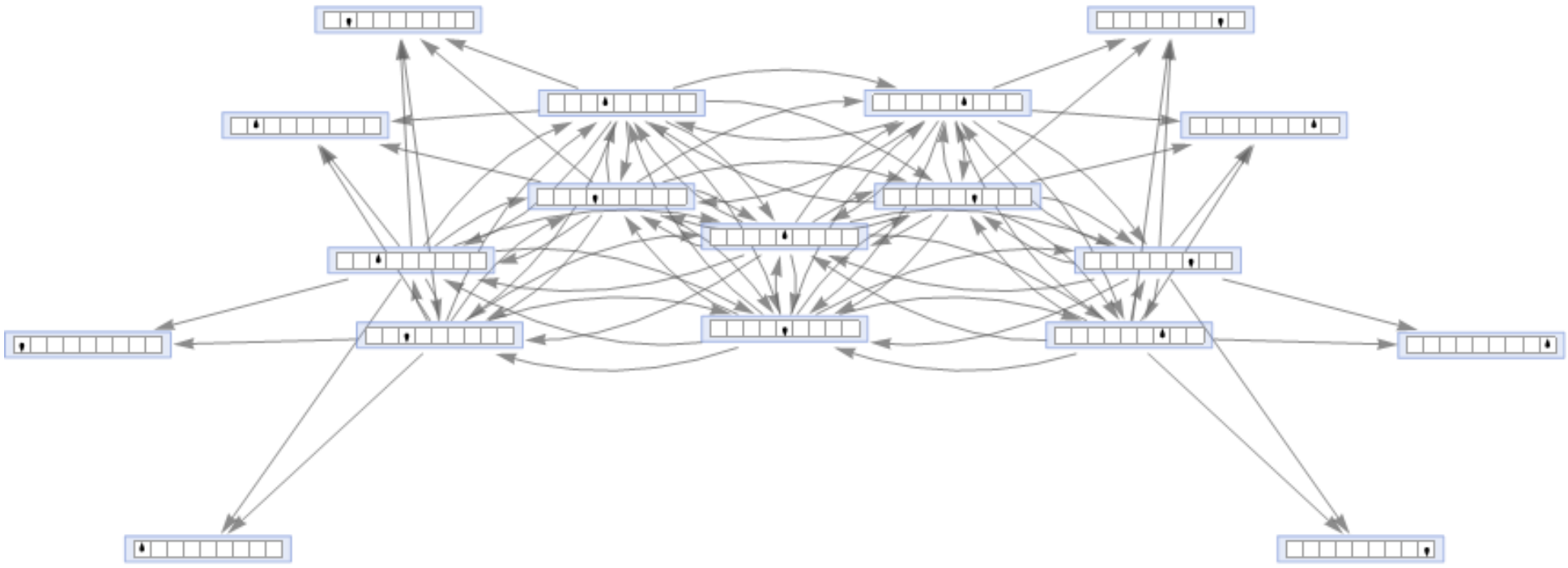

After 3 steps, removing repeated edges, etc. this gives:

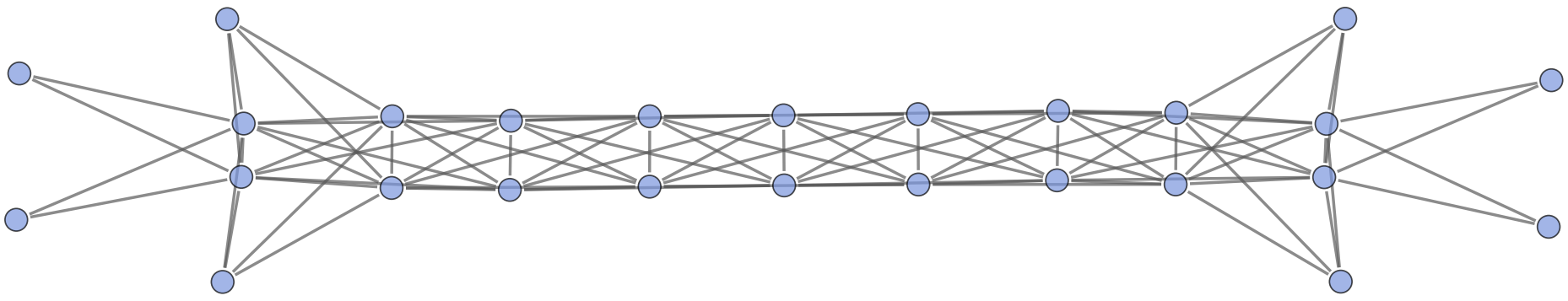

## The Limit of the Rulial Multiway Graph

What is the limiting structure of the rulial multiway graph after an infinite number of steps? The first crucial observation is that it's in a sense homogeneous: the structure of the graph around any given node is always the same (i.e. it's a vertex transitive graph). To see why this is true, recall that each node in the graph corresponds to a particular distinct configuration of the Turing machine. This node will lead to all nodes that can be obtained from it by one step of Turing machine evolution. But (assuming the head always moves ±1 square) there are always exactly 2 $s\ k$ of these. And because we are following all possible "micro-rules", we can think of ourselves as just "blindly overwriting" whatever configuration we had, so that the structure of the graph is the same independent of what configuration we're at.

As a simple example, here's what happens in the $s = 2$, $k = 1$ case, after 3 steps:

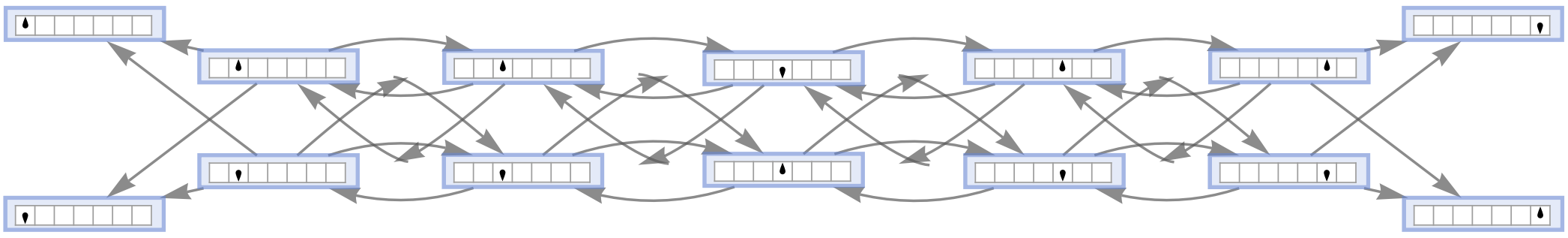

There's some trickiness at the ends, but in the central region we see as expected that each node has exactly 4 successors. If we pick out the subgraph around the blank-tape starting node it has the form:

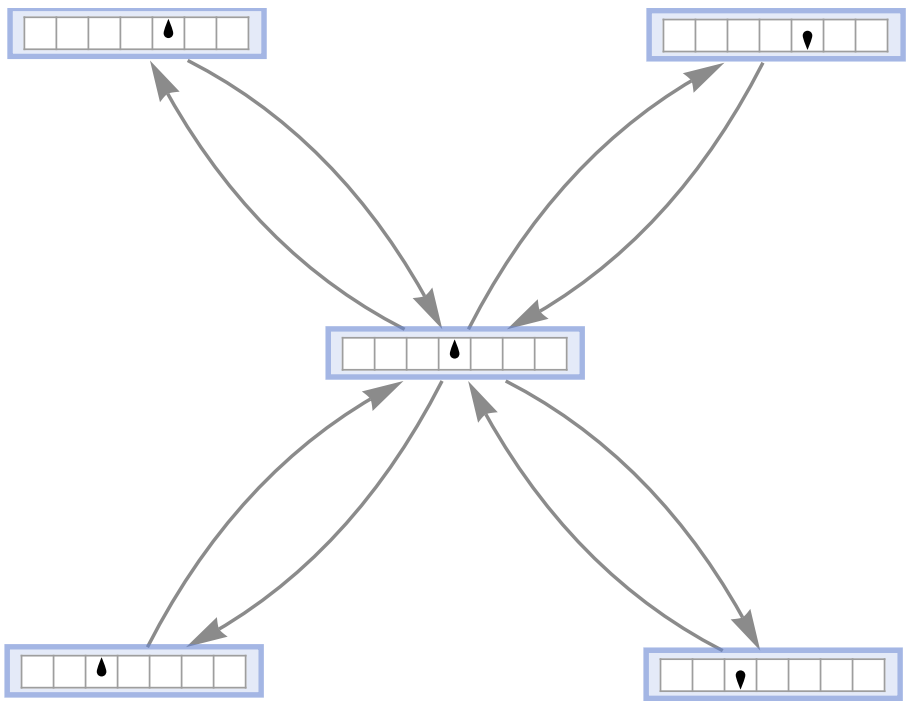

In general, the limiting neighborhood of every node is just:

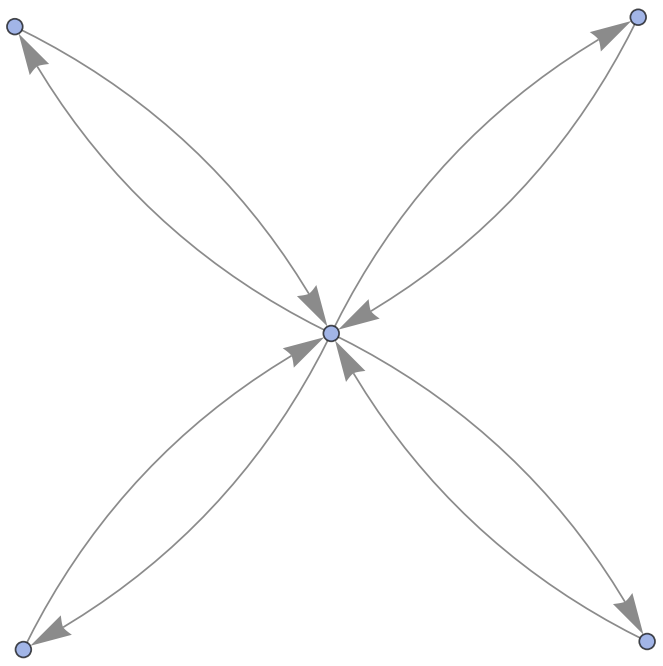

In the full graph, these neighborhoods are knitted together, giving pieces like:

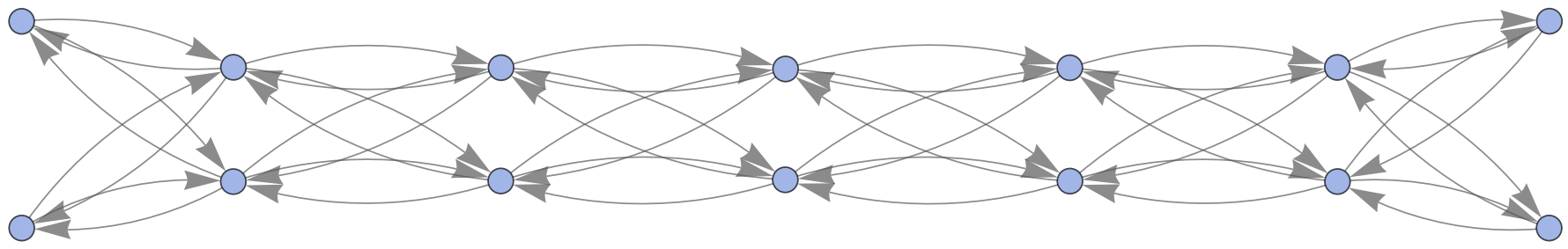

Let's now consider the case $s = 1$, $k = 2$. After 4 steps the full graph is:

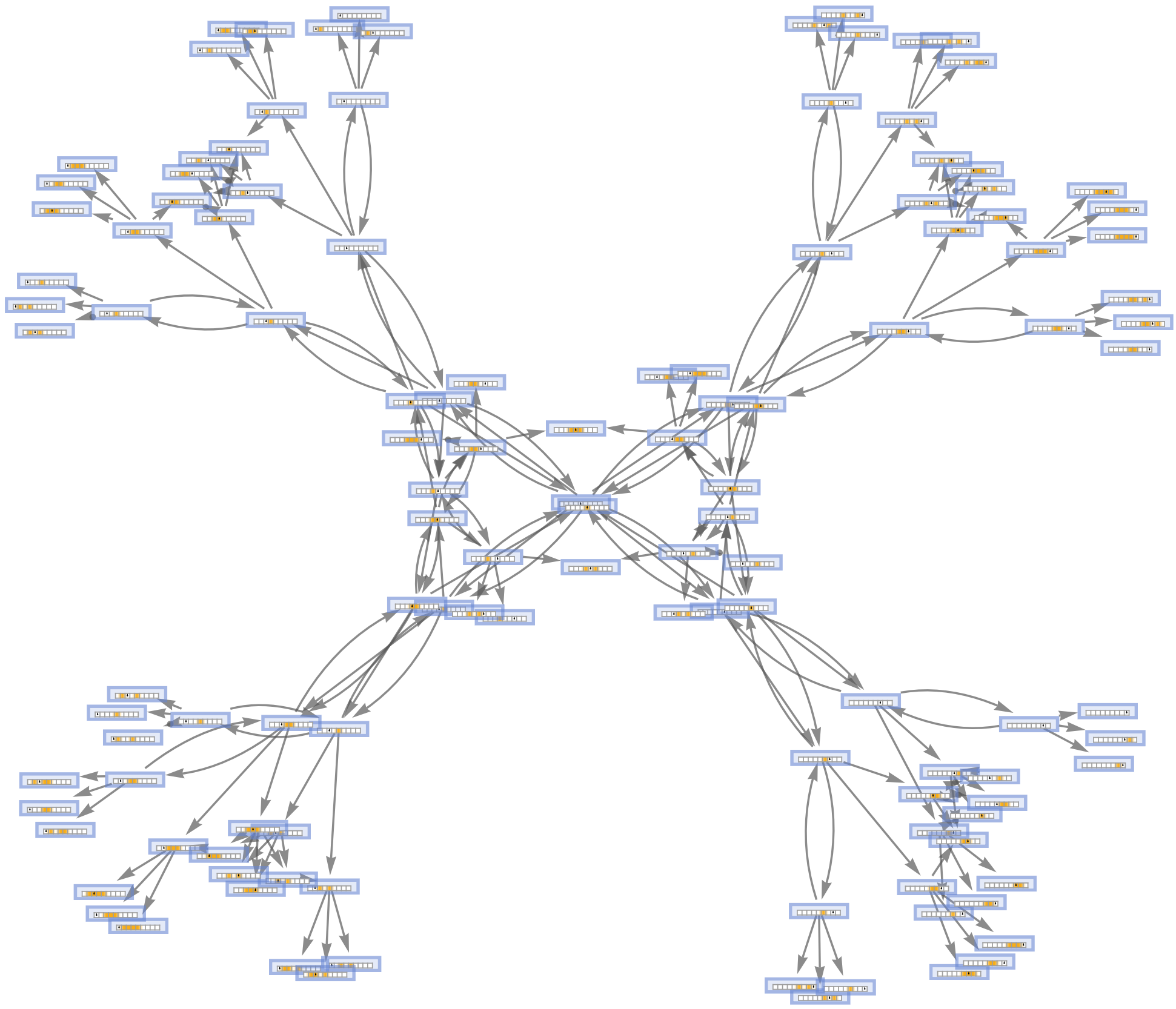

Here the neighborhood of the start node is:

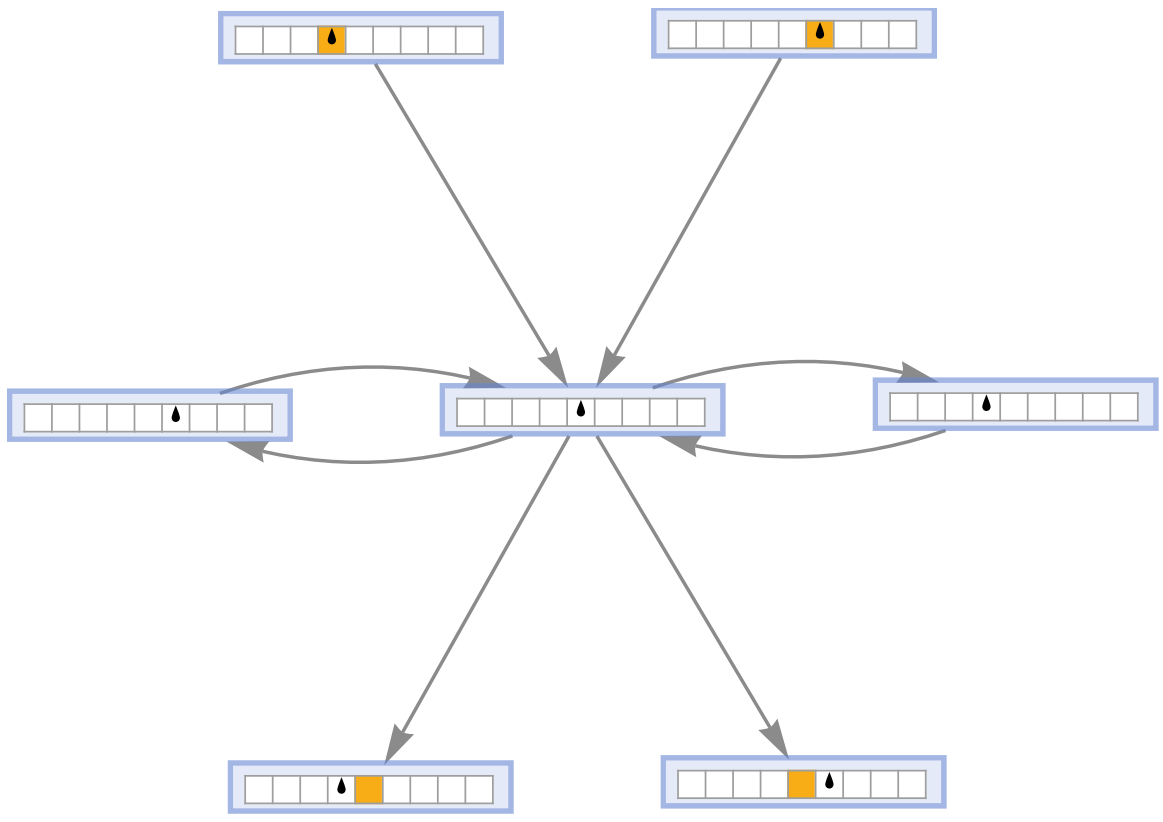

And in the limiting graph, every node will have a local neighborhood with this same structure. If we look at successively larger neighborhoods, the limiting form of these for every node will be:

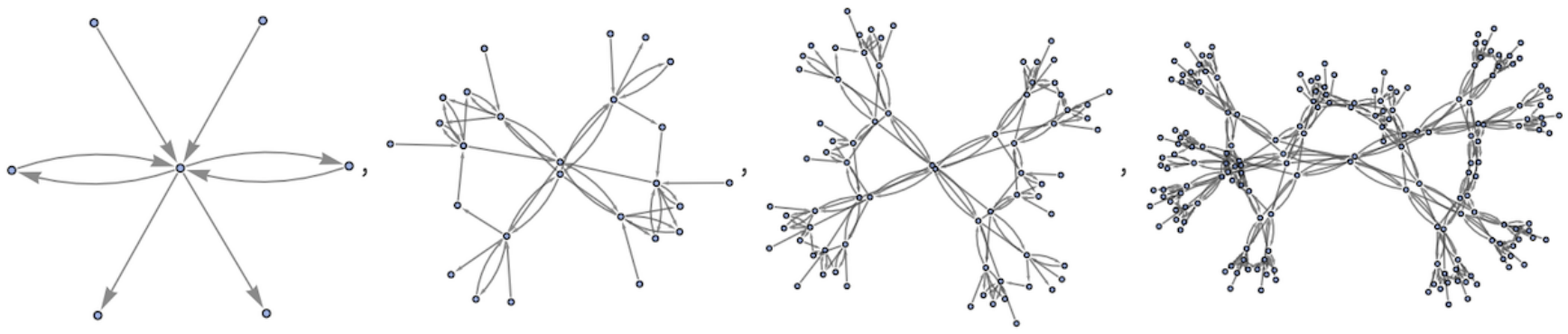

Here's a table of the number of distinct nodes in these successively larger neighborhoods for Turing machines with various values of $s$ and $k$:

| $s$ | $k$ | |
|---|---|---|
| 1 | 1 | 1, 3, 5, 7, 9, 11, 13, 15, 17, 19, 21, 23, 25, 27, 29, 31, … |
| 2 | 1 | 1, 5, 10, 14, 18, 22, 26, 30, 34, 38, 42, 46, 50, 54, 58, 62, … |
| 3 | 1 | 1, 7, 15, 21, 27, 33, 39, 45, 51, 57, 63, 69, 75, 81, 87, 93, … |
| 4 | 1 | 1, 9, 20, 28, 36, 44, 52, 60, 68, 76, 84, 92, 100, 108, 116, 124, … |
| 1 | 2 | 1, 5, 18, 50, 124, 288, 640, 1384, 2928, 6112, 12 608, 25 824, 52 544, 106 496, 215 040, 433 280, … |
| 2 | 2 | 1, 9, 36, 100, 248, 576, 1280, 2768, 5856, 12 224, 25 216, 51 648, 105 088, 212 992, 430 080, … |
| 3 | 2 | 1, 13, 54, 150, 372, 864, 1920, 4152, 8784, 18 336, 37 824, 77 472, 157 632, 319 488, … |
| 4 | 2 | 1, 17, 72, 200, 496, 1152, 2560, 5536, 11 712, 24 448, 50 432, 103 296, … |
| 1 | 3 | 1, 7, 39, 153, 543, 1809, 5787, 18 117, 55 755, 170 181, 515 727, 1 557 873, … |
| 2 | 3 | 1, 13, 78, 306, 1086, 3618, 11 574, 36 234, 111 510, 340 362, … |
| 3 | 3 | 1, 19, 117, 459, 1629, 5427, 17 361, 54 351, 167 265, … |

In the limit $t \to \infty$ the number of nodes reached in all cases goes like:

$$M_t \sim k^t$$

For $k = 1$, one finds (for $t > 1$):

$$M_t = (2\,t + 1)\,s$$

For $t = 1$ one has:

$$M_1 = 2\,s\,k + 1$$

If one ignores directedness in the graph, and just counts the total number of neighbors out to distance $t$, the results for $k = 1$ are the same as before, but in other cases they are different:

| $s$ | $k$ | |
|---|---|---|
| 1 | 2 | 1, 7, 26, 74, 180, 412, 900, 1924, 4028, 8348, 17 116, 34 908, 70 780, … |
| 2 | 2 | 1, 13, 52, 148, 360, 824, 1800, 3848, 8056, 16 696, 34 232, 69 816, … |
| 3 | 2 | 1, 19, 78, 222, 540, 1236, 2700, 5772, 12 084, 25 044, 51 348, … |
| 4 | 2 | 1, 25, 104, 296, 720, 1648, 3600, 7696, 16 112, … |
| 1 | 3 | 1, 11, 63, 273, 975, 3273, 10 443, 32 685, 100 443, … |
| 2 | 3 | 1, 21, 126, 546, 1950, 6546, 20 886, … |
| 3 | 3 | 1, 31, 189, 819, 2925, 9819, … |

(These results asymptotically seem larger by a factor $\frac{k^2}{2\,k-1}$.)

So for $k = 1$, the limiting rulial multiway graph behaves like a 1-dimensional space, but for all $k \geq 2$, it behaves like an infinite-dimensional space, in which the volumes of geodesic balls grow exponentially with volume.

## Finite Tapes

So far we've always assumed that our Turing machine tape is unbounded. But in some ways it's easier to see what's going on if we instead limit the tape, so we have a finite total number of states ($n\,s\,k^n$). One thing we can do is to consider a cyclic tape.

For $s = k = 1$ with successively longer tapes we then get:

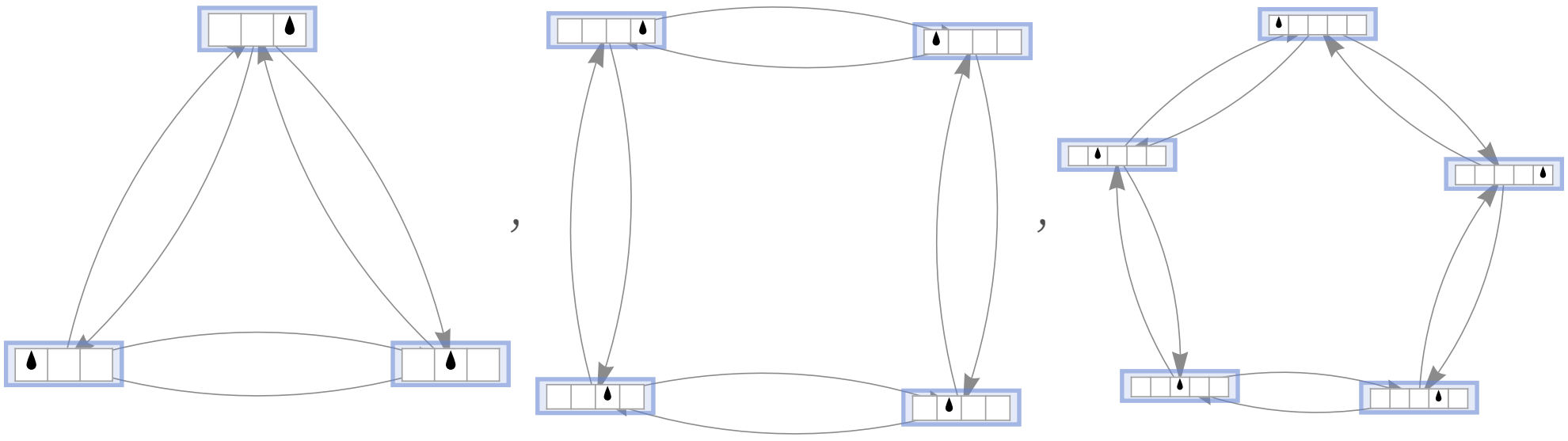

For $s = 2$, $k = 1$ we get:

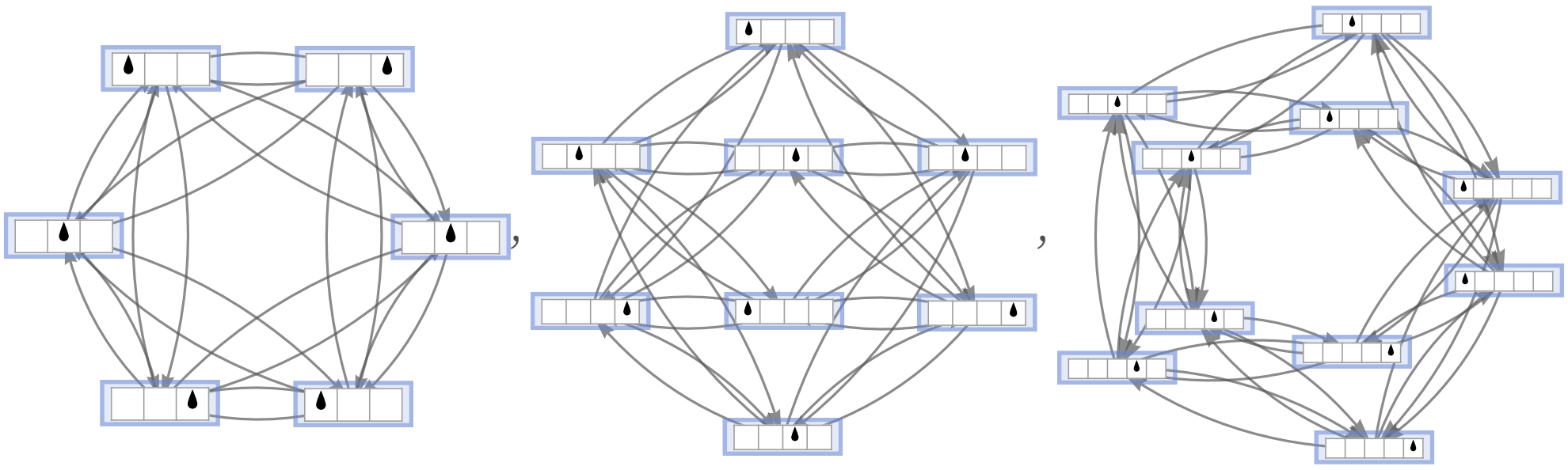

For tape size 10, the results for successive values of $s$ are:

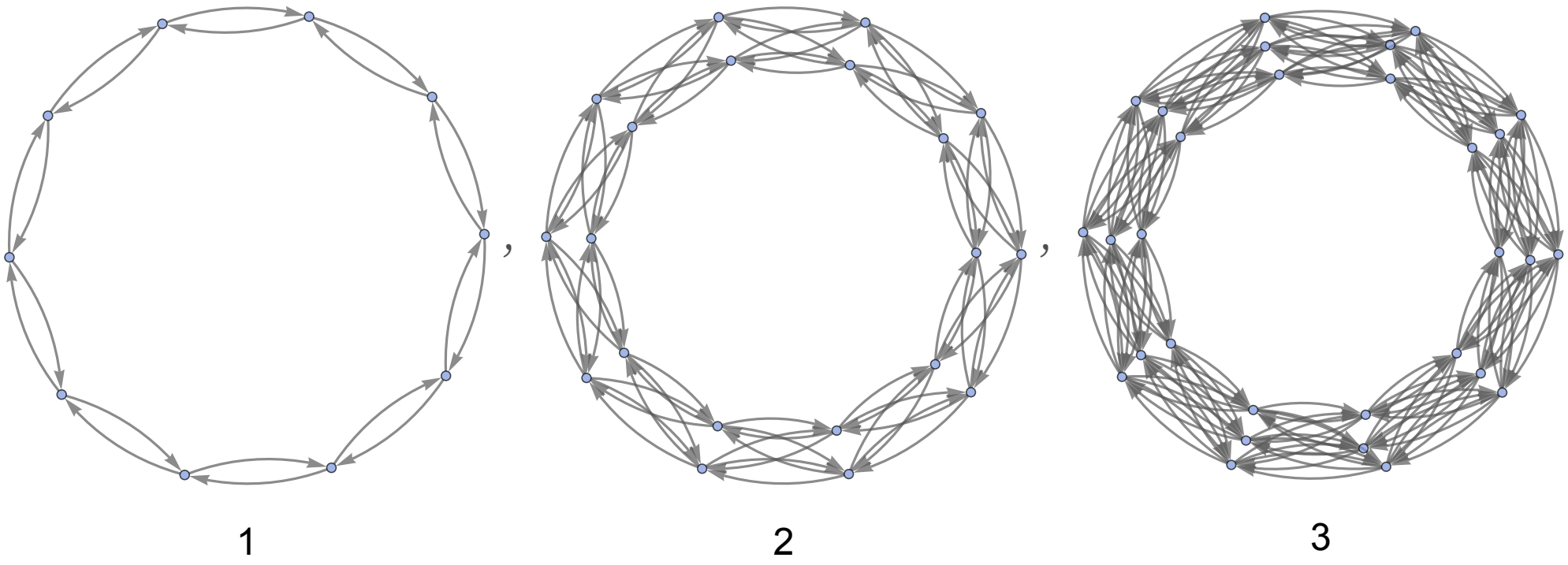

If instead of having a cyclic tape, we just have a finite tape, and insist that the head never goes off either end, then for $s = k = 1$ we get:

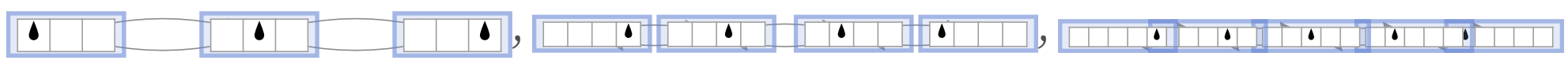

For $s = 2$, $k = 1$ we get

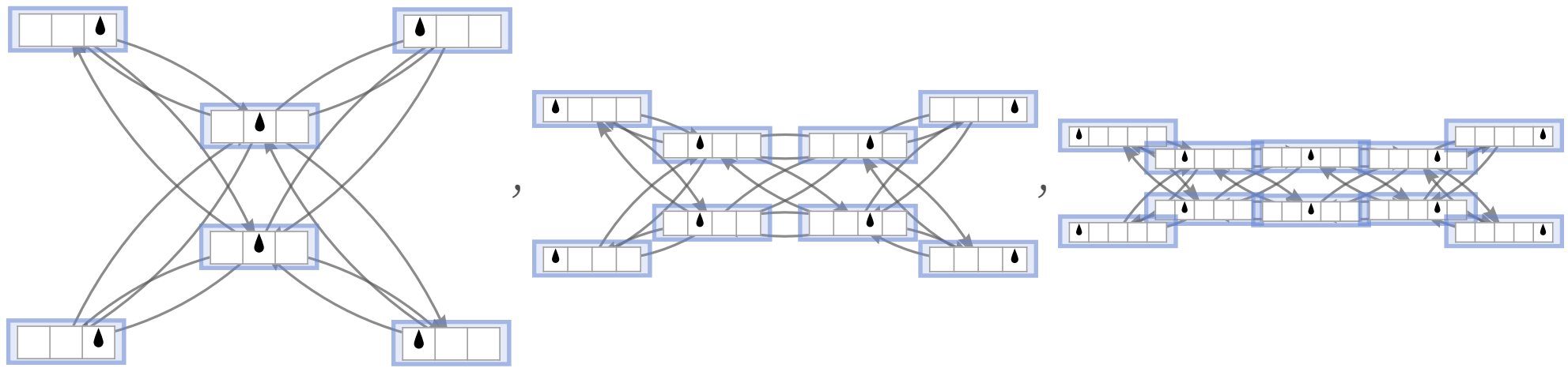

and the overall behavior for successive $s$ is exactly as we saw above for unbounded tapes:

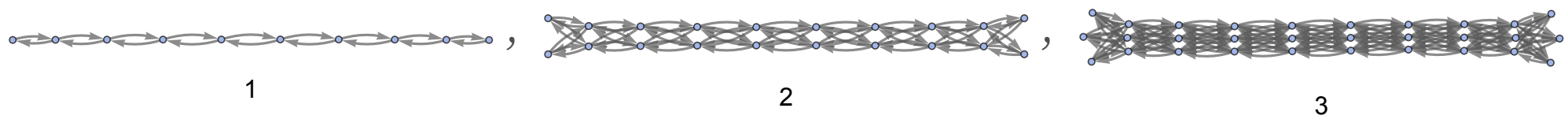

For $s = 1$, $k = 2$, this is what happens in the cyclic case for length 3:

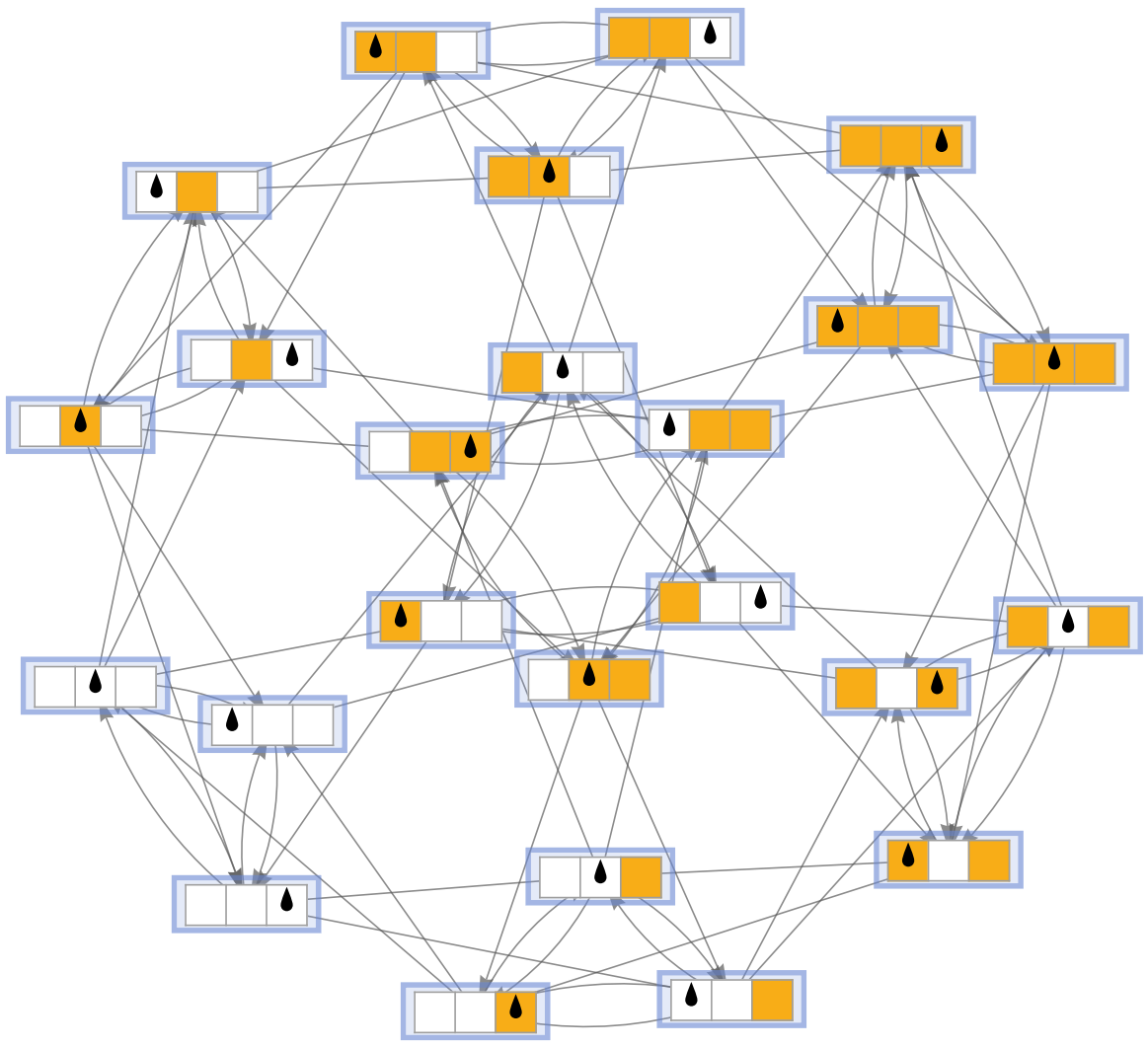

The results for lengths 1 through 6 are:

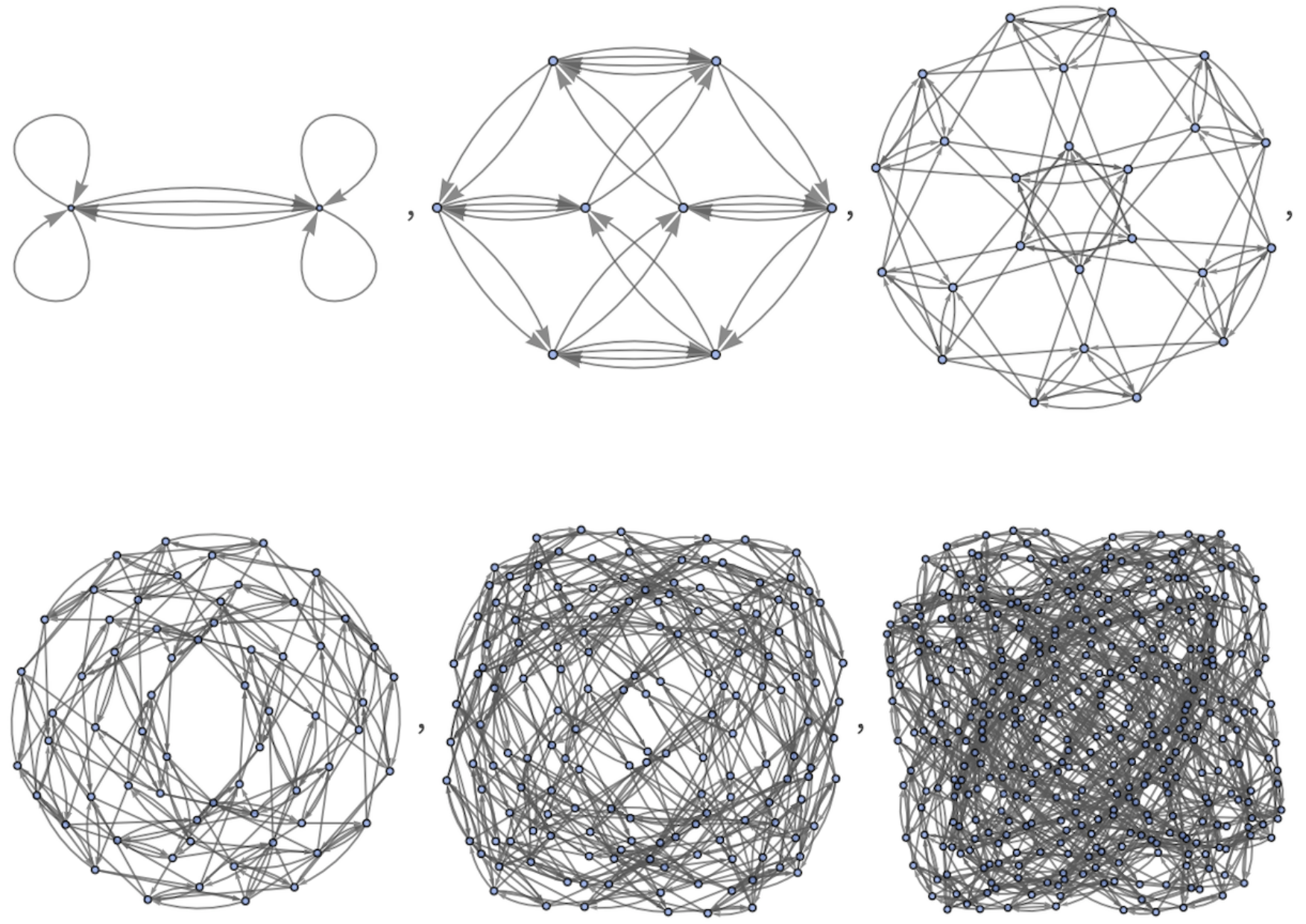

Rendering these in 3D makes the connection with progressively higher-dimensional hyper-cubes slightly clearer:

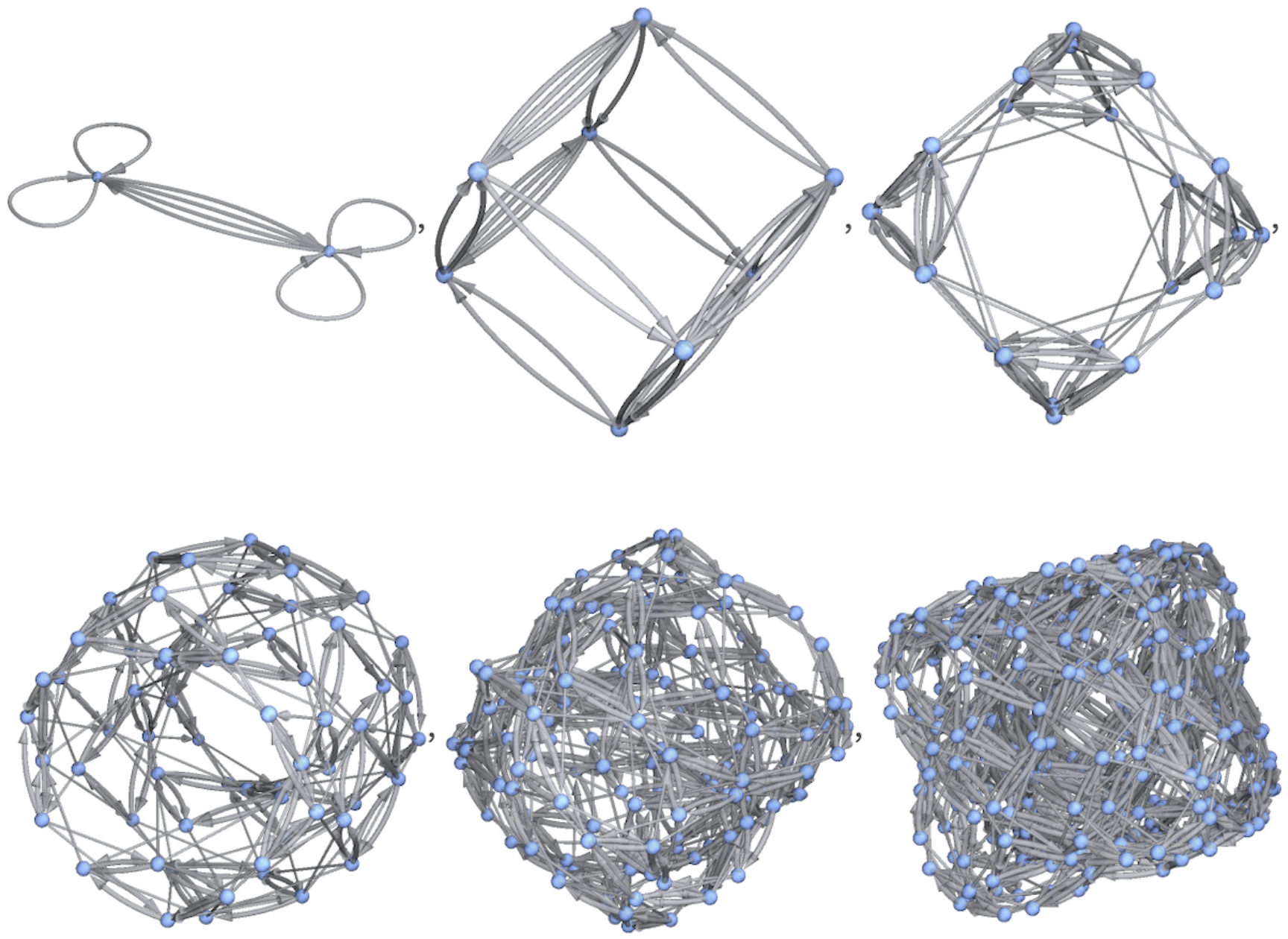

The non-cyclic case for lengths 2 through 6 (length 1 is trivial):

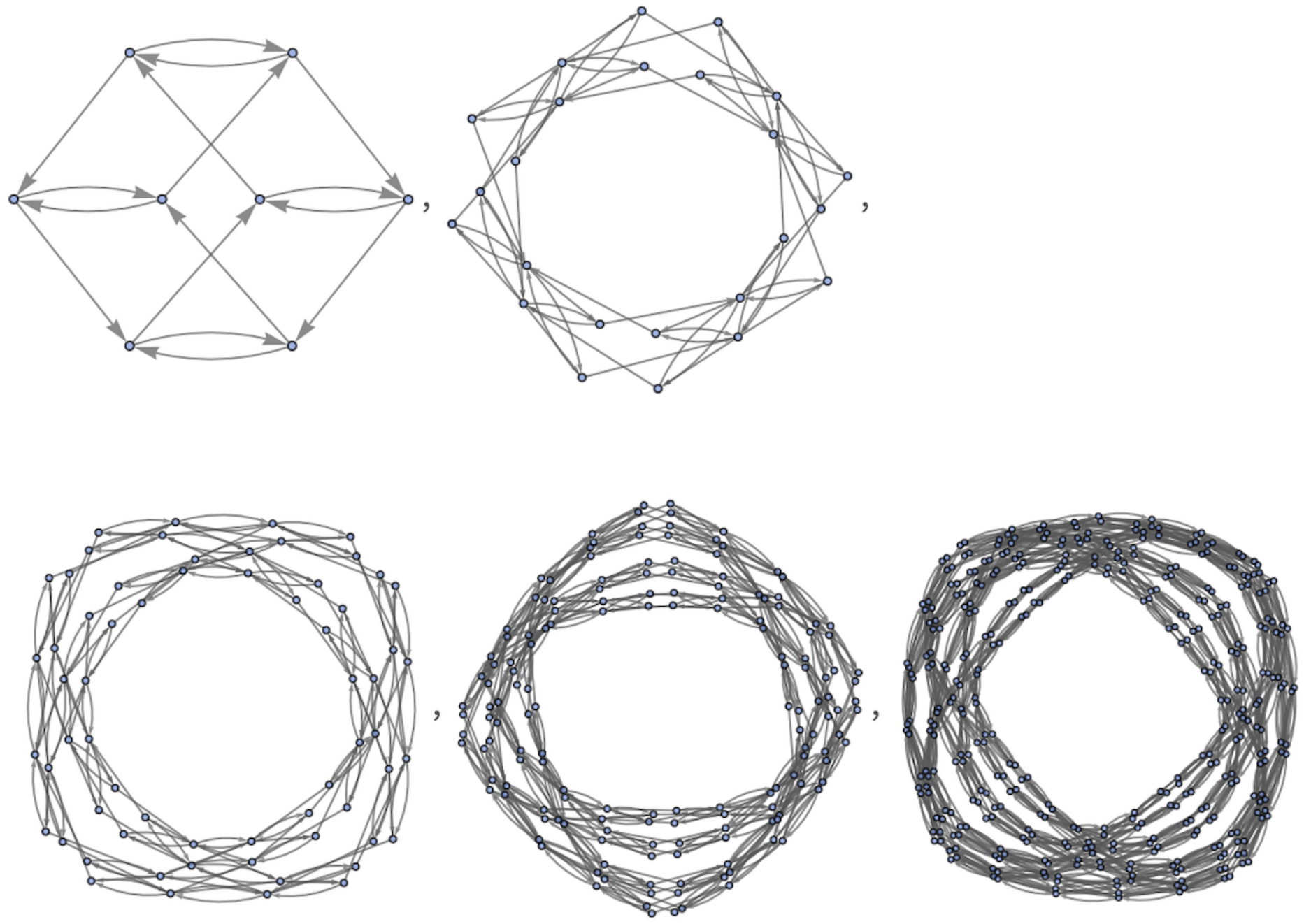

Rendered in 3D these become:

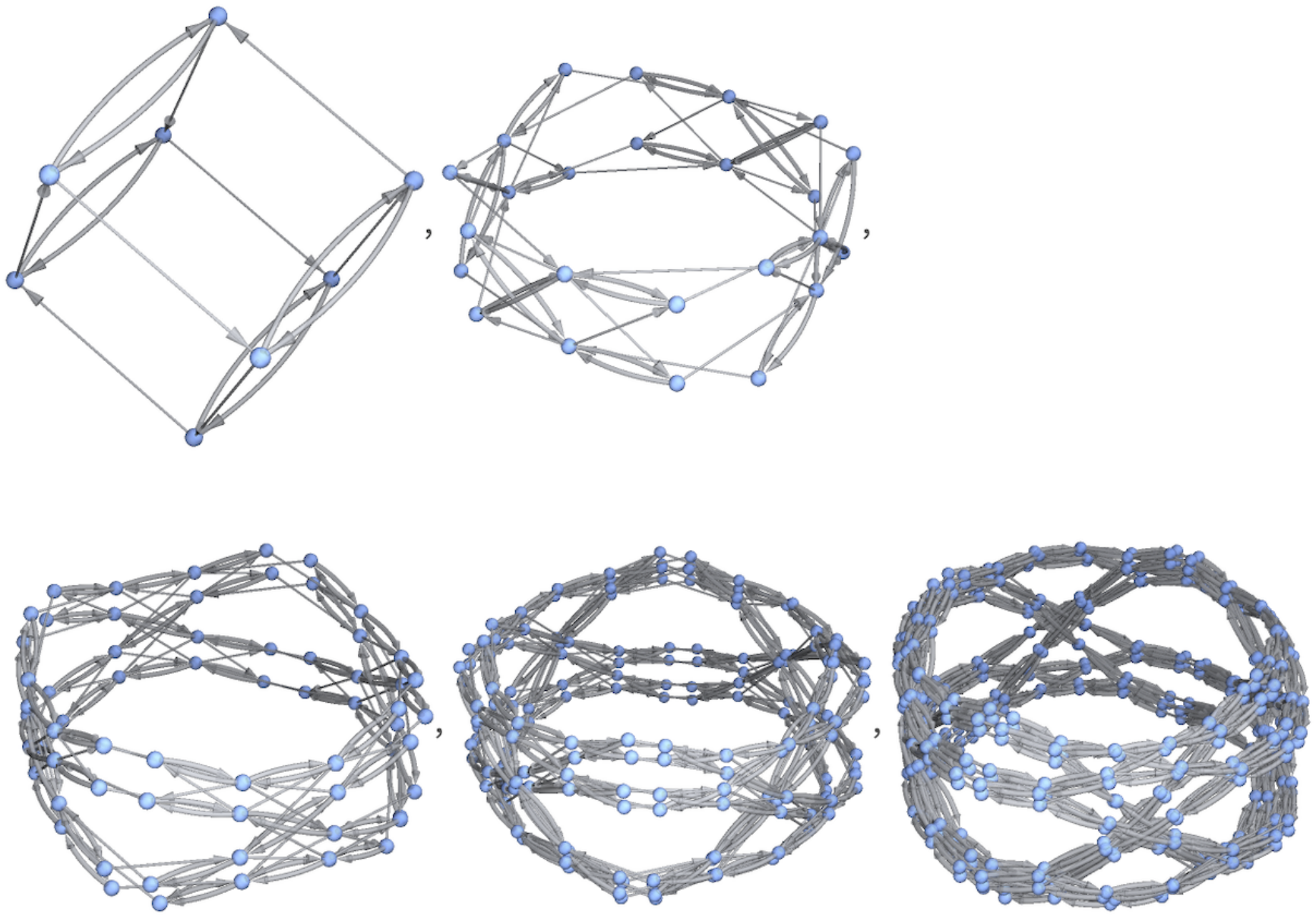

In the case $s = 2$, $k = 2$ for a length-3 cyclic tape we get:

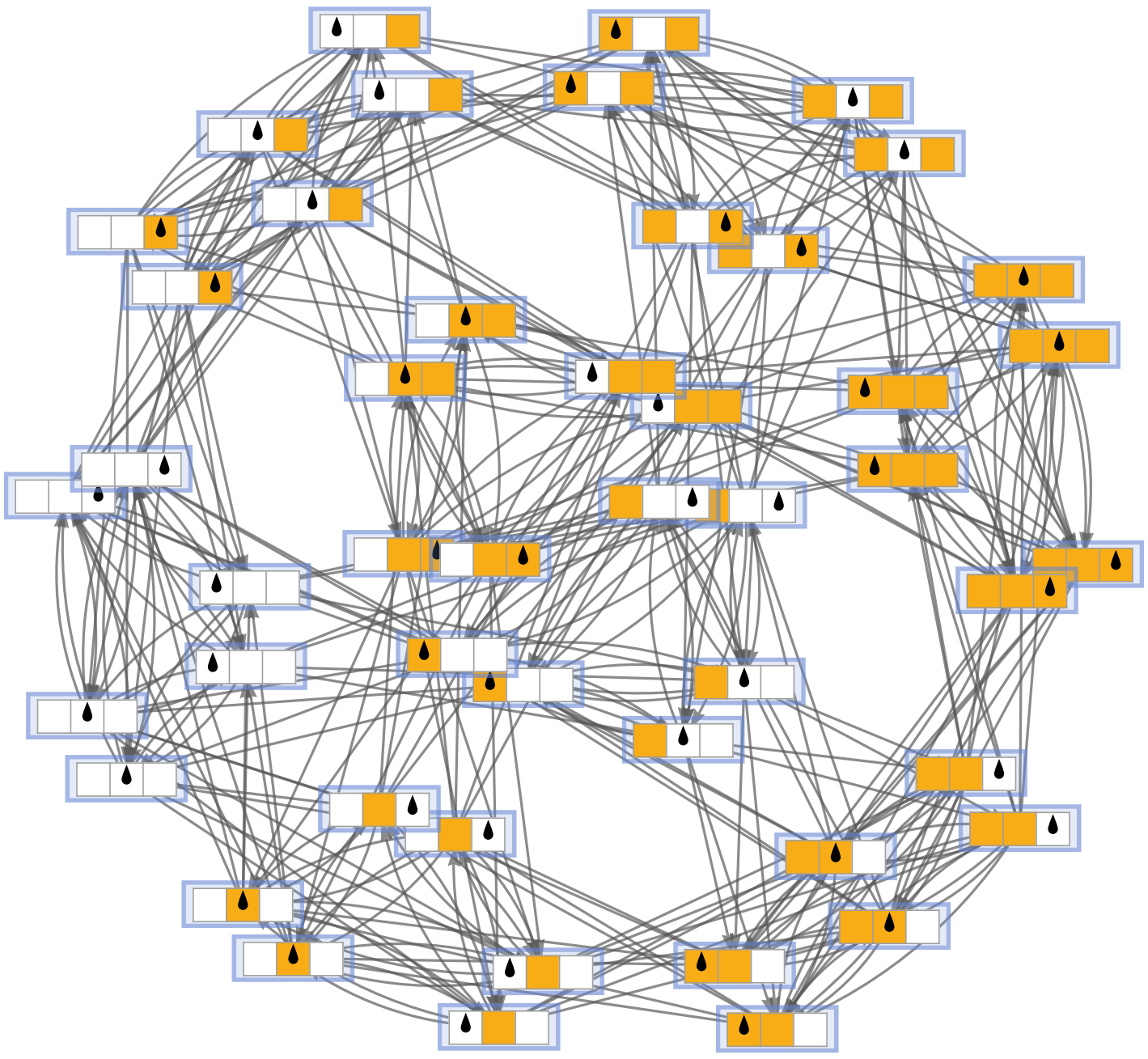

Rendering the results for lengths 1 through 6 in 3D gives:

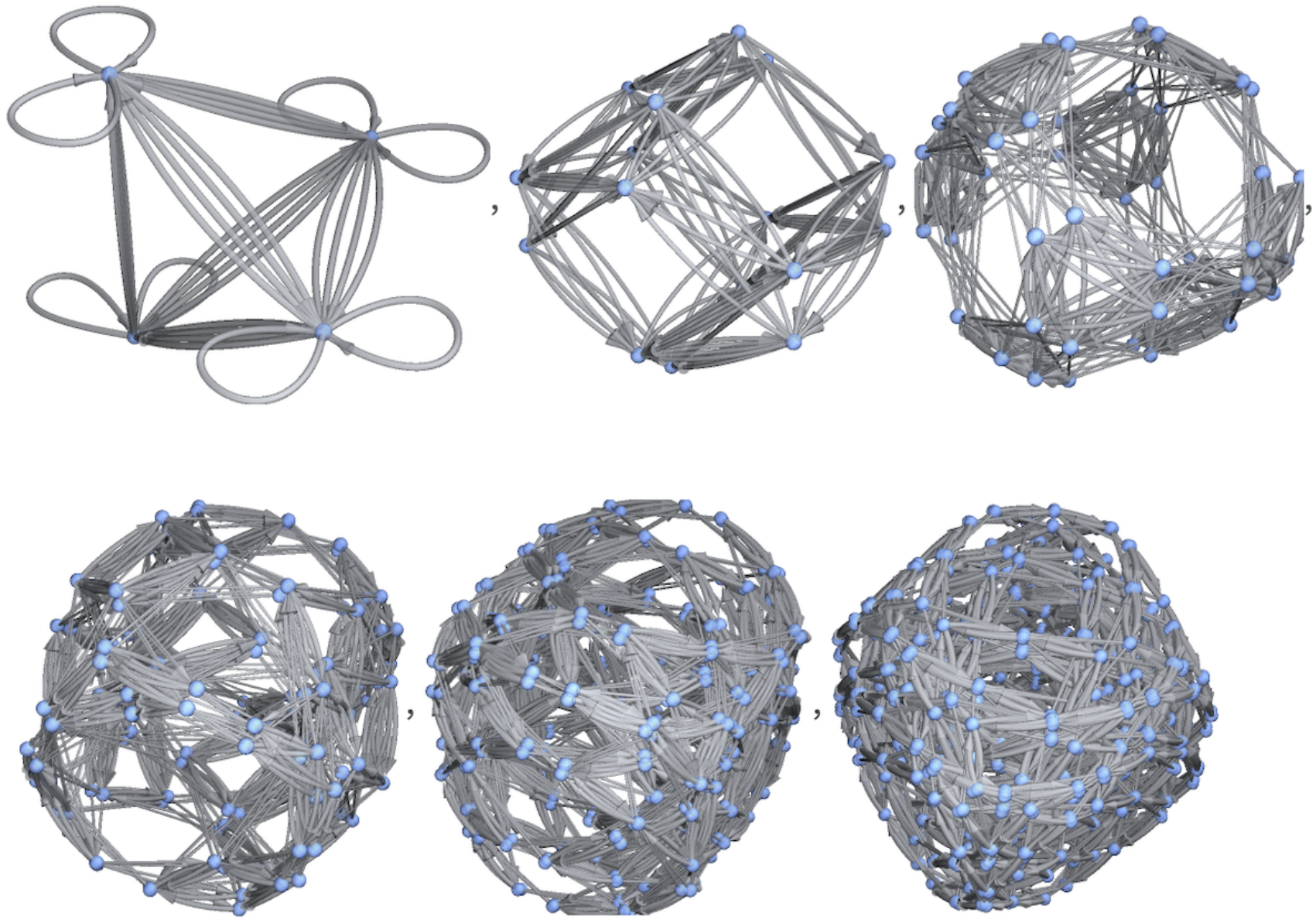

In the non-cyclic case the results for lengths 2 through 6 in 3D are (the length-1 case is trivial):

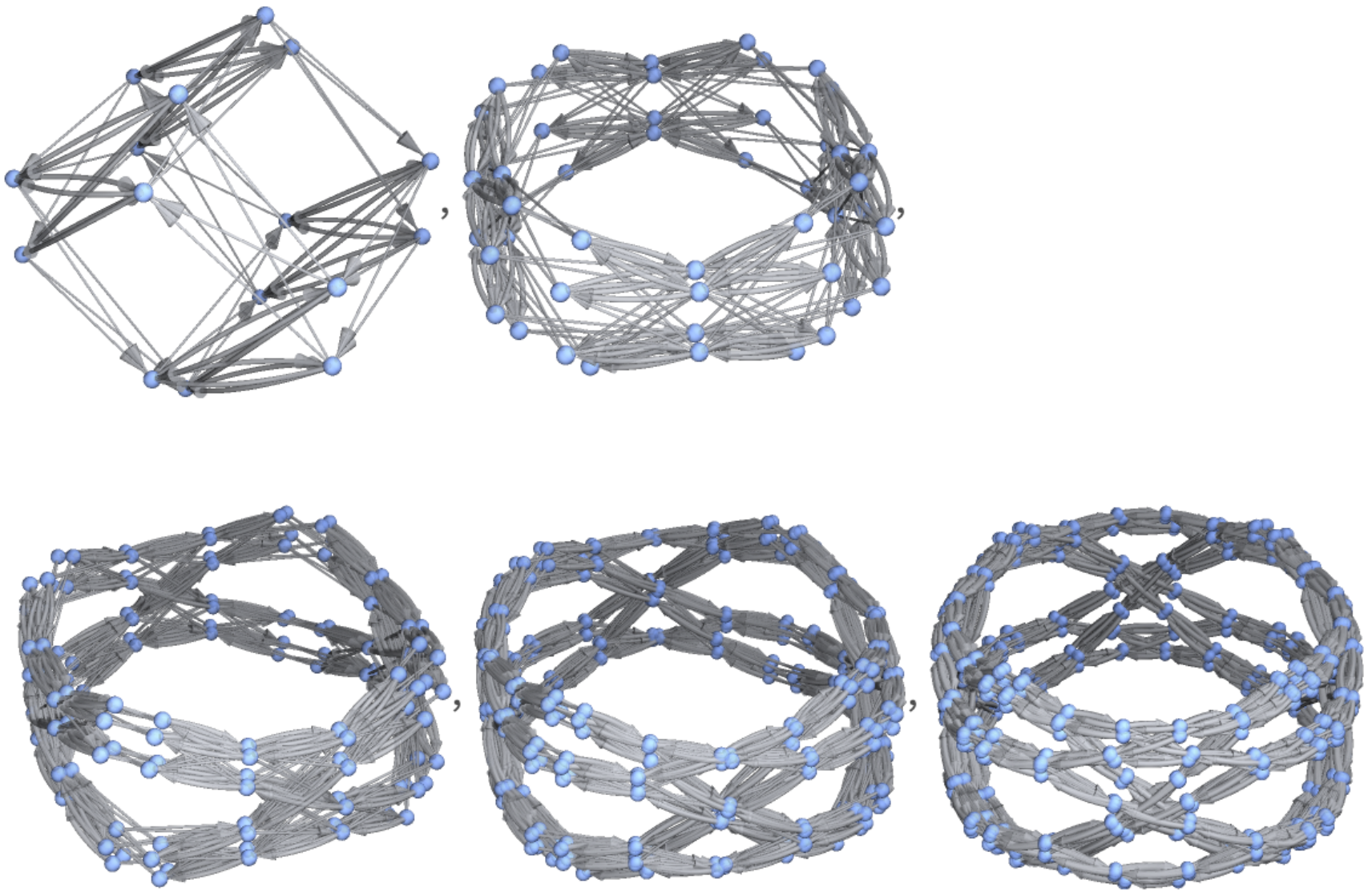

## The Turing Machine Group

It turns out that there's a nice mathematical characterization of rulial multiway graphs for Turing machines: they're just Cayley graphs of groups that we can call "Turing machine groups". Why is this? Basically it's because the possible configurations of a Turing machine have a direct correspondence with transformations that can act on these configurations. And in particular, one can pick out certain transformations that correspond to individual transitions in a non-deterministic Turing machine, and use these as generators in a presentation of the group.

Let's start by considering the case of Turing machines with finite cyclic tapes. If the tape has length $n$, the total number of possible configurations of the machine is $n\, s\, k^n$. (Note that if the tape is finite but not cyclic then one doesn't get a group.)

Assume for now $s = 1$, $k = 2$. Then for $n = 3$, there are 24 possible configurations, and the rulial multiway graph is:

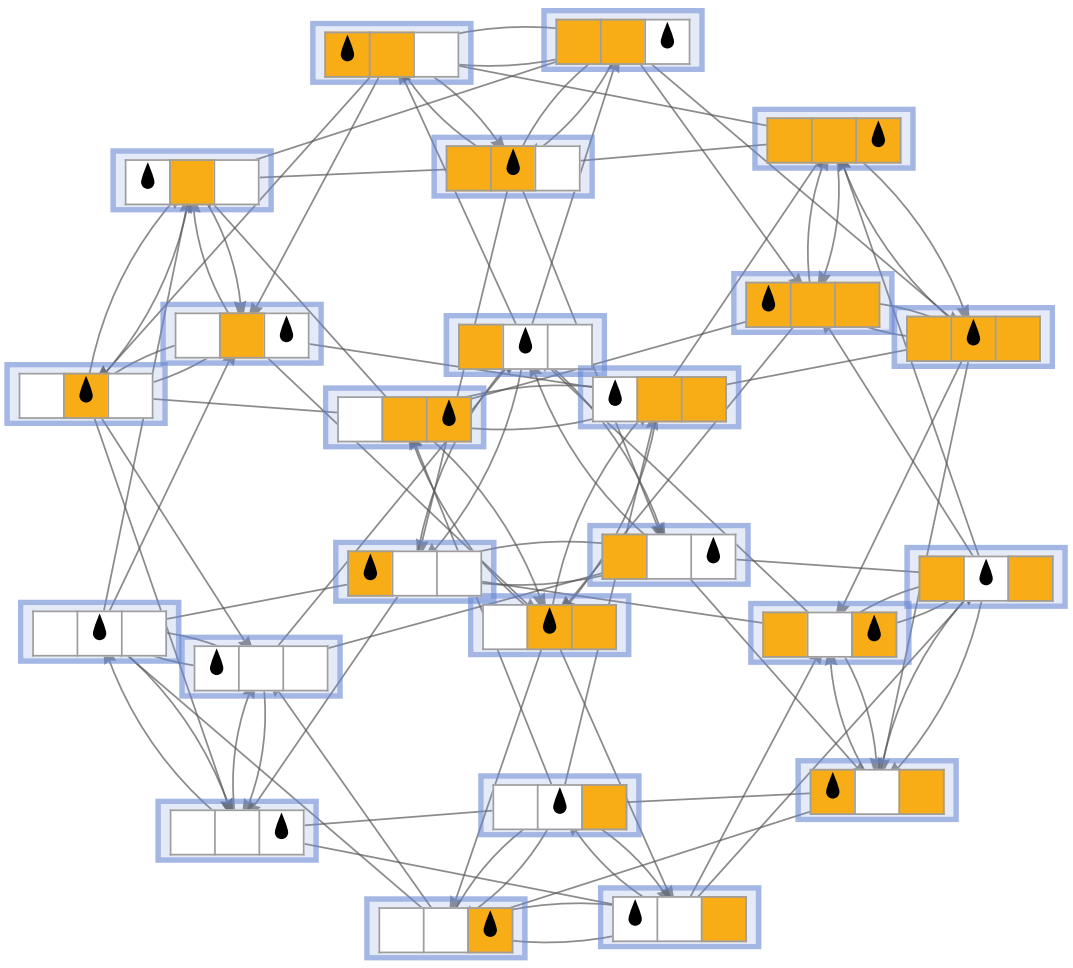

But this graph turns out to be a Cayley graph for the finite group $A_4 \times \mathbb{Z}_2$:

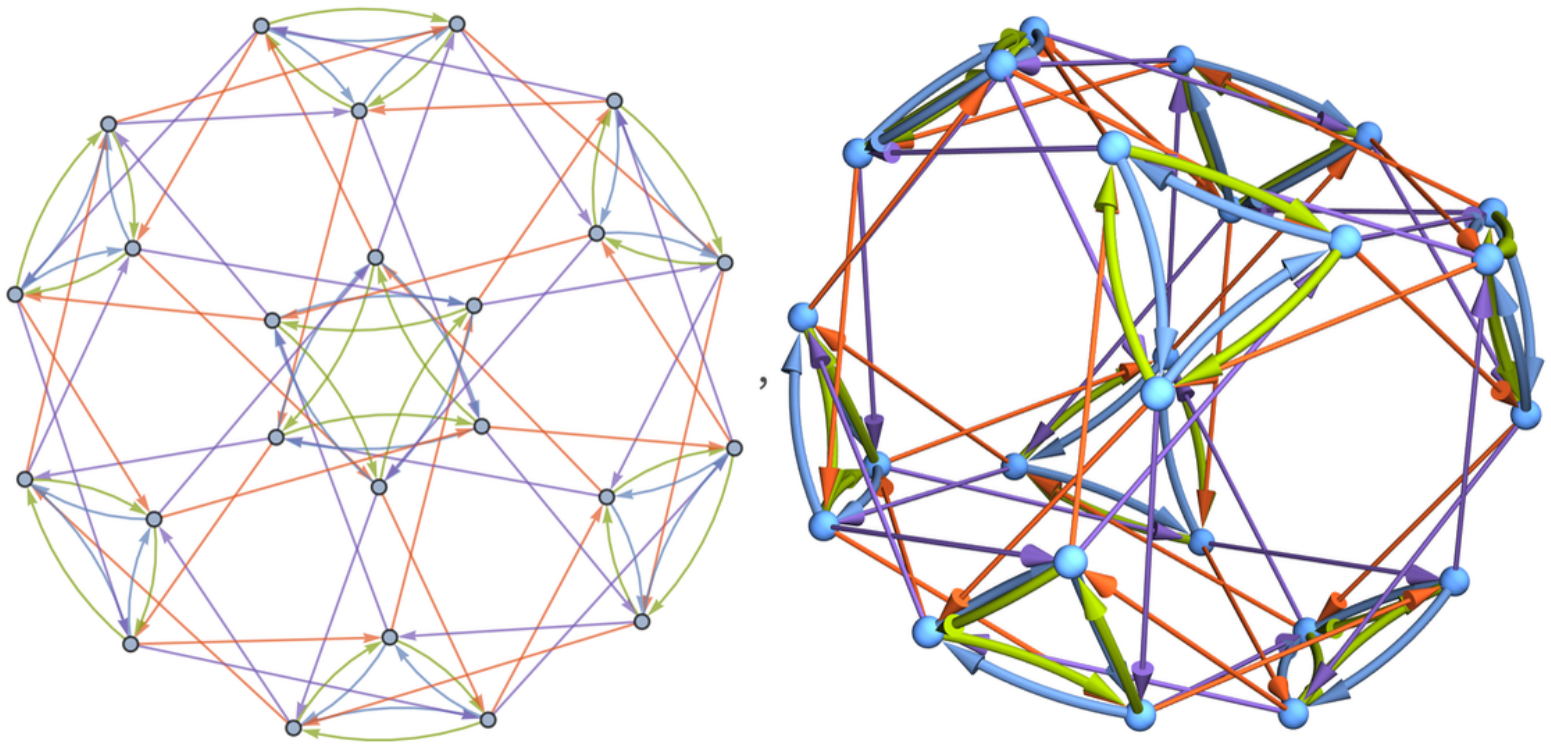

To construct this Cayley graph let's represent the configurations of the Turing machine by pairs of integers $\{i,u\}$, where $0 \leq i \leq n-1$ gives the position of the head, and the bits of $u$ (with $0 \leq u \leq 2^n - 1$) give the values on the tape. (Since $s = 1$, we don't have to worry about the state of the head.) With this representation of the configurations, consider the "multiplication" operation:

```
f[{i_, u_}, {j_, v_}] := {Mod[i + j, n], BitXor[BitShiftRight[u, j], v]}
```

When this operation acts on the configurations it defines a group. Here's the multiplication table for $n = 3$:

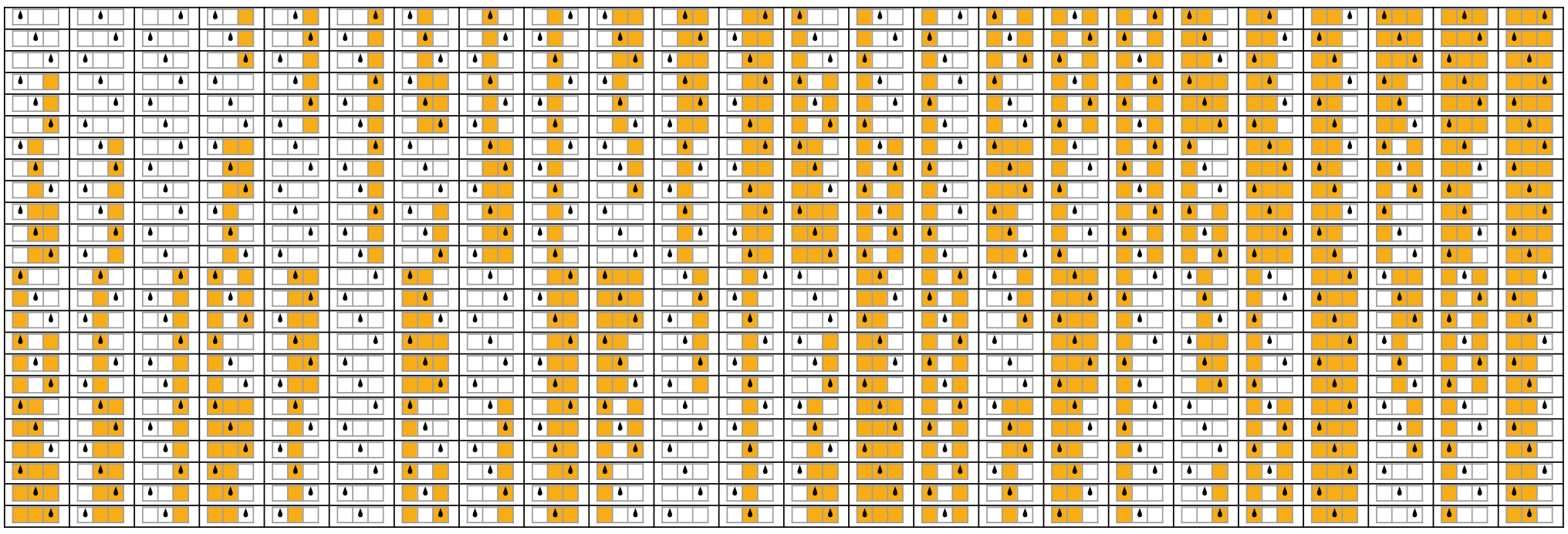

Or equivalently:

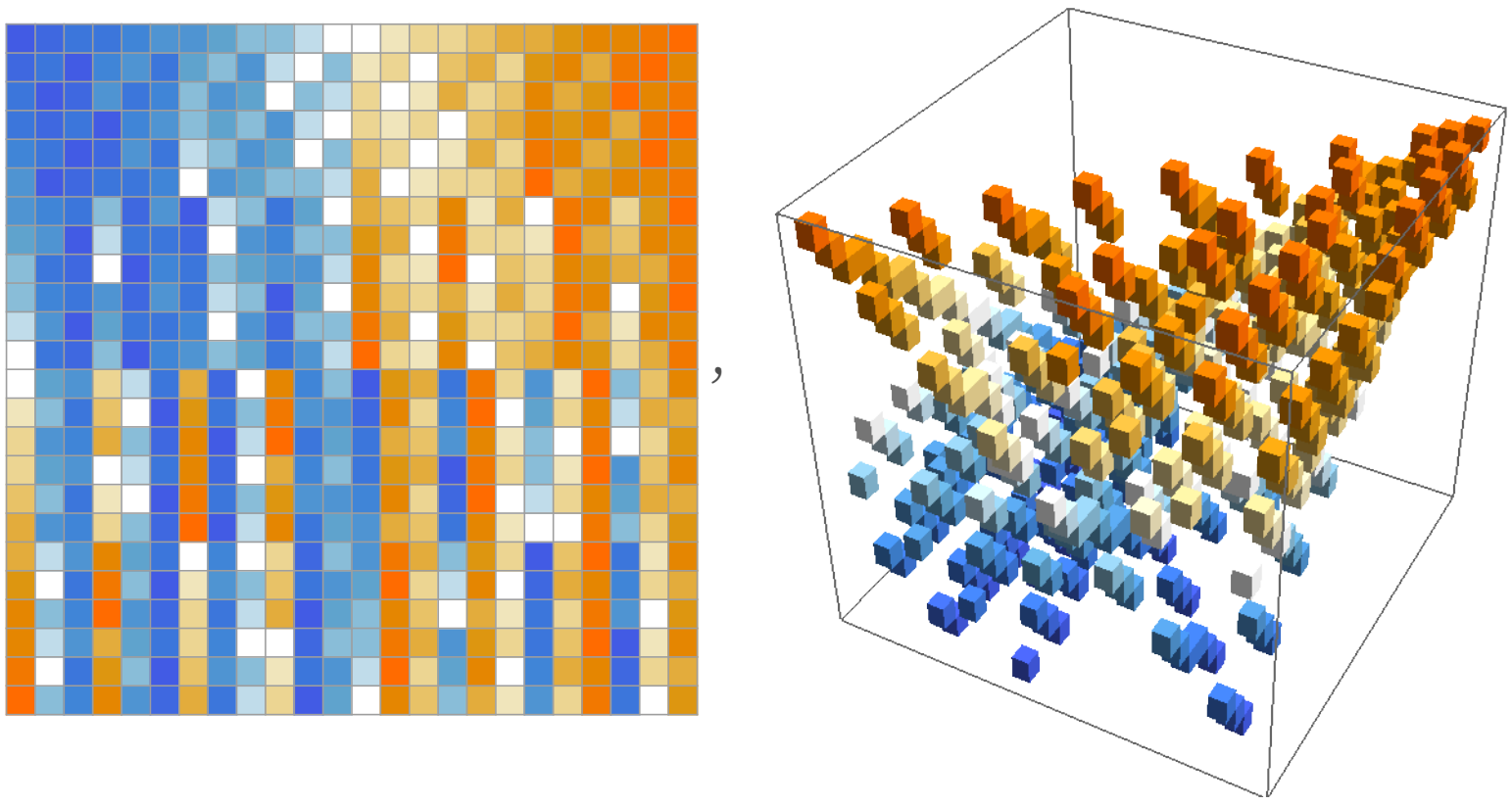

But now consider the four elements {±1,0}, {±1,1} (position –1 wraps around on the cyclic tape):

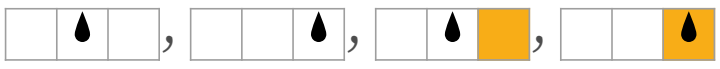

We can consider these as the result of applying the 4 possible Turing machine transitions

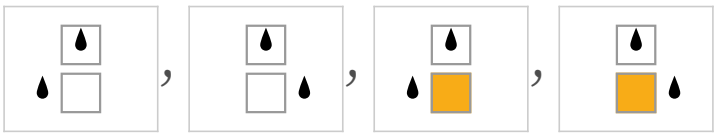

to the configuration (corresponding to {0,0}):

And now by treating these elements as generators (and effectively applying them not just to the initial configuration, but to any configuration), we get as the Cayley graph of the group exactly the rulial multiway graph above.

It’s worth noting that the 4 elements we’ve used don’t correspond to the minimal set of generators for the group. Two elements suffice. And for example, we can use {1,0} and {0,1}, which can be thought of respectively as moving the head right, and flipping the color of one cell (again relative to the “identity” configuration {0,0}):

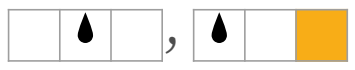

With these generators, we get the Cayley graph:

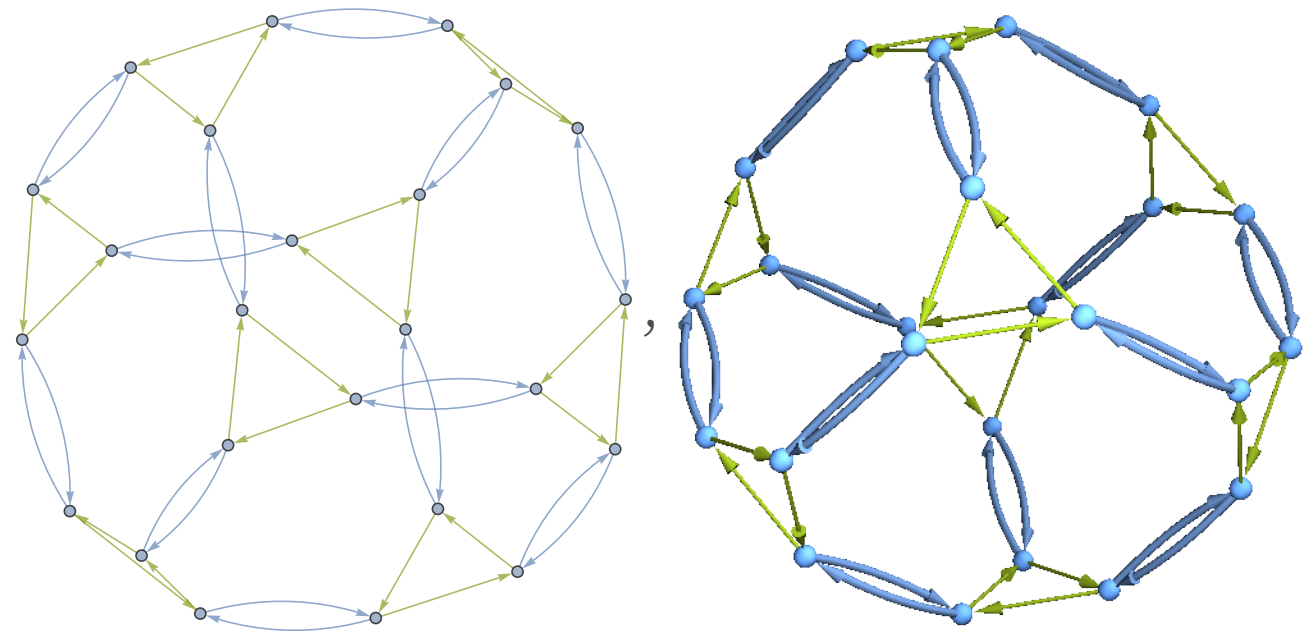

How else can we characterize the group? For any tape size $n$ we can write it in terms of explicit permutations:

```
PermutationGroup[{Cycles[{{1, 2}}], Cycles[{Range[1, 2 n – 1, 2], Range[2, 2 n, 2]}]}]
```

(For $n = 3$, the group can be generated by the permutations {{2,1},{3,4,5,6,1,2}}.)

We can also represent it symbolically in terms of generators and relations. Calling our “move-right” generator $R$, and “bit-flip” generator $F$, the group then satisfies at least the relations:

$$R^n = F^2 = 1,\ RFR^{-1}F = FRFR^{-1}$$

OK, so does the group have a name? For $n = 2$, it’s the 8-element dihedral group $D_4$ and for $n = 3$, it’s the 24-element group $A_4 \times \mathbb{Z}_2$. For larger $n$, there doesn’t seem to be a standard name. But given our derivation we can just call it $\mathrm{TM}^n_{1,2}$. And we can express it as a semidirect product (or wreath product):

$$\mathrm{TM}^n_{1,2} = \mathbb{Z}_n \ltimes (\mathbb{Z}_2)^n$$

The normal subgroup $(\mathbb{Z}_2)^n$ here represents states of the tape, and corresponds to a Boolean $n$-cube. The cyclic group $\mathbb{Z}_n$ represents the position of the head, and acts on the Boolean $n$-cube by rotating its coordinates.

For any $n$, we can use just two generators, producing the sequence of Cayley graphs:

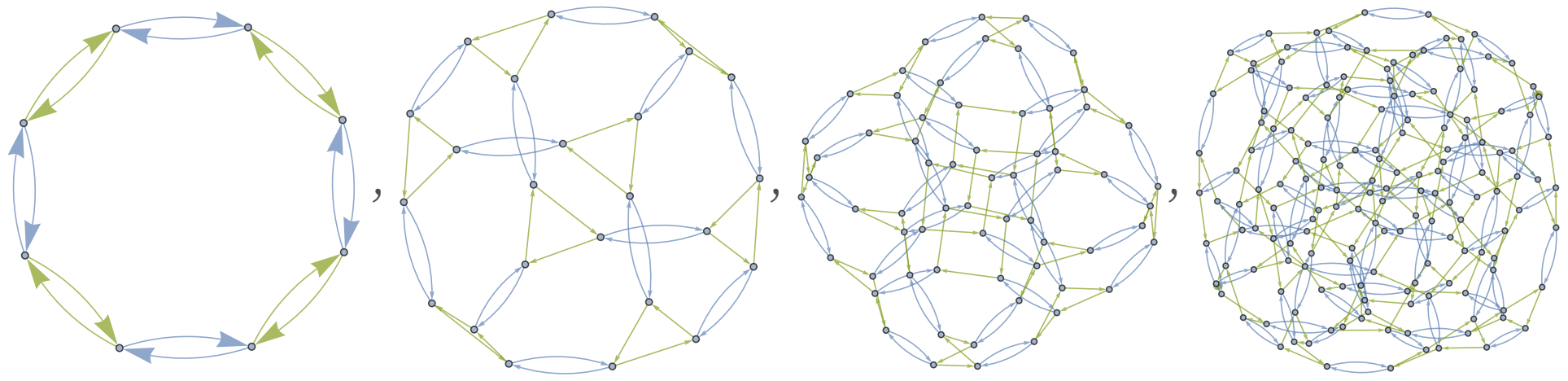

Undirected versions of these are exactly the cube-connected cycle graphs that have arisen in studying communications networks:

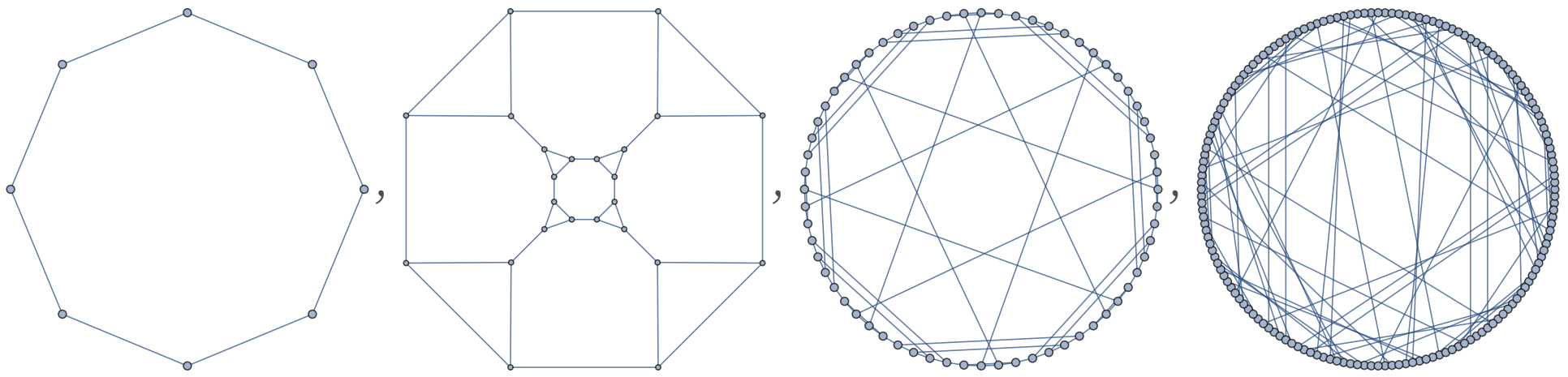

So now what about the limit $n \to \infty$? Now, the group is no longer finite, but we've still got the relations

$F^2 = 1,\ RFR^{-1}F = FRFR^{-1}$

and in the end, we can see that the group can be described as a semidirect product:

$\mathrm{TM}_{1,2} = \mathbb{Z} \ltimes (\mathbb{Z}_2)^{\mathbb{Z}}$

In a sense, the group—and its Cayley graph—are dominated by the infinite-dimensional Boolean hypercube. But there's more going on. And perhaps there's a useful characterization of the limit than can be derived by the methods of modern geometric theory.

For arbitrary $s$ and $k$, we can potentially generalize to get:

$\mathrm{TM}_{s,k} = (\mathbb{Z}_s \times \mathbb{Z}) \ltimes (\mathbb{Z}_k)^{\mathbb{Z}}$

# Causal Graphs for Deterministic Turing Machines

In a deterministic Turing machine, every step involves one updating event—and the causal graph can be drawn by just joining successive locations of the head, and successive points where the head returns to a square where it has been before:

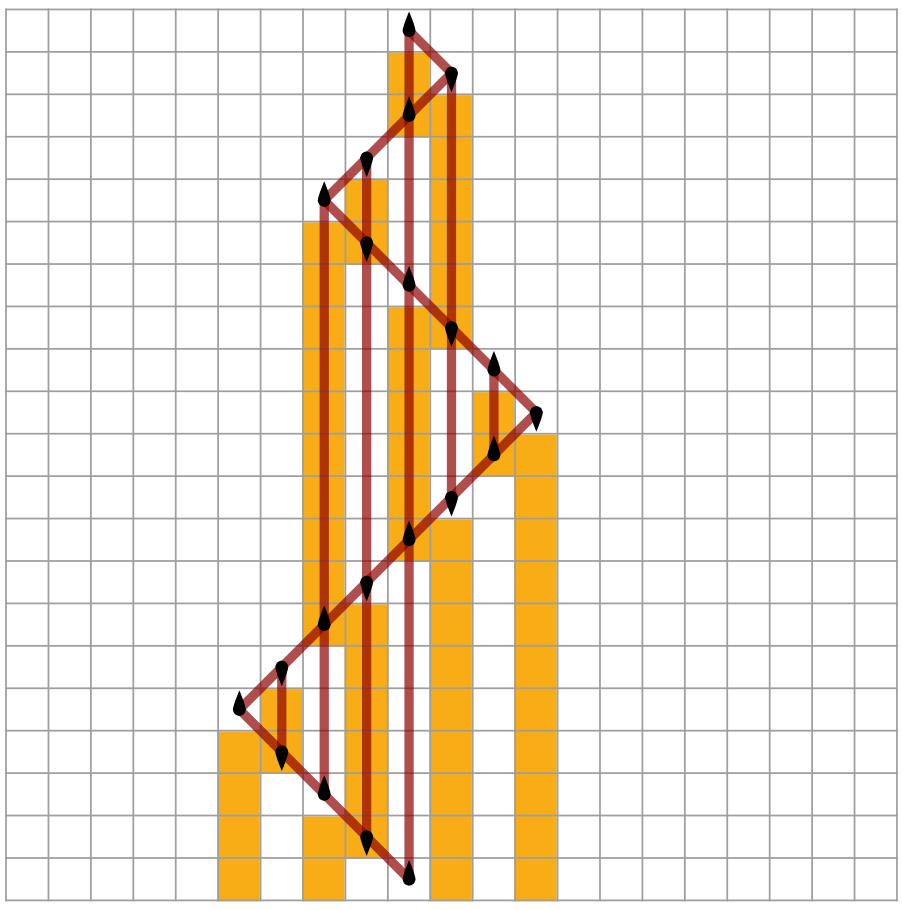

Continuing for a few more steps, the causal graph for this particular Turing machine becomes:

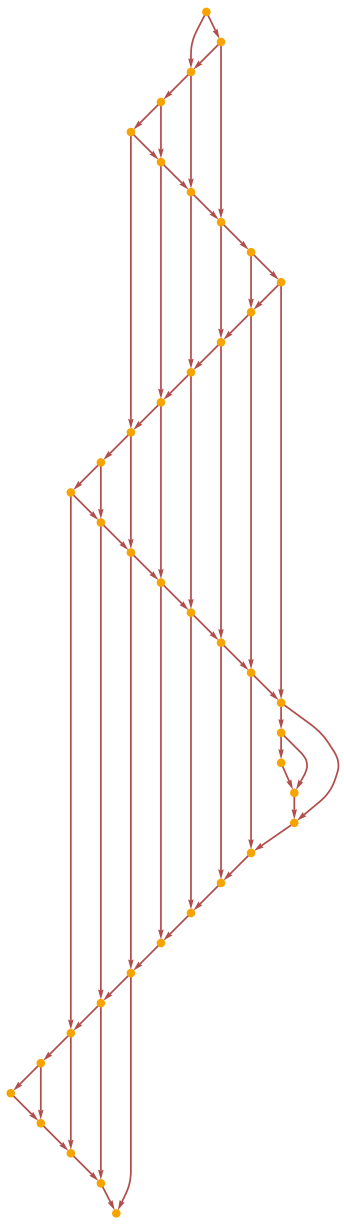

Continuing for more steps, and redrawing the graph, we see that we get a simple grid:

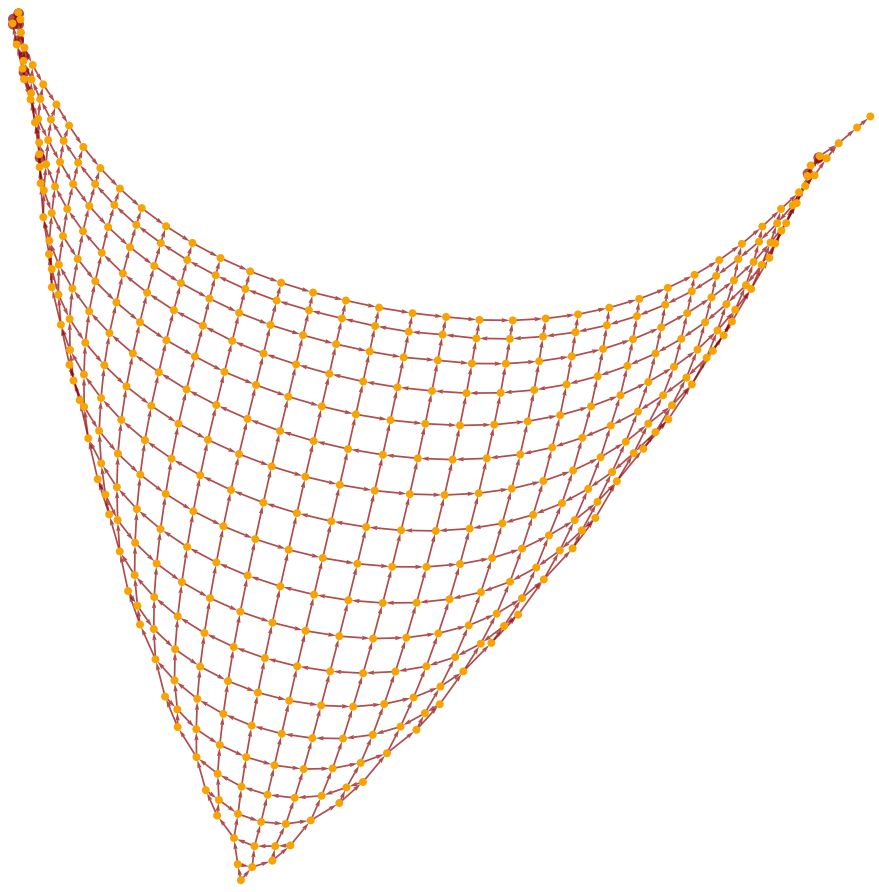

Of $s = 2$, $k = 2$ Turing machines, the one with the most exotic causal graph is the "binary counter" machine 1953:

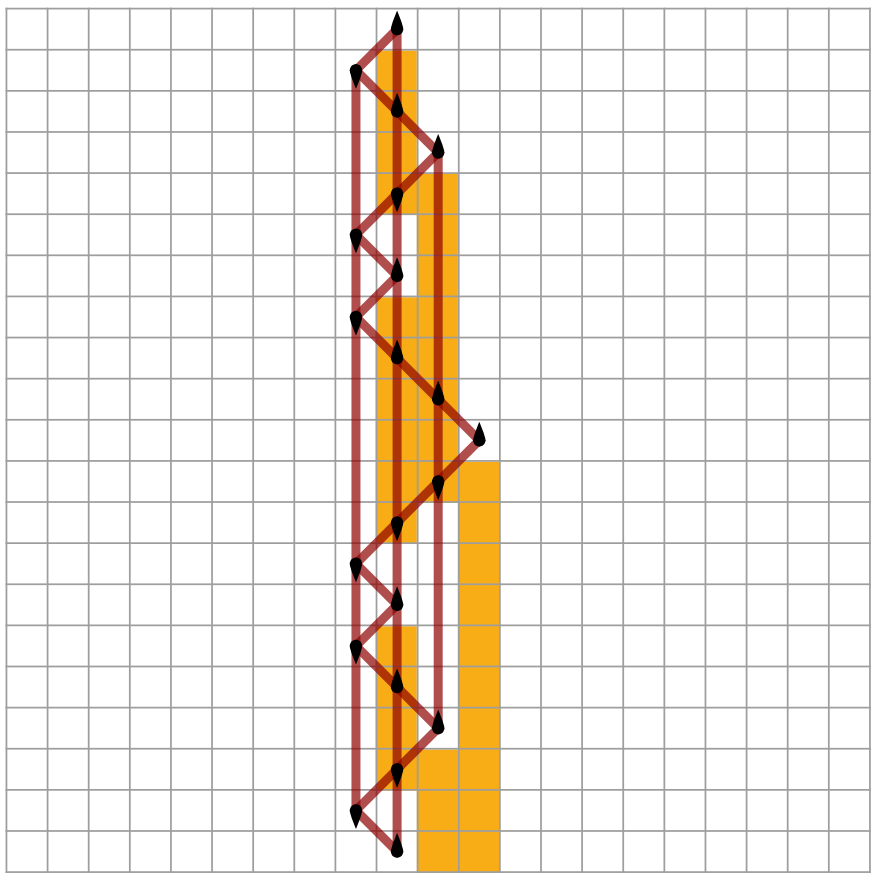

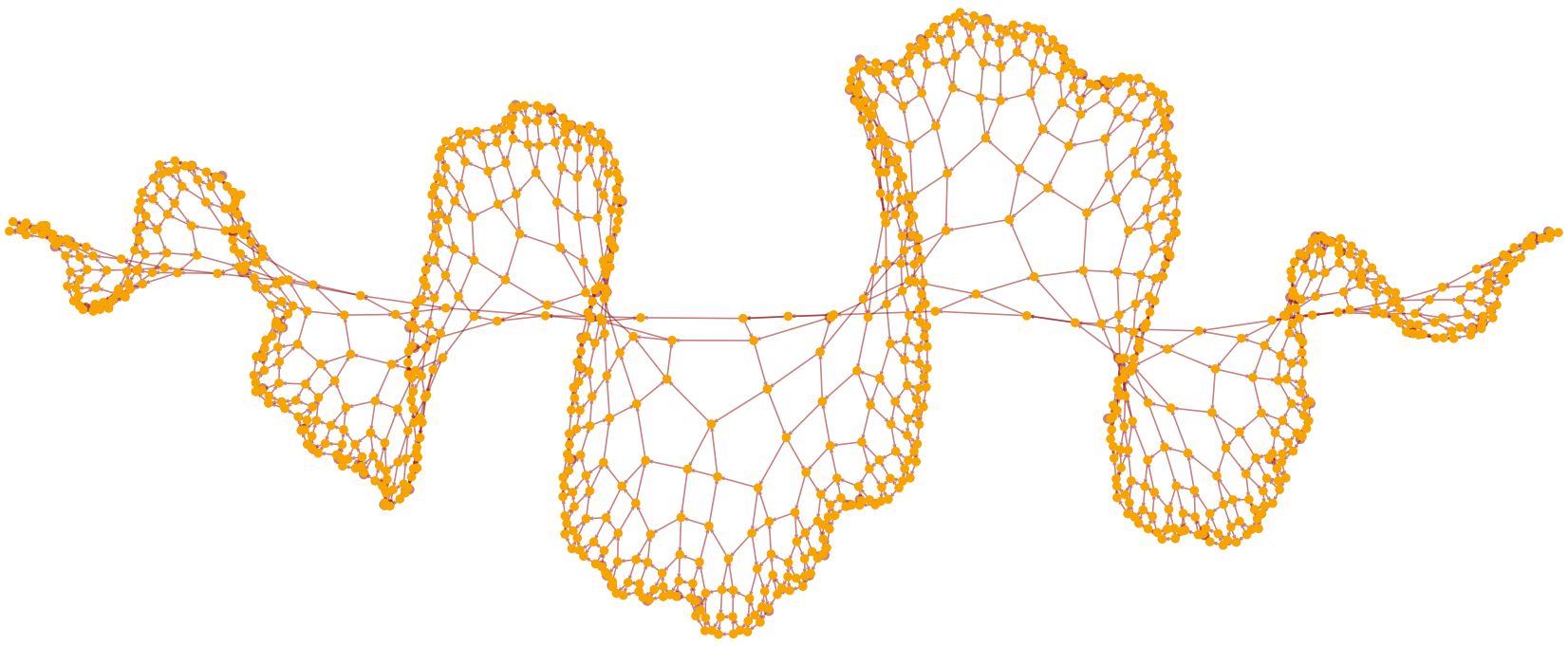

The causal graph gives a good “overall map” of the behavior of a Turing machine. Here are a few Turing machines (with $s = 3$, $k = 2$ and $s = 4$, $k = 2$) from *A New Kind of Science* (compare also the $s = 2$, $k = 3$ universal Turing machine):

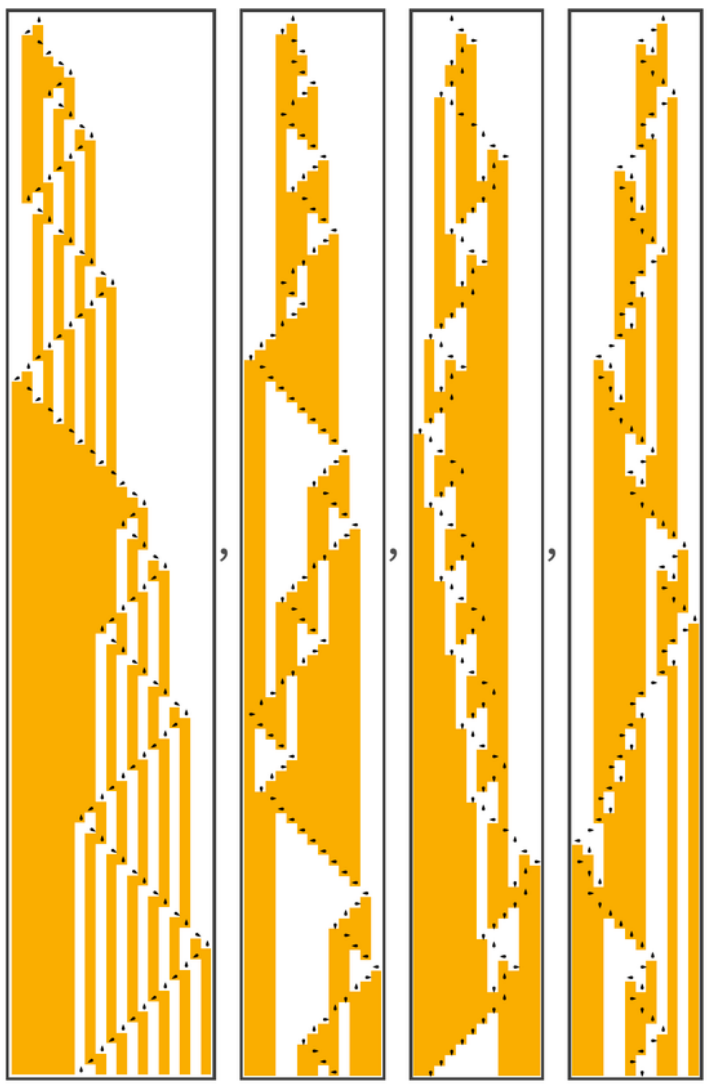

And here are their respective causal graphs:

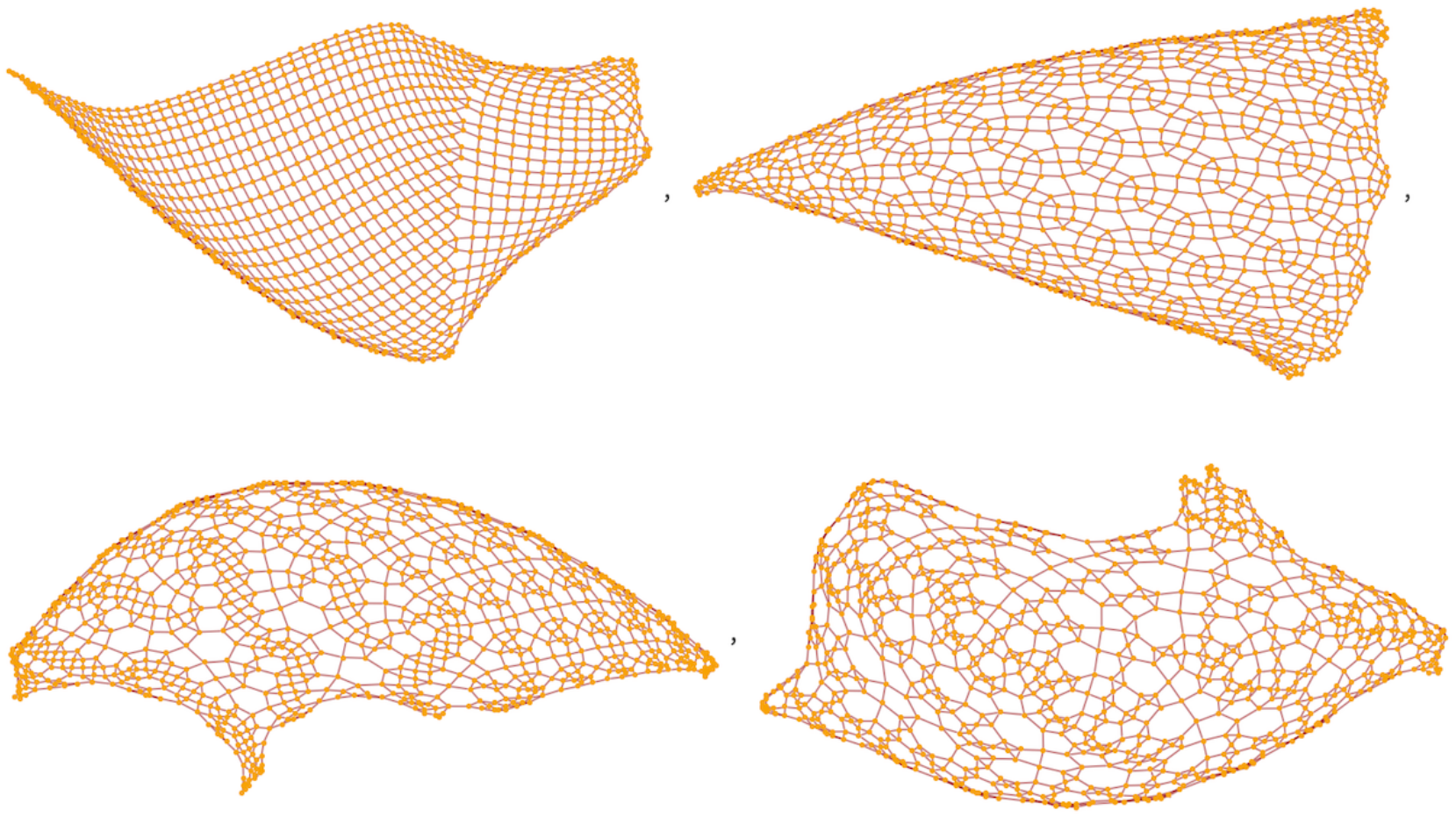

## Rulial Multiway Causal Graphs

What can we say about the causal graph associated with a rulial multiway system? The first important observation is that rulial multiway graphs always exhibit causal invariance, since by including transitions associated with all possible rules one is inevitably including both rules and their inverses, with the result that every branching of edges in the rulial multiway graph is always associated with a corresponding merging.

It’s fairly easy to see this for two steps of $s = 2$, $k = 2$ Turing machines. The rulial multiway states graph here is:

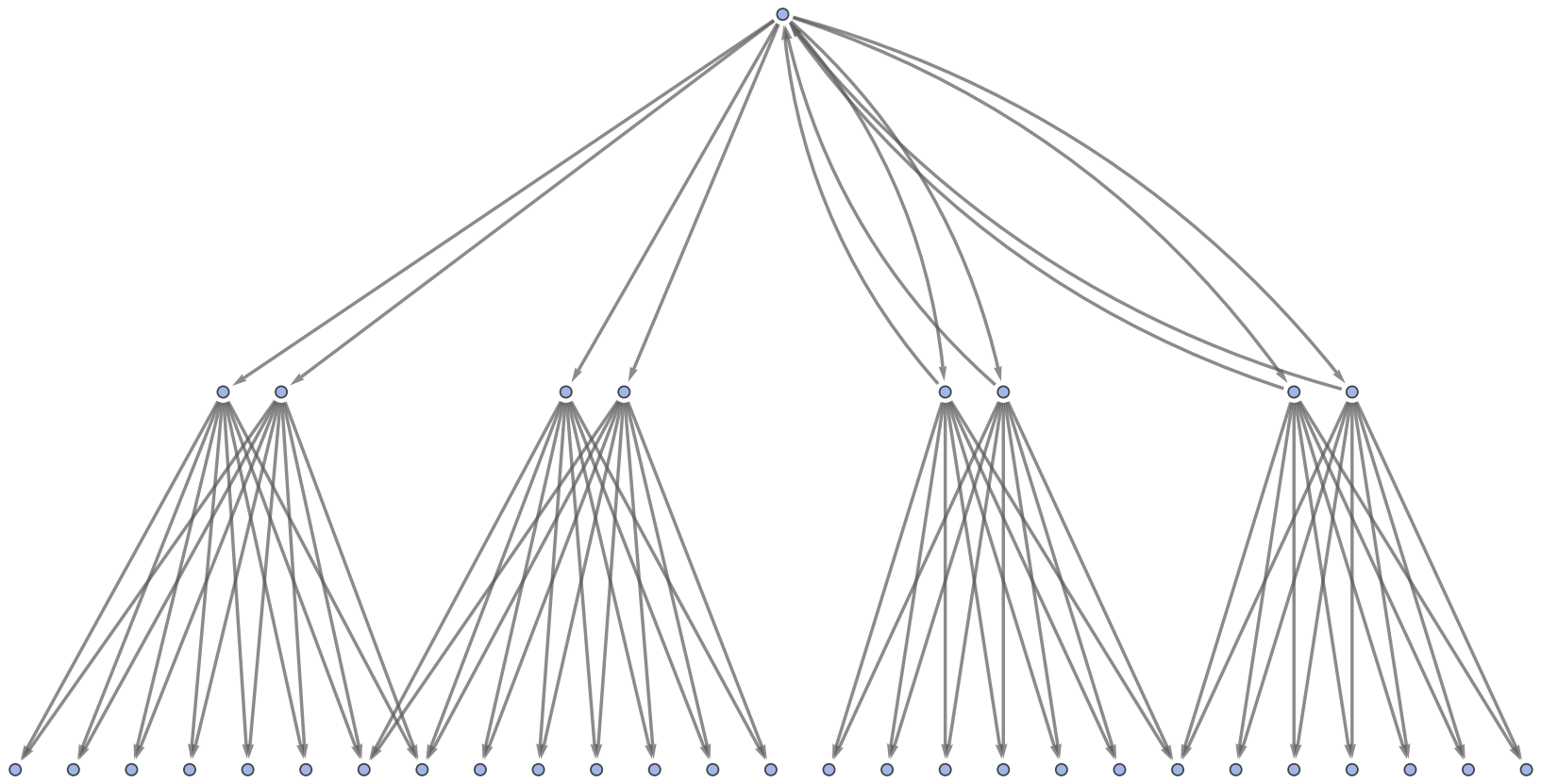

This can also be rendered:

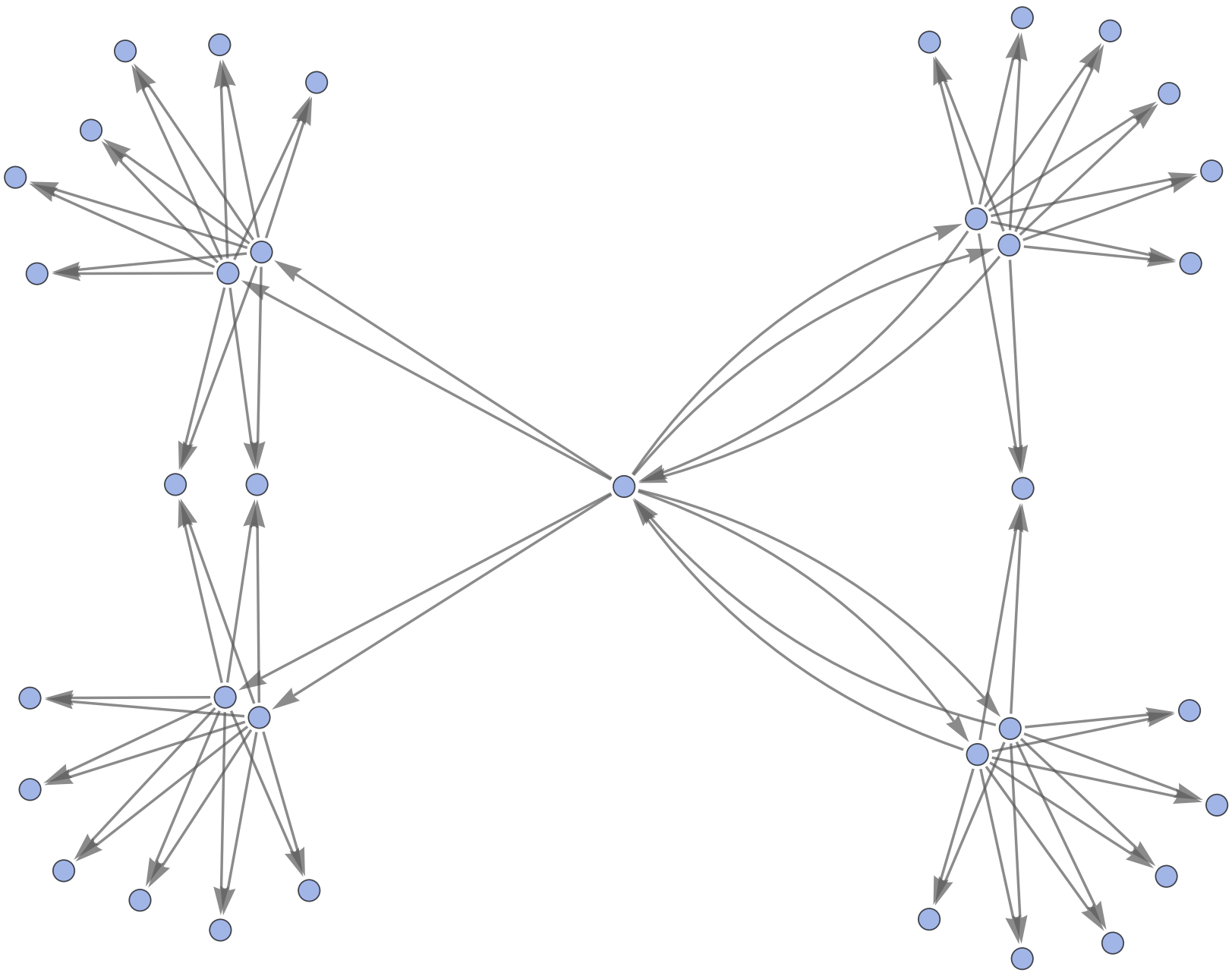

Explicitly showing events we get:

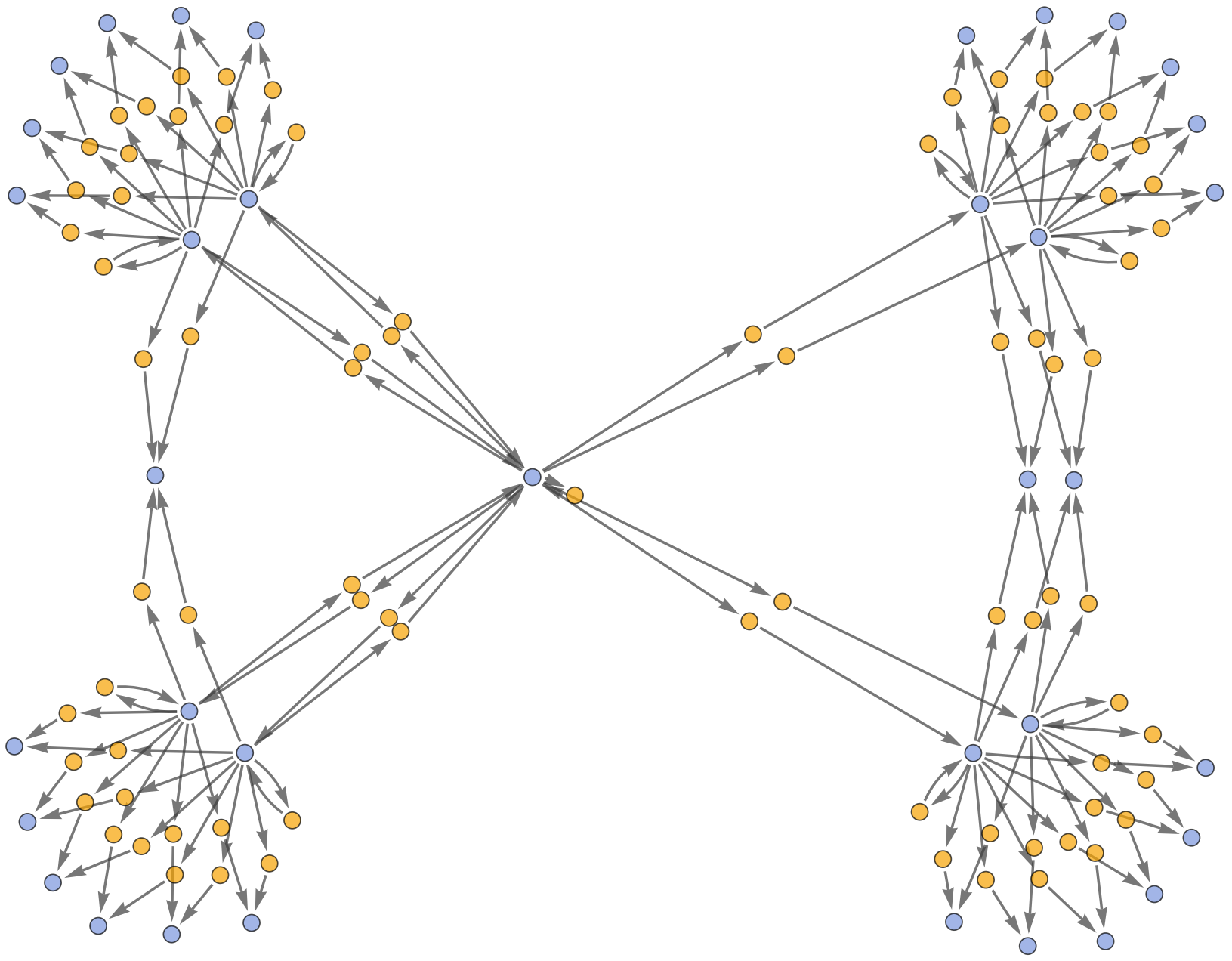

Including causal connections we get

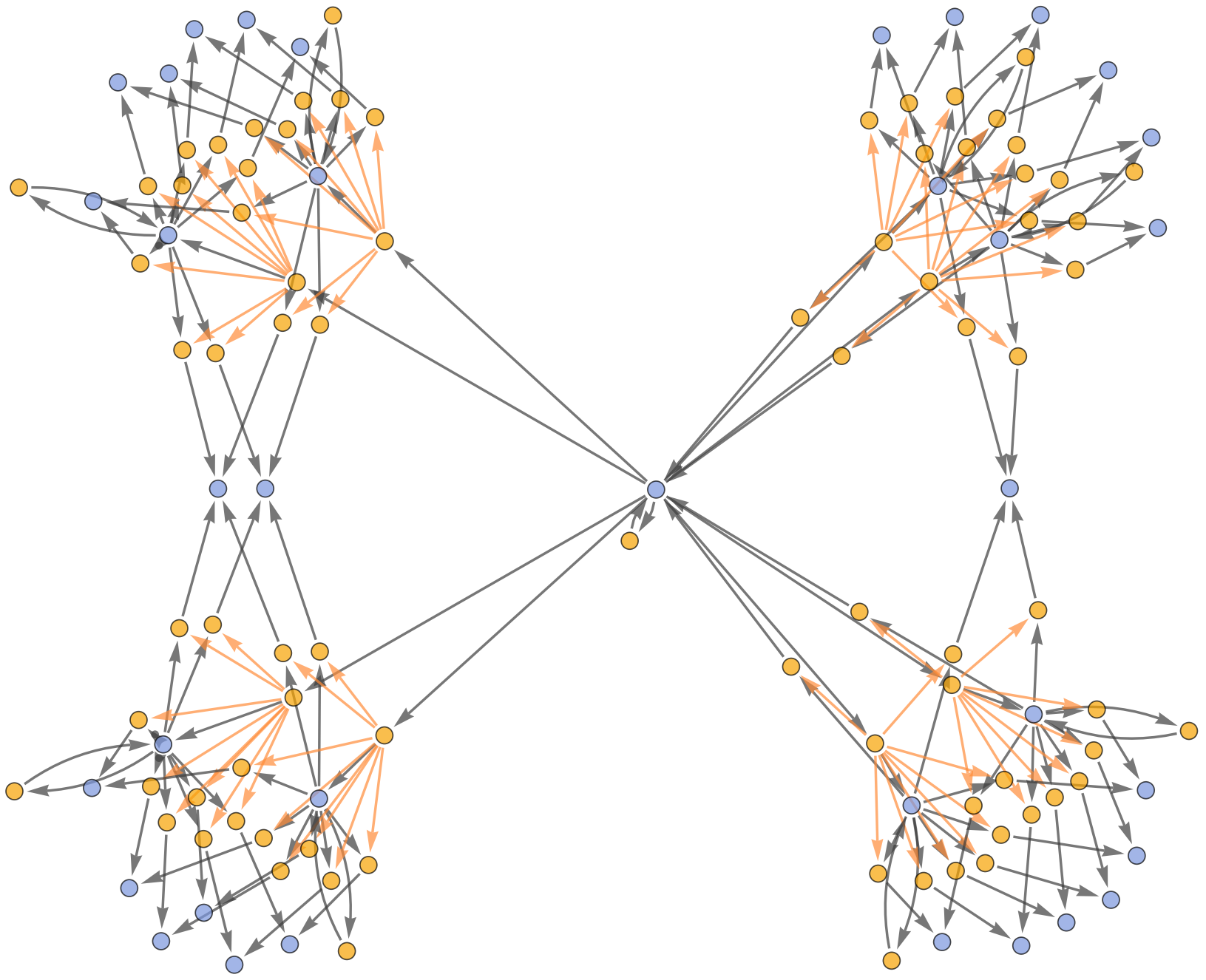

or after 3 steps:

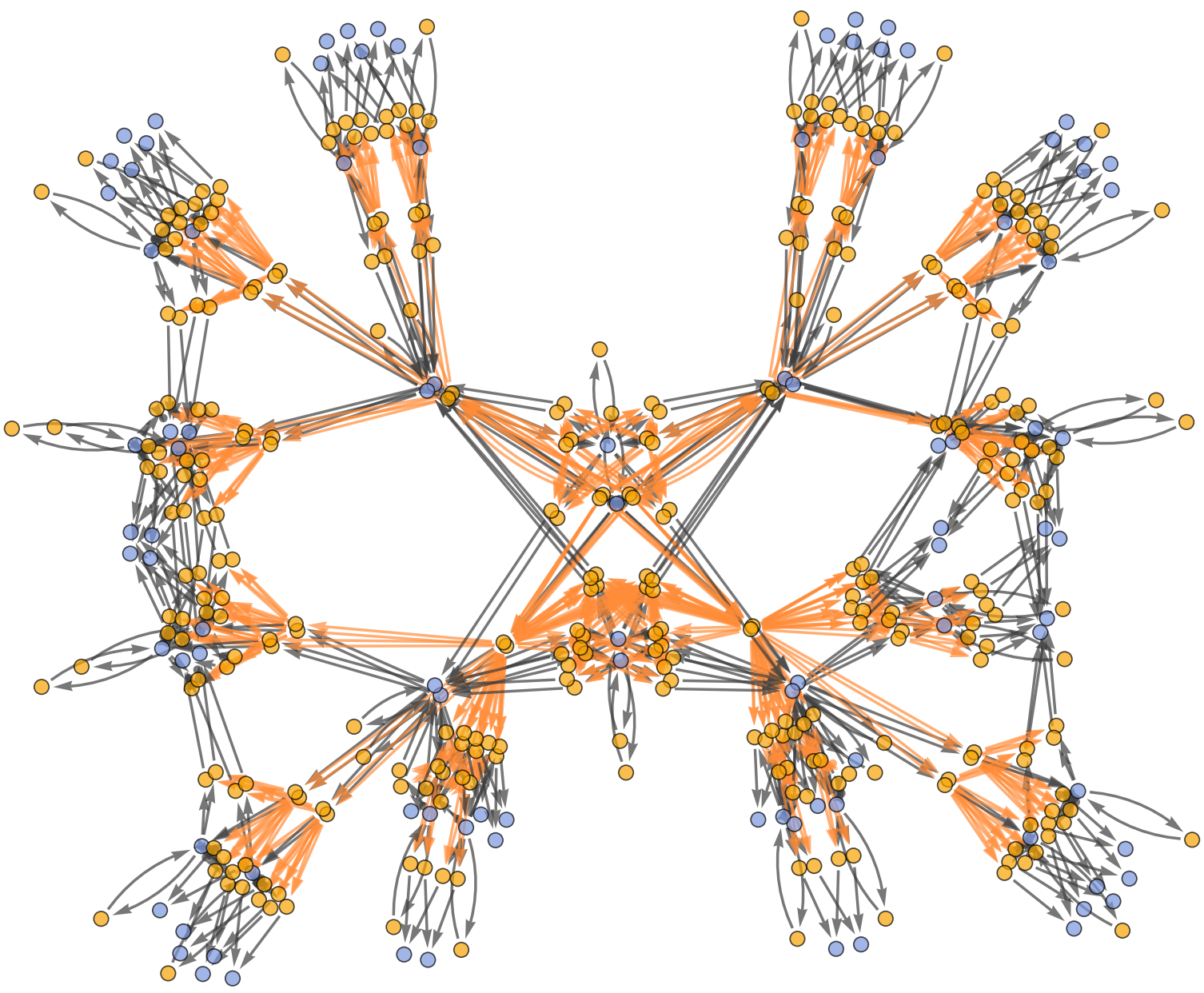

The pure rulial multiway causal graph in this case is then:

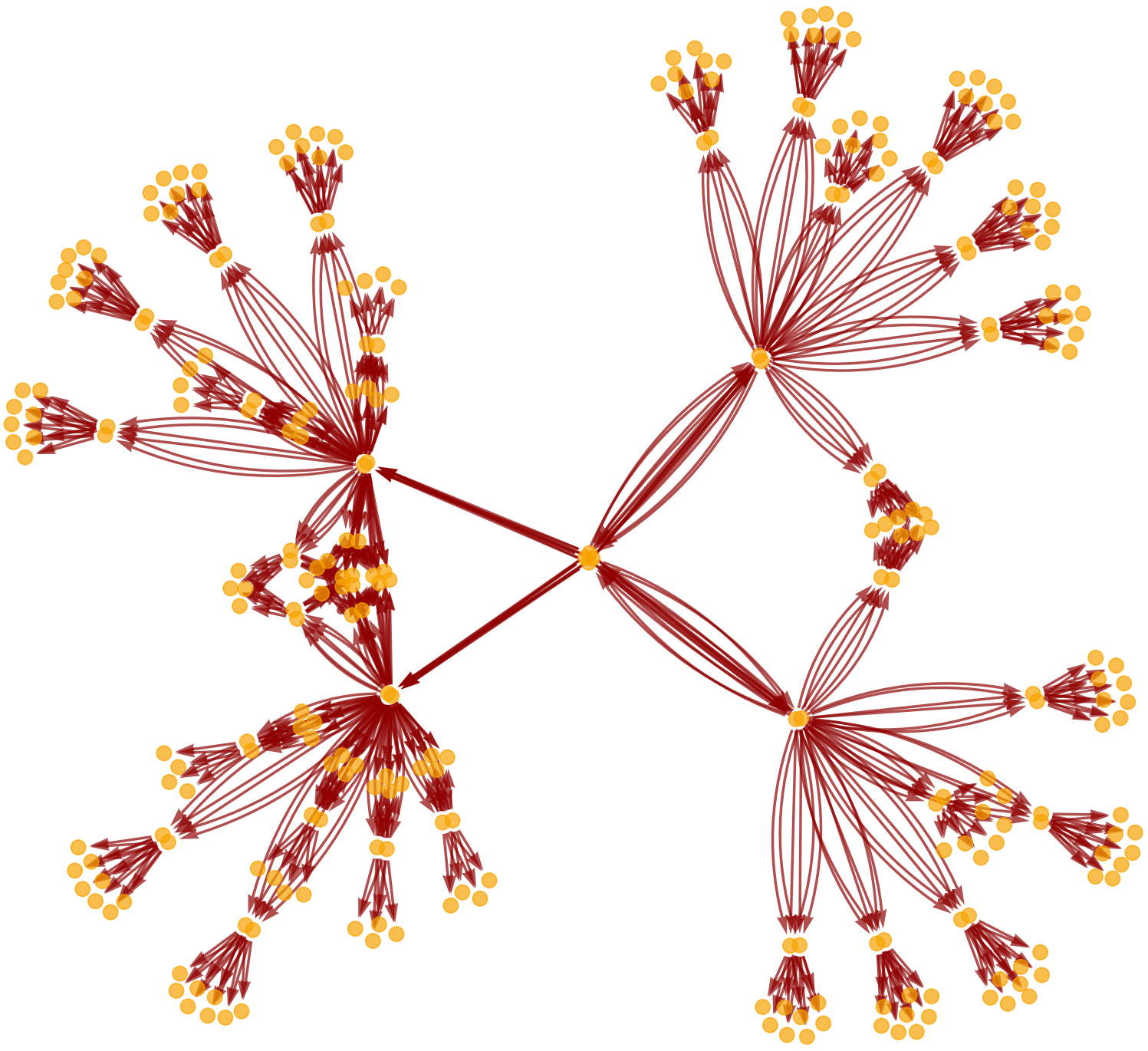

In layered form this becomes:

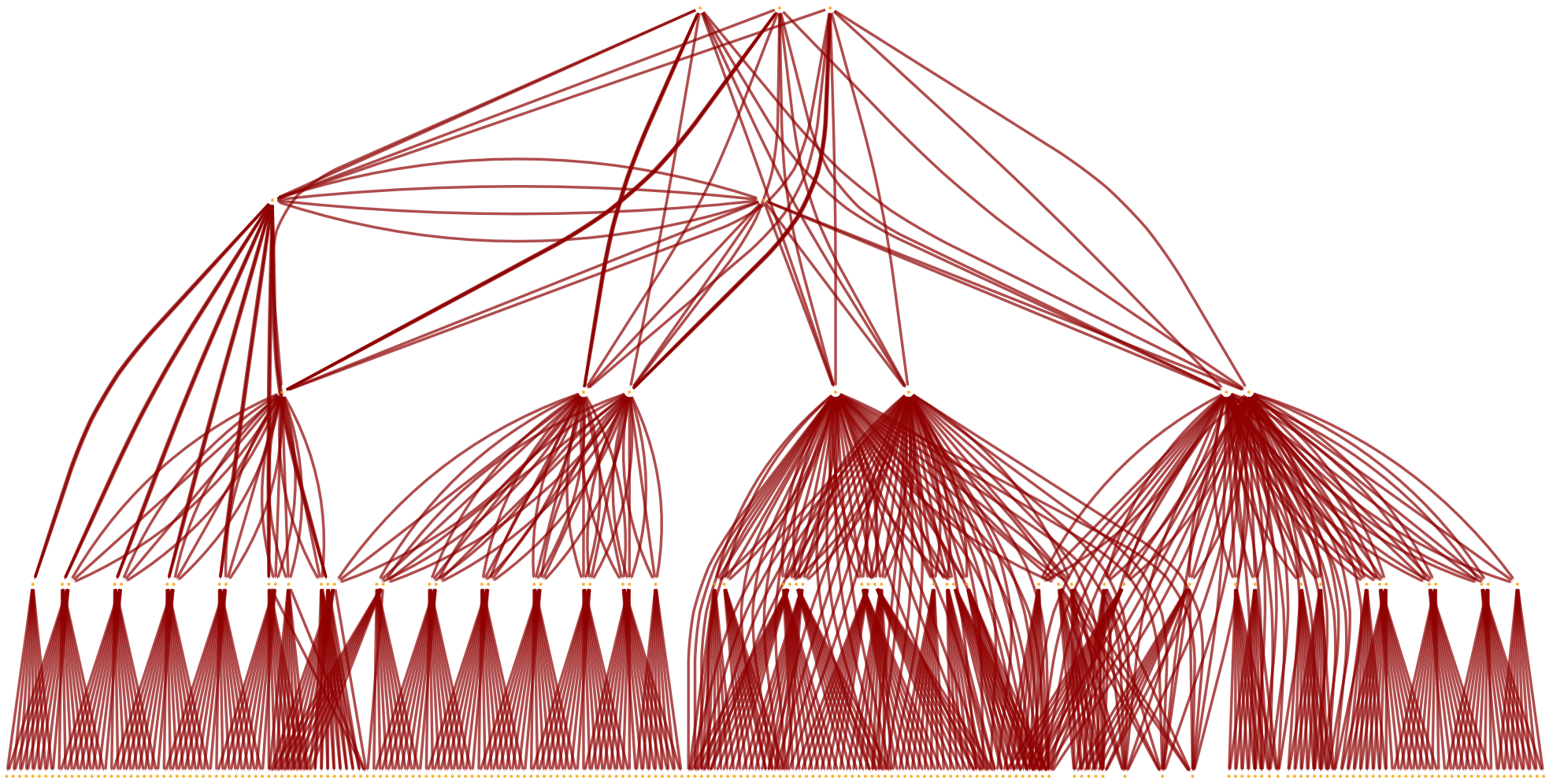

After 5 steps the growth in the number of nodes reached going from the root grows like:

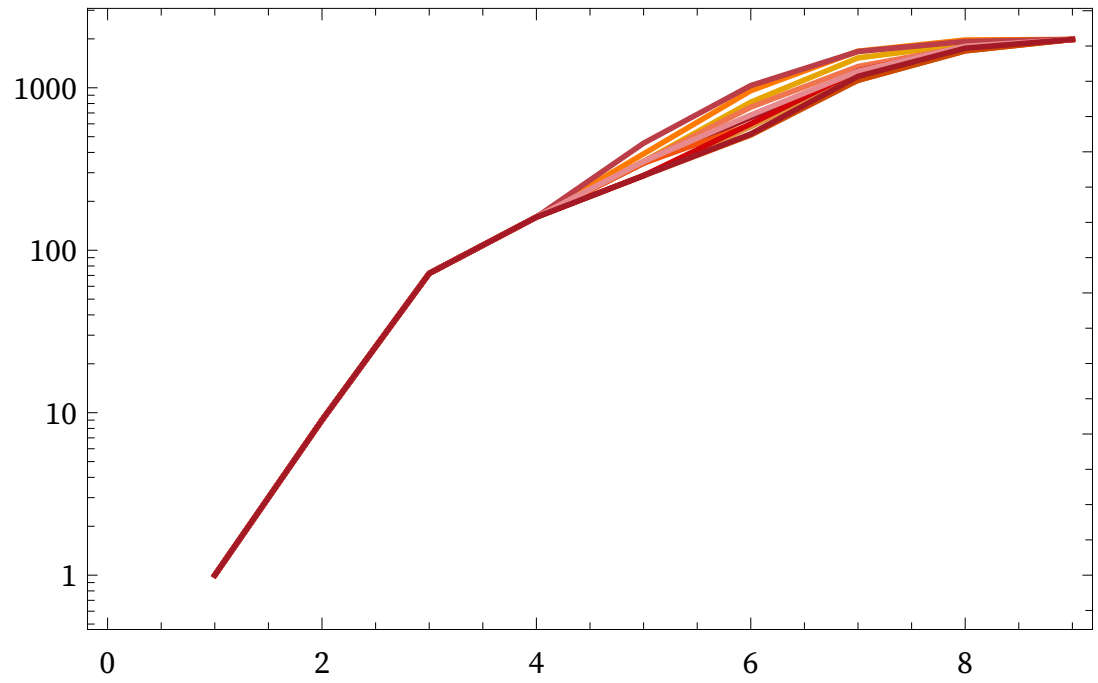

Causal invariance implies that this multiway causal graph is ultimately composed of a large number of interwoven copies of a single causal graph. After any given number of steps of evolution, the various copies of this causal graph will have "reached different stages". Here are the results after 3 steps:

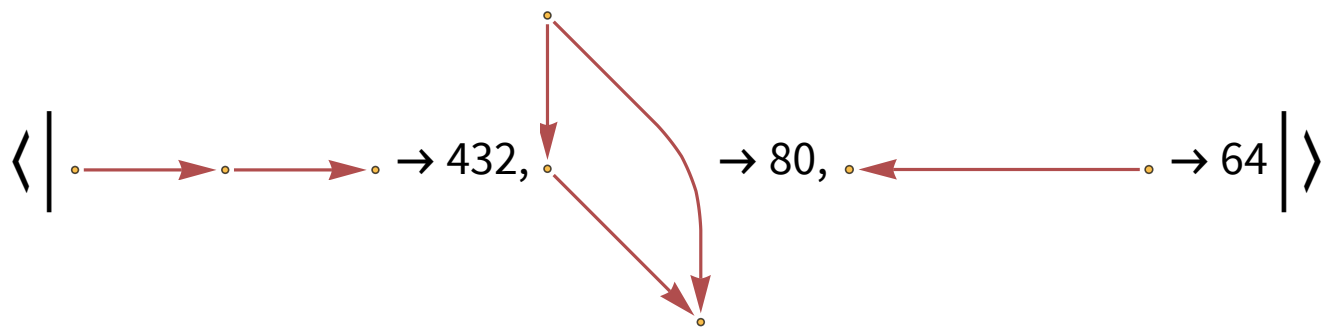

And after 4 steps:

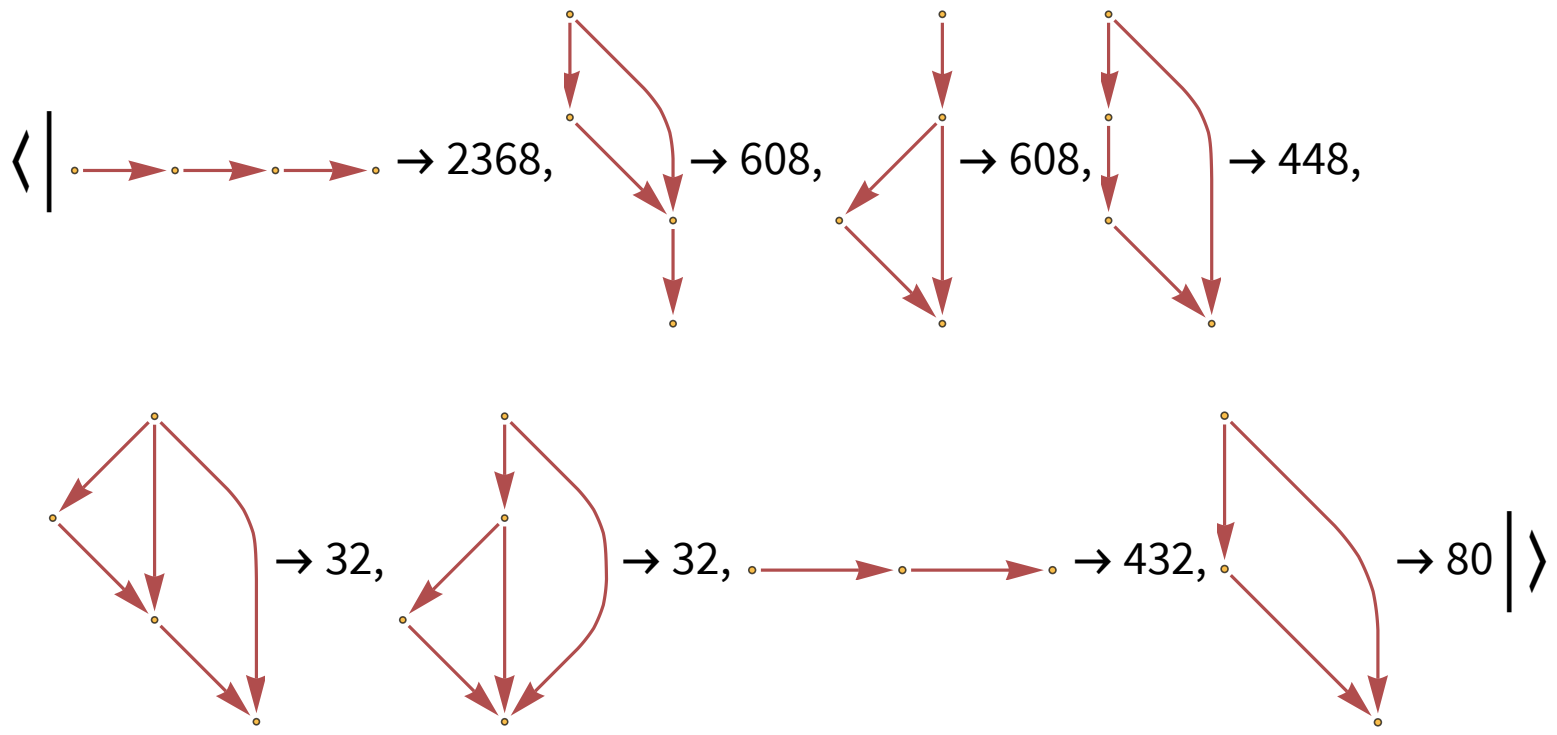

One can also look at rulial multiway causal graphs for other sets of Turing machines. For $s = 1$, $k = 1$ one has (here after 5 steps)

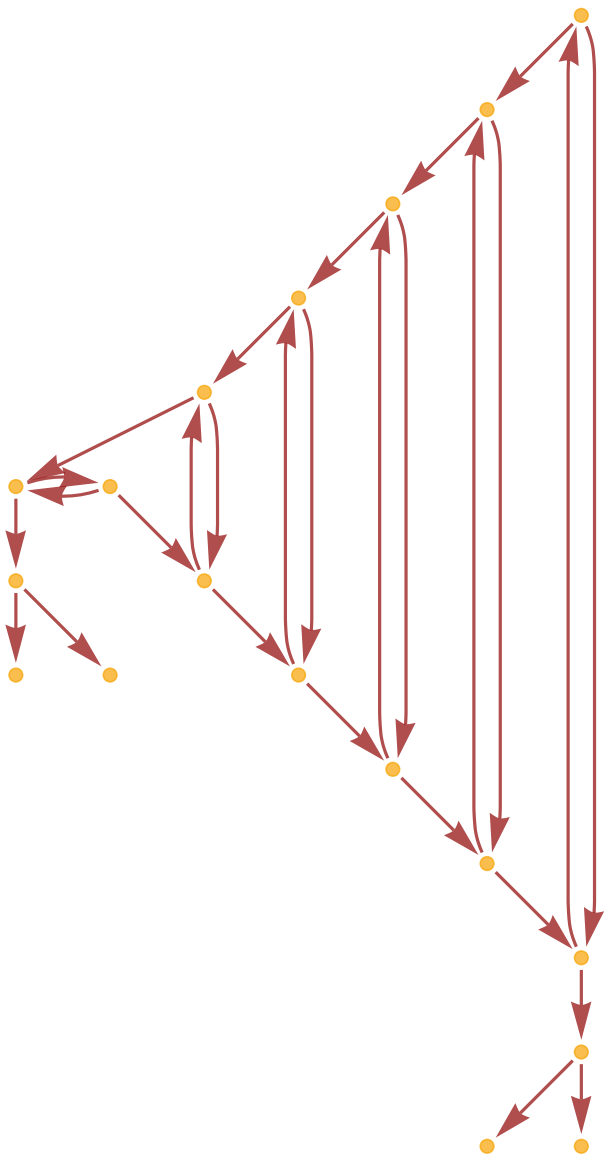

which is equivalent to:

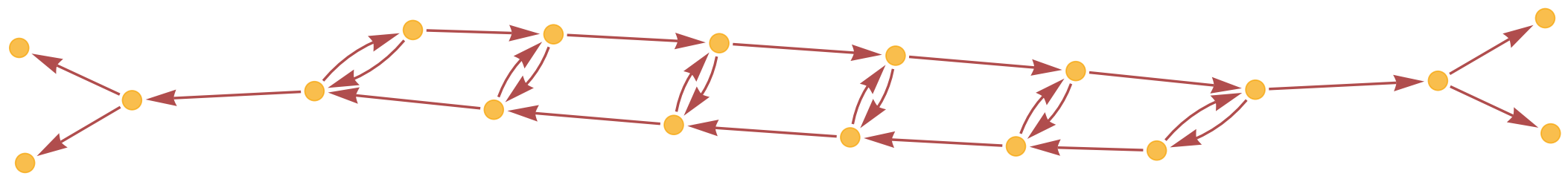

The individual causal graphs in this case are immediately all the same, and are just:

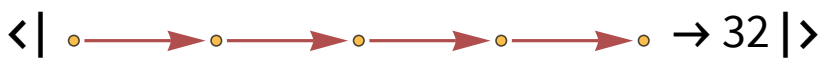

For $s = 2$, $k = 1$ one already has after 3 steps

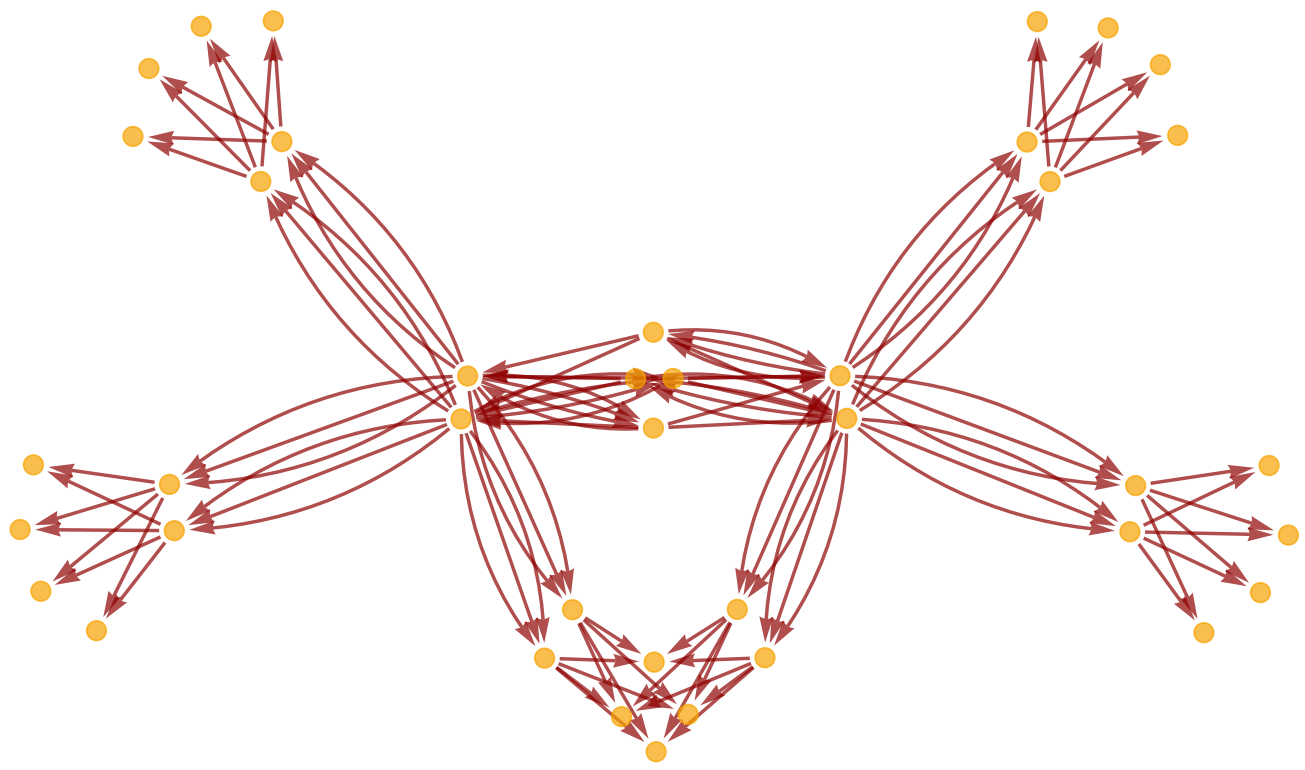

or in layered form:

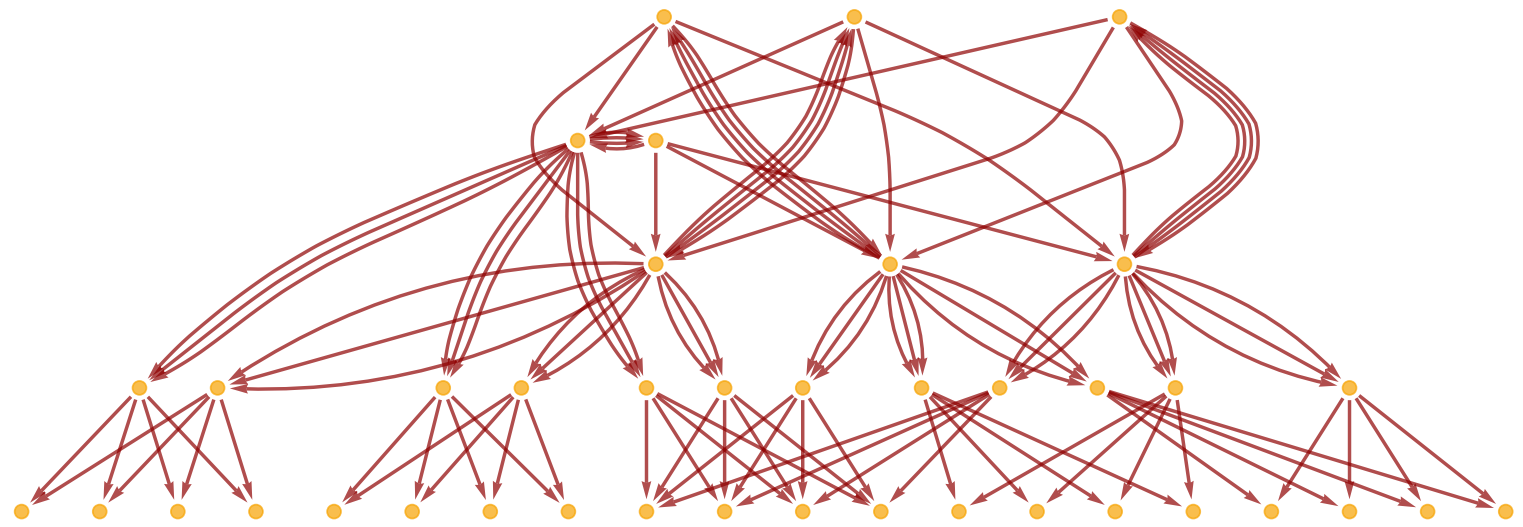

After 5 steps this becomes

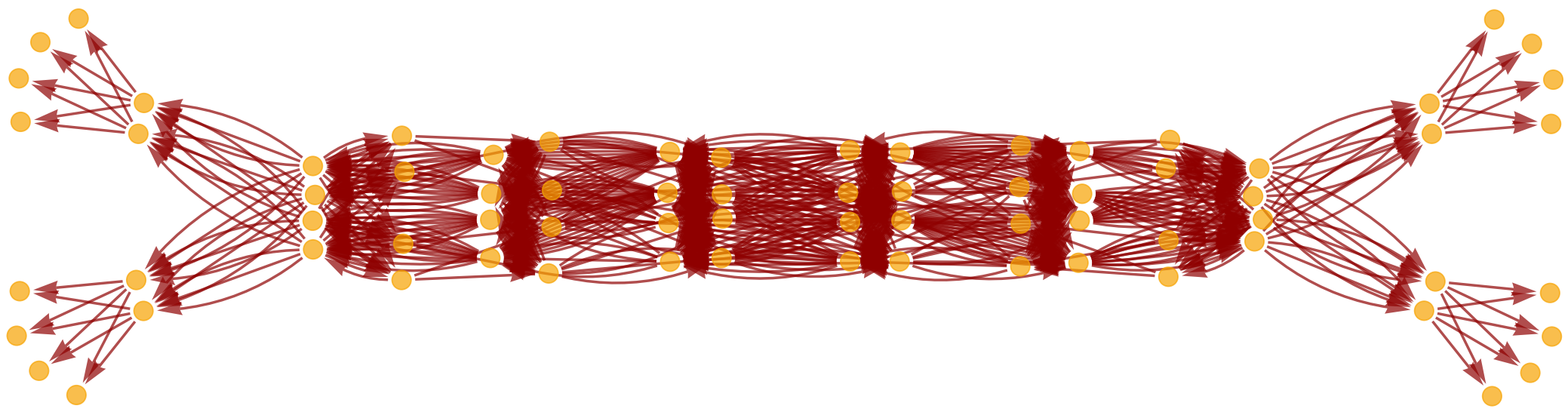

or in layered form:

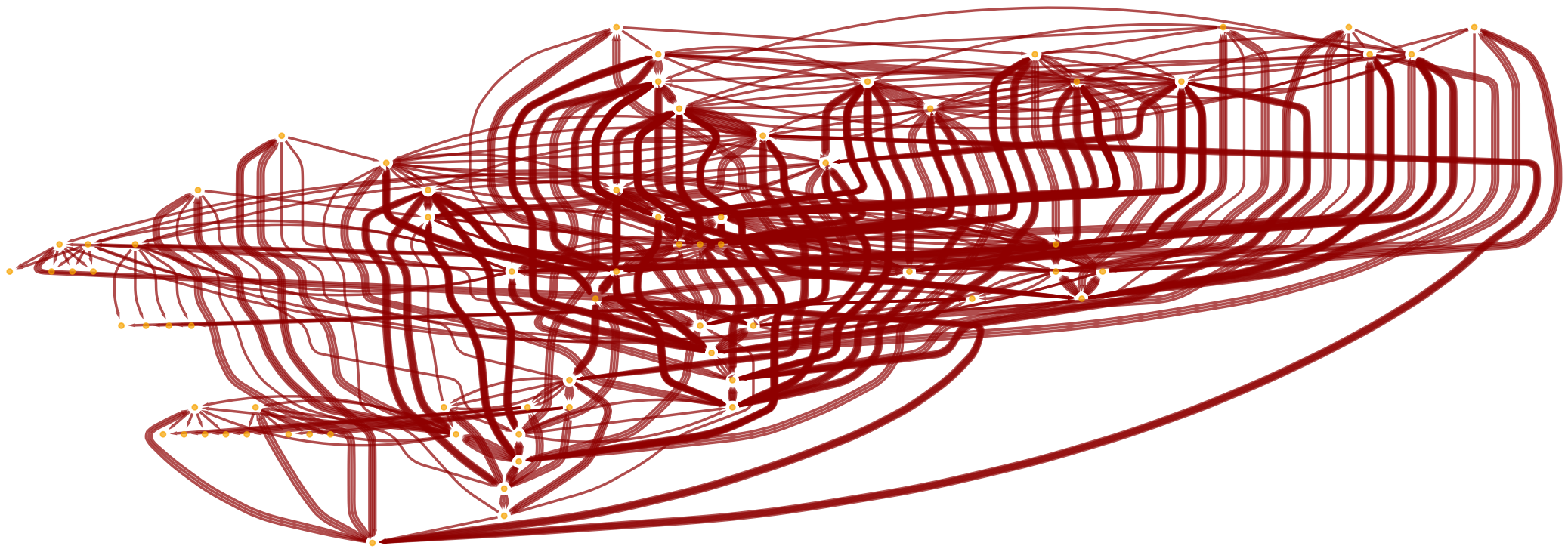

In this case, the individual causal graphs after 5 steps are:

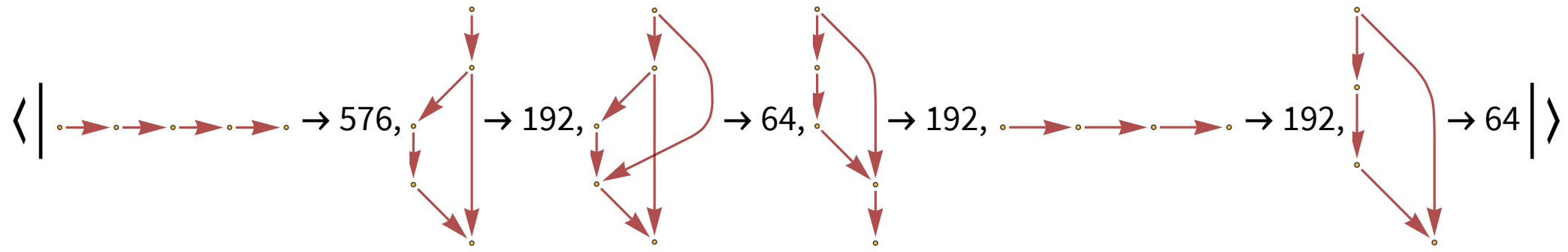

For $s = 1$, $k = 2$ one has very similar results to the case $s = 2$, $k = 2$. After 3 steps the causal graph is:

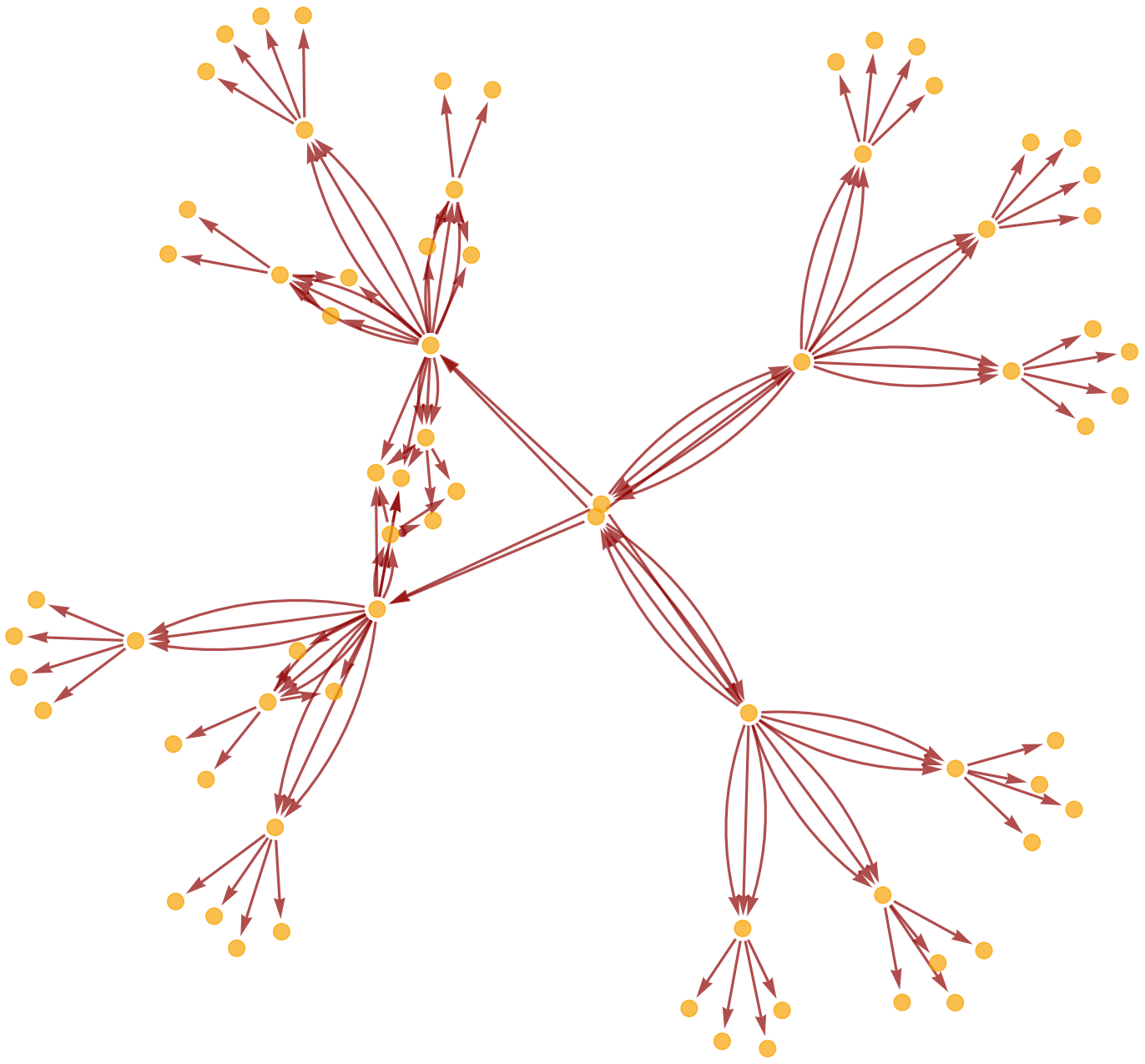

Or after 5 steps:

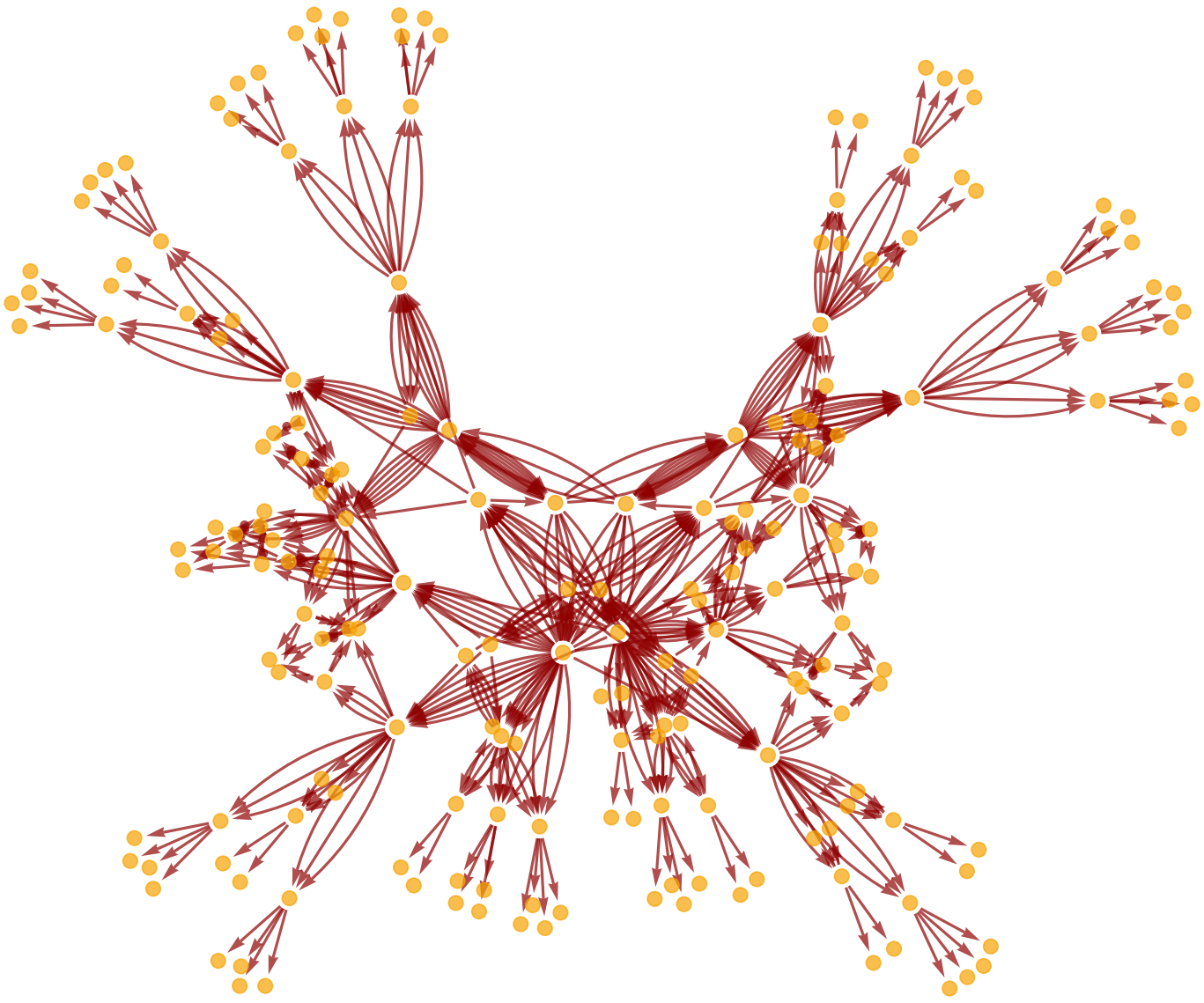

Removing multiple edges this is:

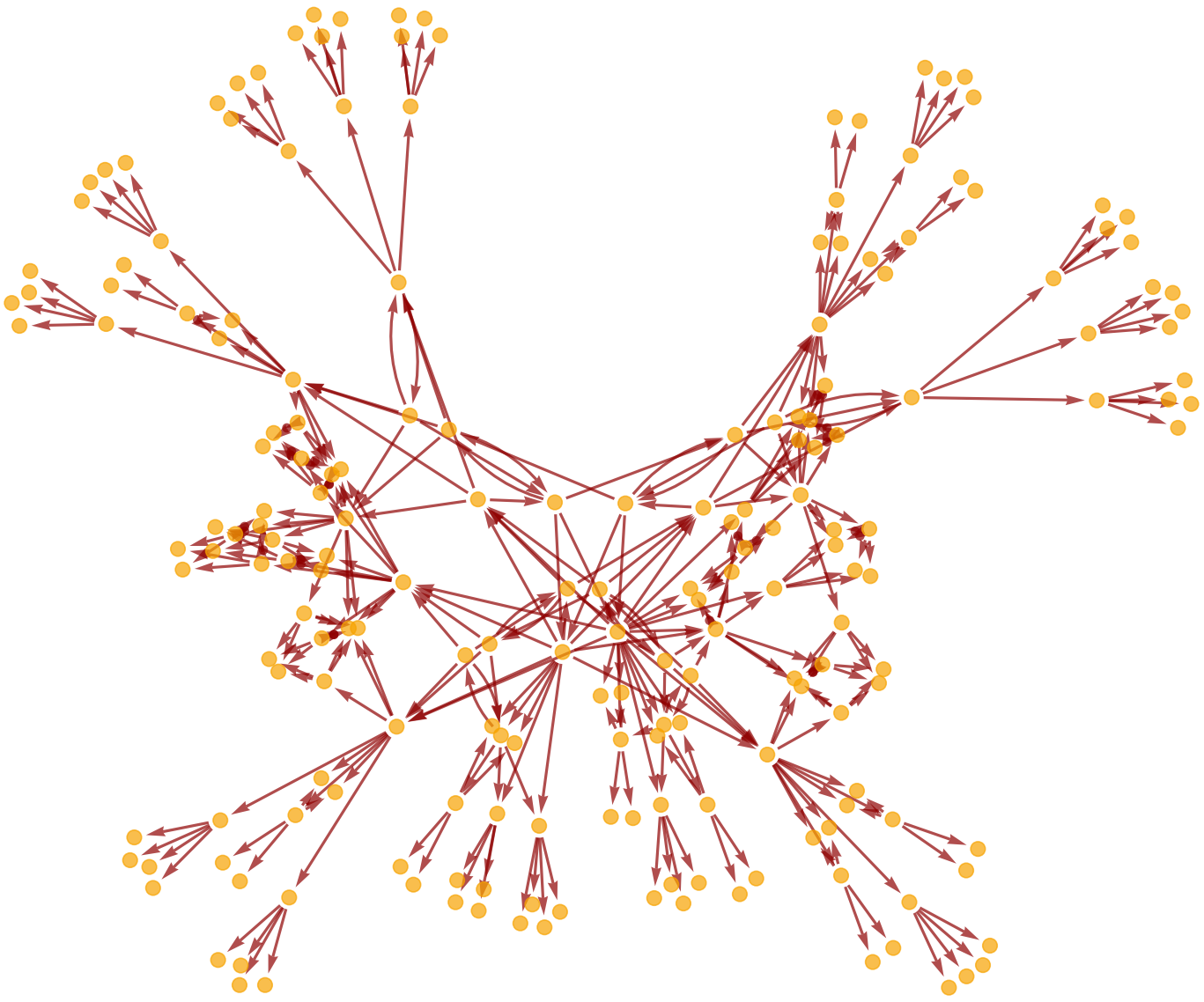

The individual causal graphs in this case are:

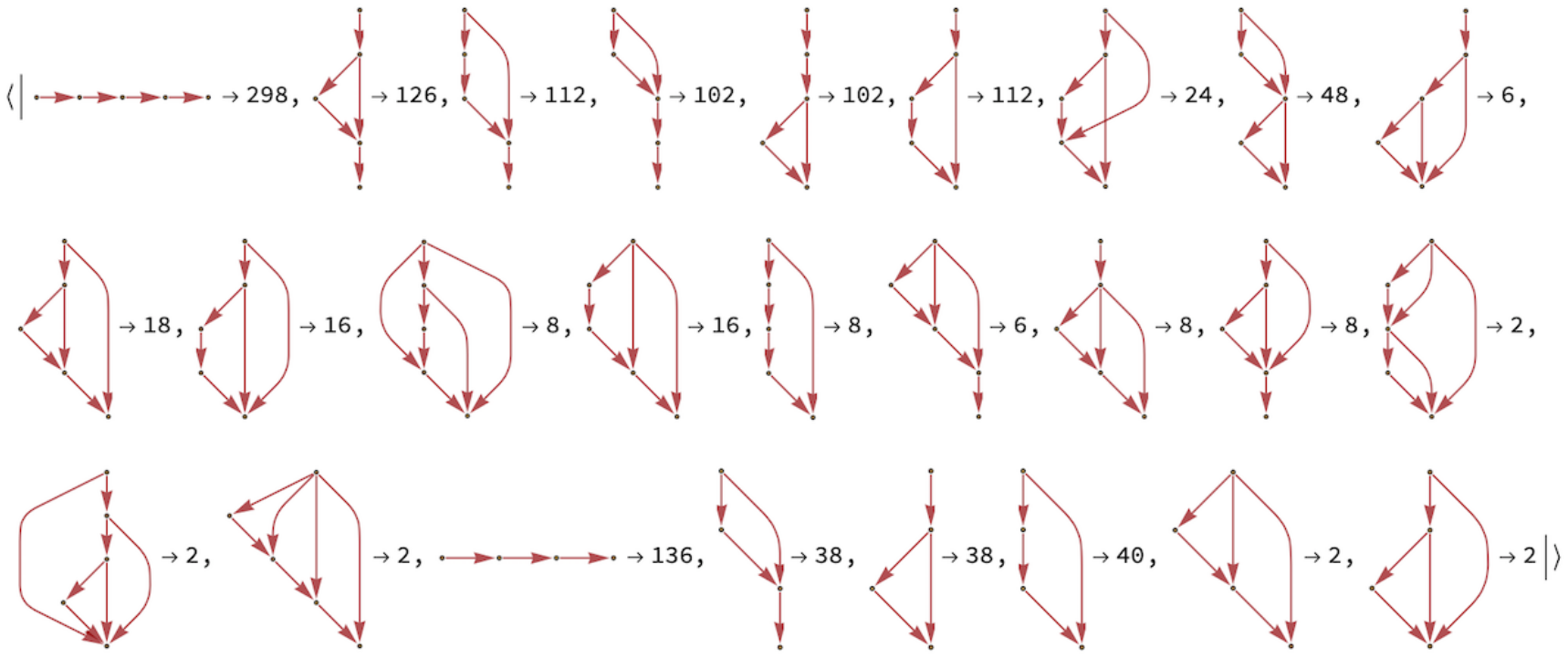

## Rulial Graphs

Just as for ordinary multiway graphs can study branchial graphs which represent their transversals, so similarly for rulial multiway graphs one can study rulial graphs which represent their transversals. The layered way we have drawn multiway graphs corresponds to a particular choice of foliation—and with this choice, we can immediately generate rulial graphs.

For $s = 2$, $k = 1$ one has after 2 steps

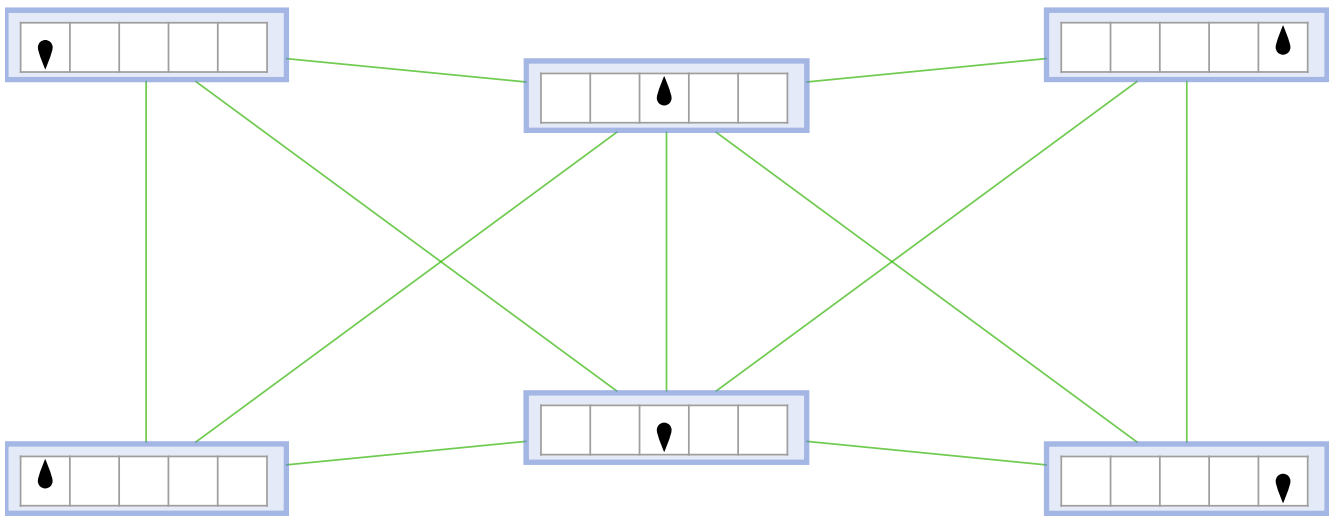

while after 3 steps one gets:

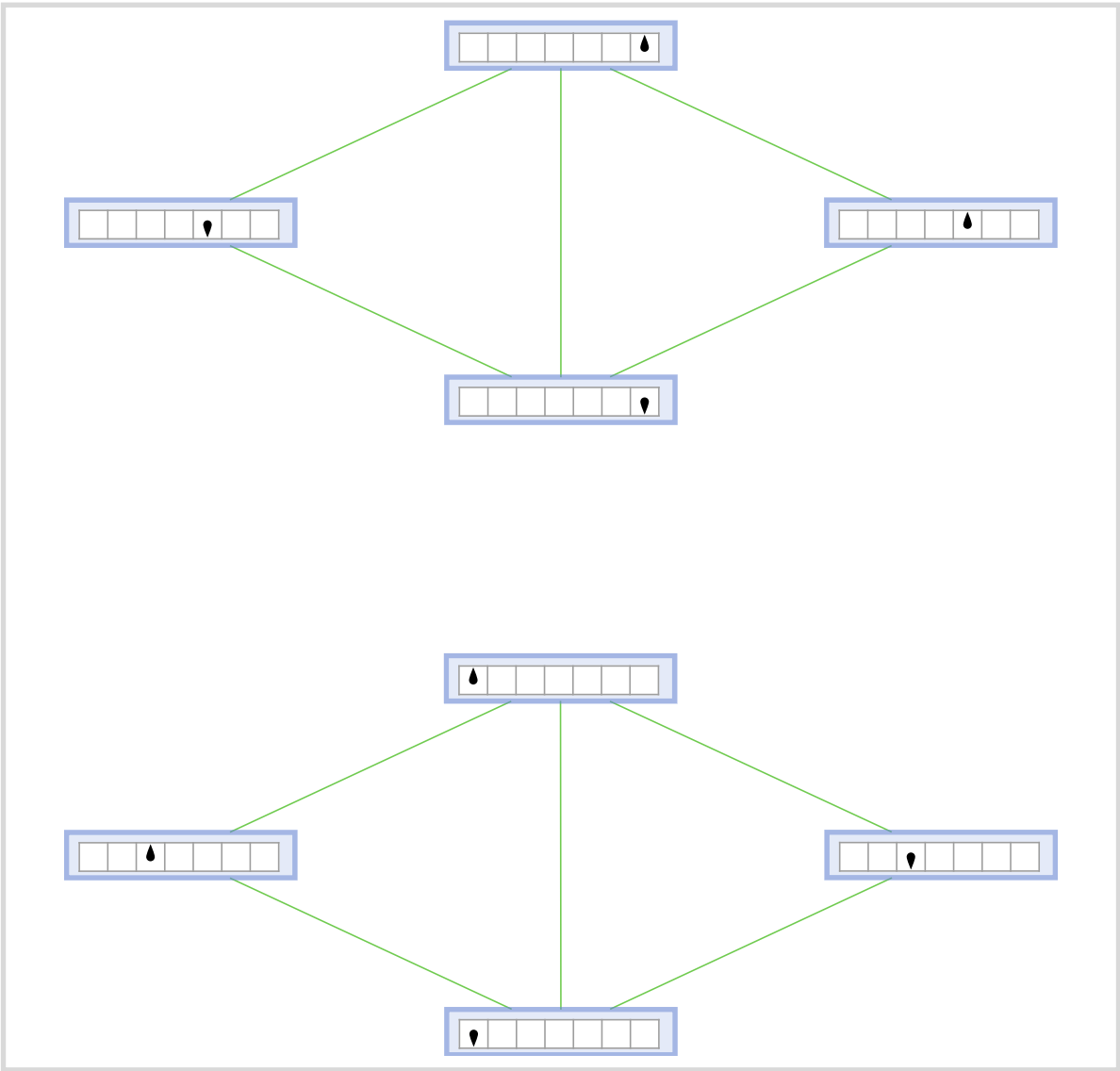

For $s = 1$, $k = 2$ one gets after 2 steps:

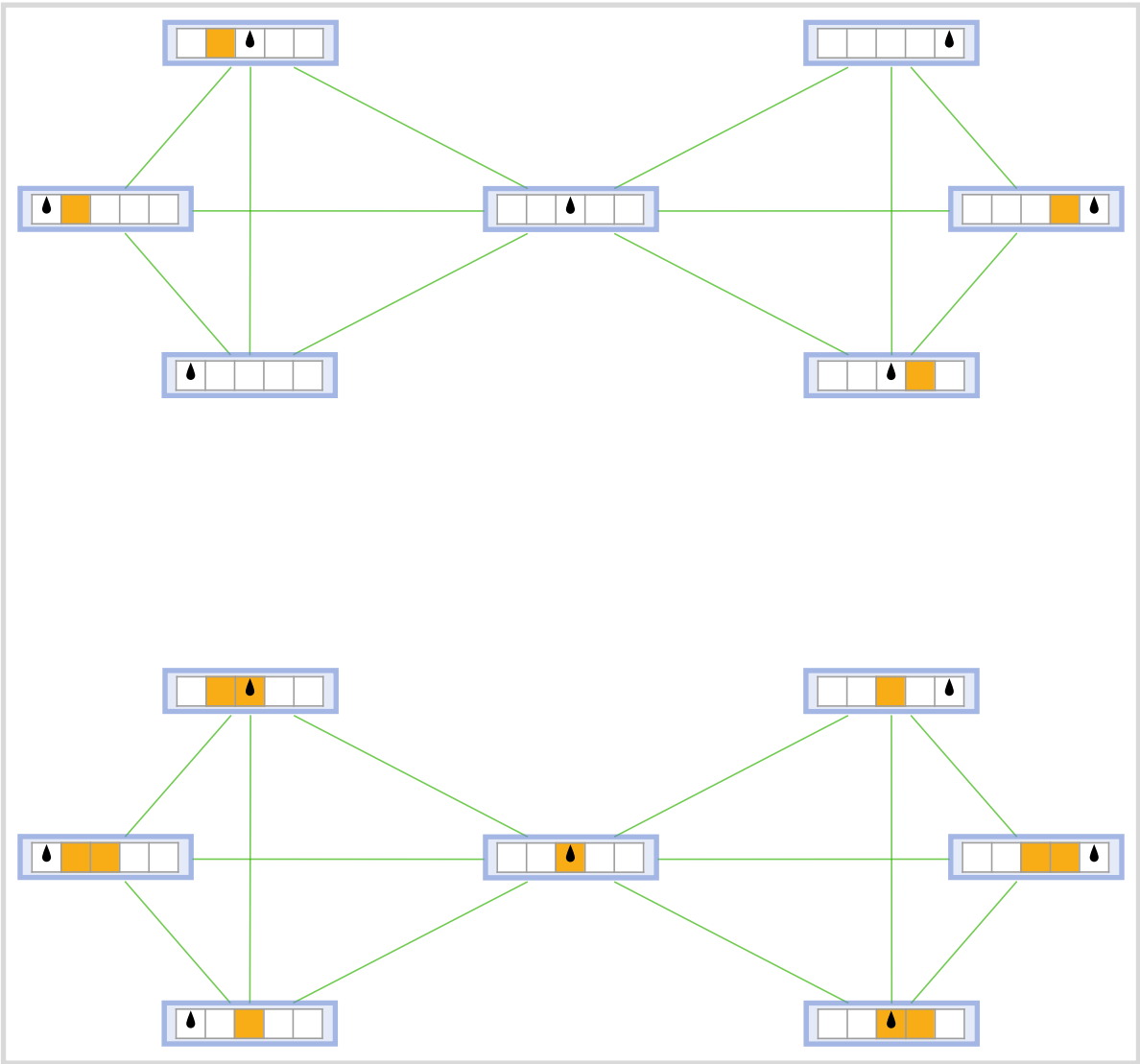

while after 3 steps one has:

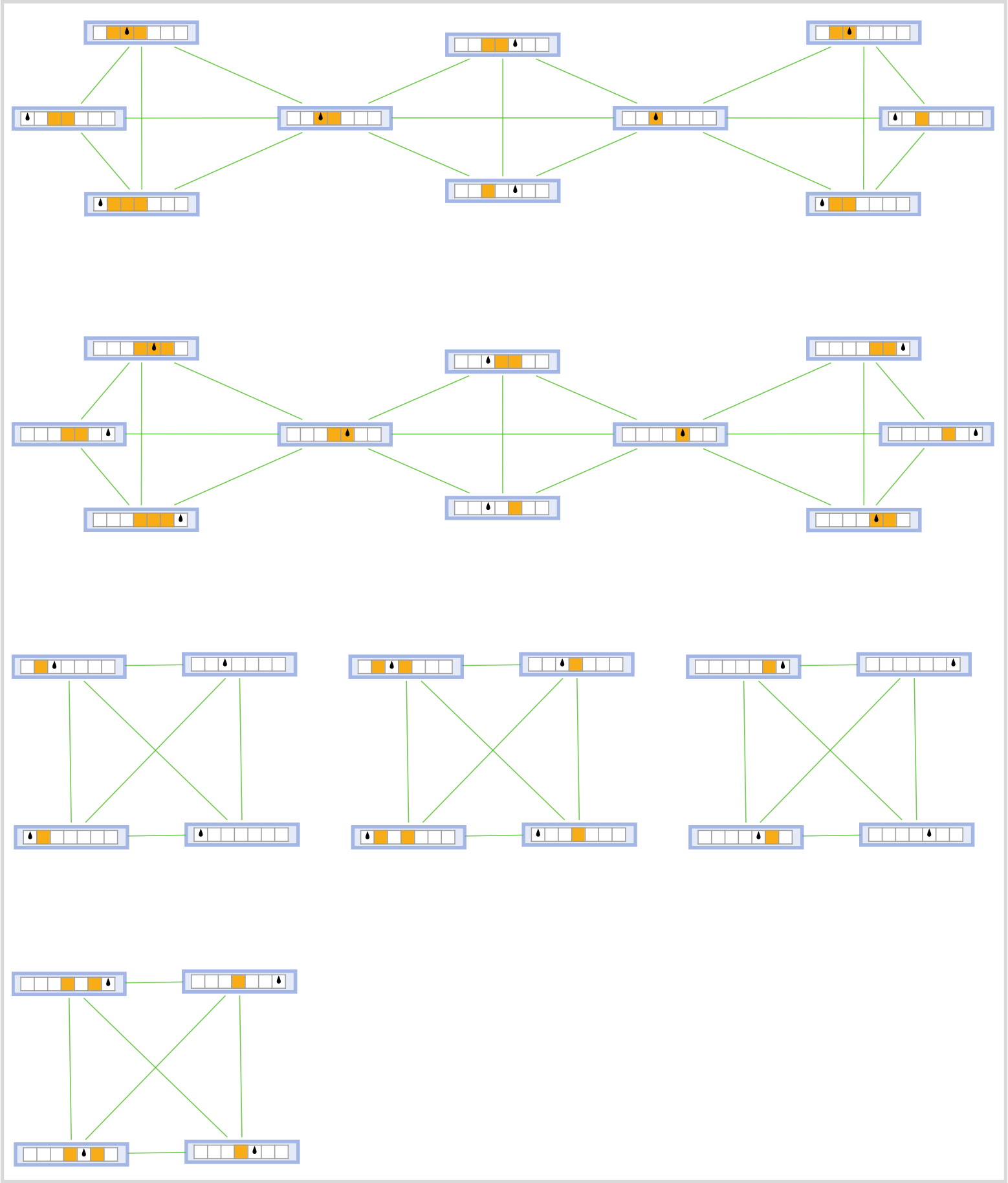

The sequence of results for steps 1 through 4 is:

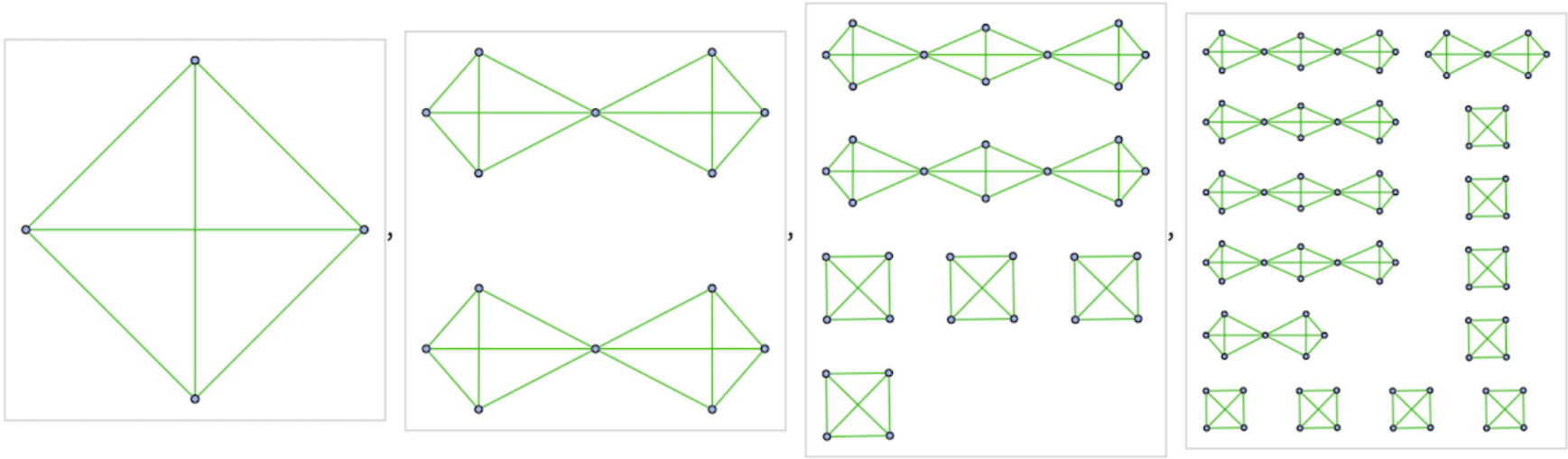

For $s = 2$, $k = 2$ the corresponding results are:

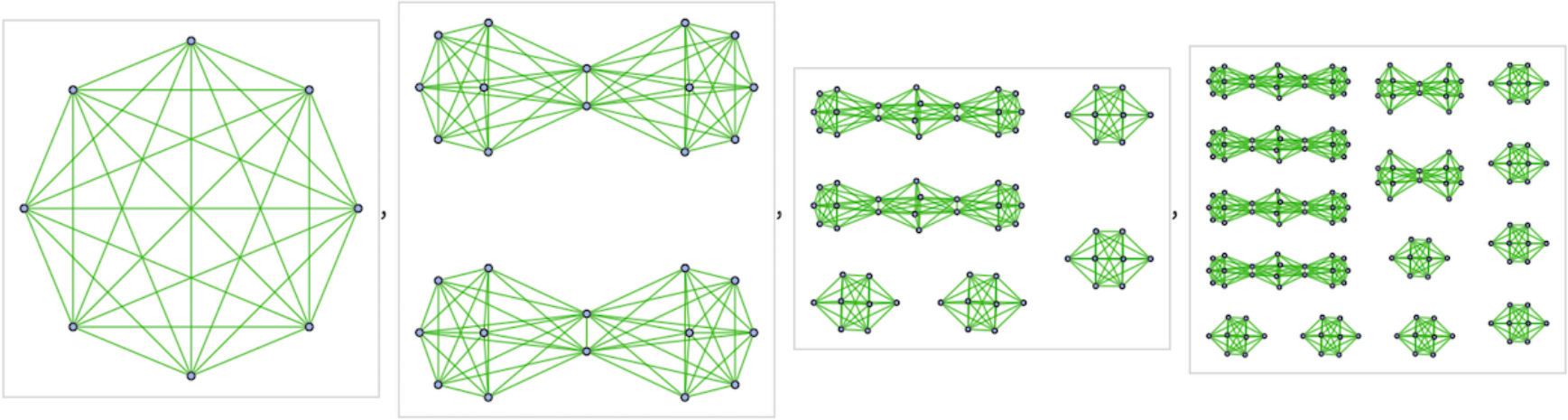

What do these pictures mean? Just as branchial graphs in ordinary multiway systems can be thought of as "entanglement maps" for states (interpreted in our models as quantum states) in ordinary "multiway space", so here these rulial graphs can be thought of as "entanglement maps" for states in rulial space. In other words, they are a kind of map of which Turing machine configurations are "evolutionarily close" to which other ones, in the sense that they can be reached with only a few different choices of rules.

## Deterministic Turing Machine Paths in Rulial Space

The rulial multiway graph defines all paths that can be followed by all non-deterministic Turing machines. At each node in this graph there is therefore an outgoing edge corresponding to any possible Turing machine transition from the configuration corresponding to that node. But what if one considers just a single deterministic Turing machine? Its evolution is then a single path within the rulial multiway graph.

Consider for example the $s = 2$, $k = 2$ Turing machine:

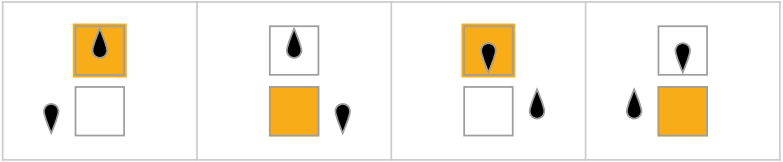

This machine evolves from a blank tape according to:

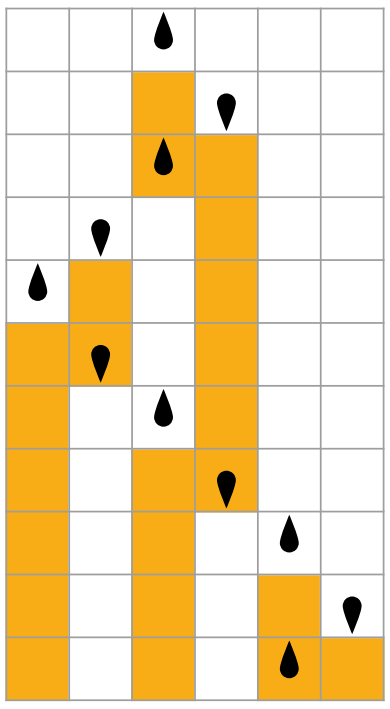

This corresponds to a path in the rulial multiway graph:

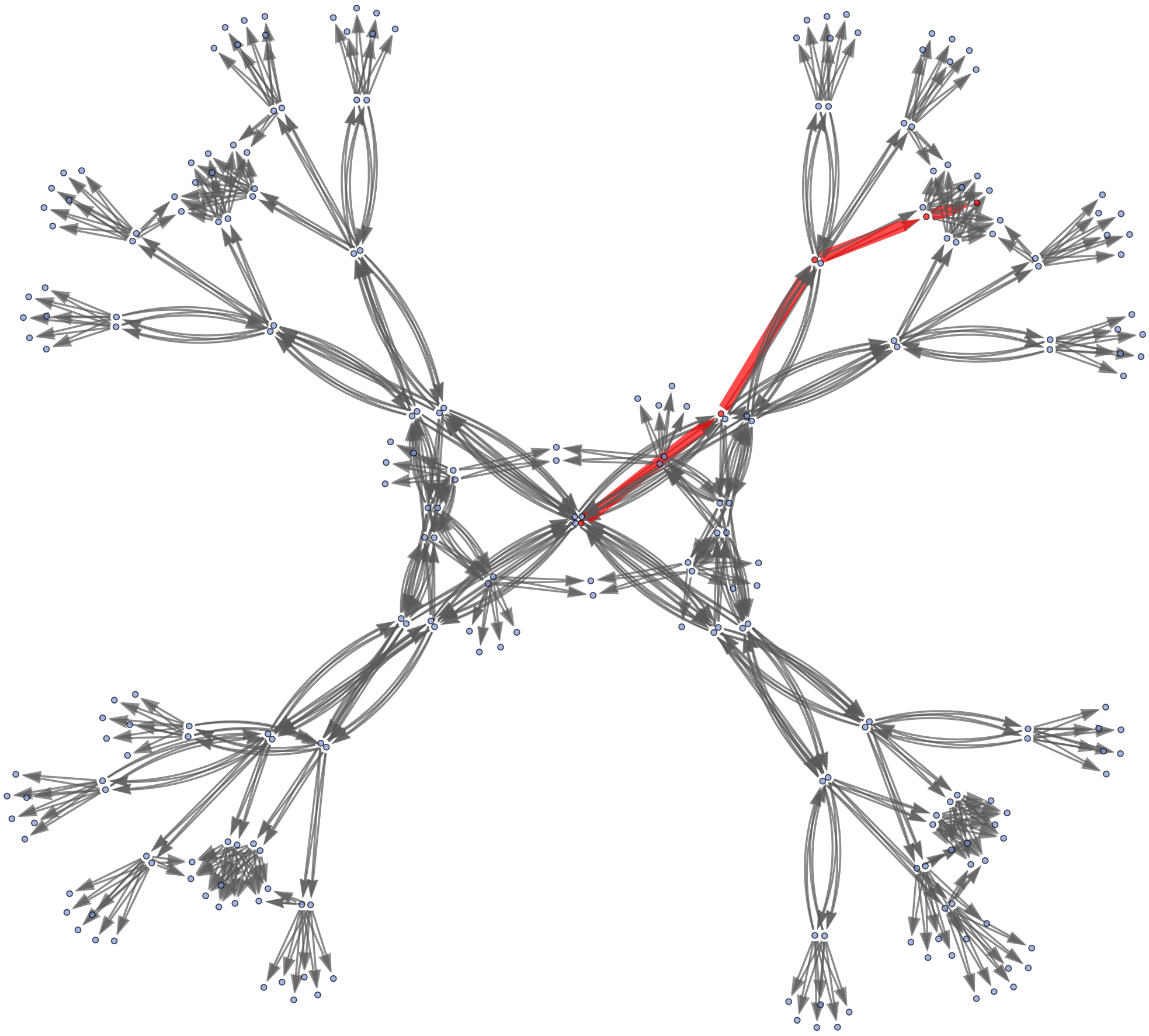

With a different initial condition

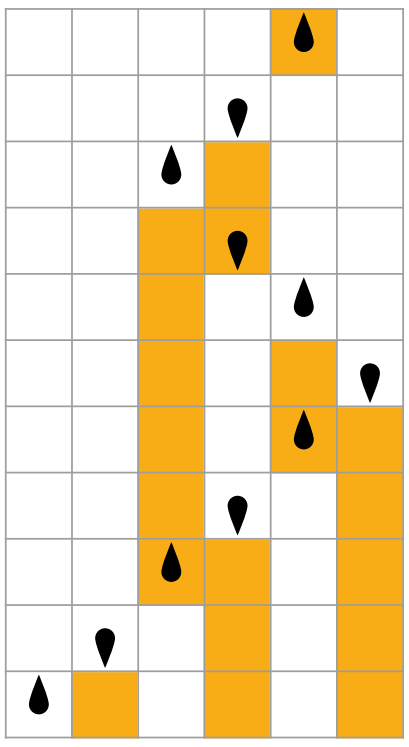

the path can be very different:

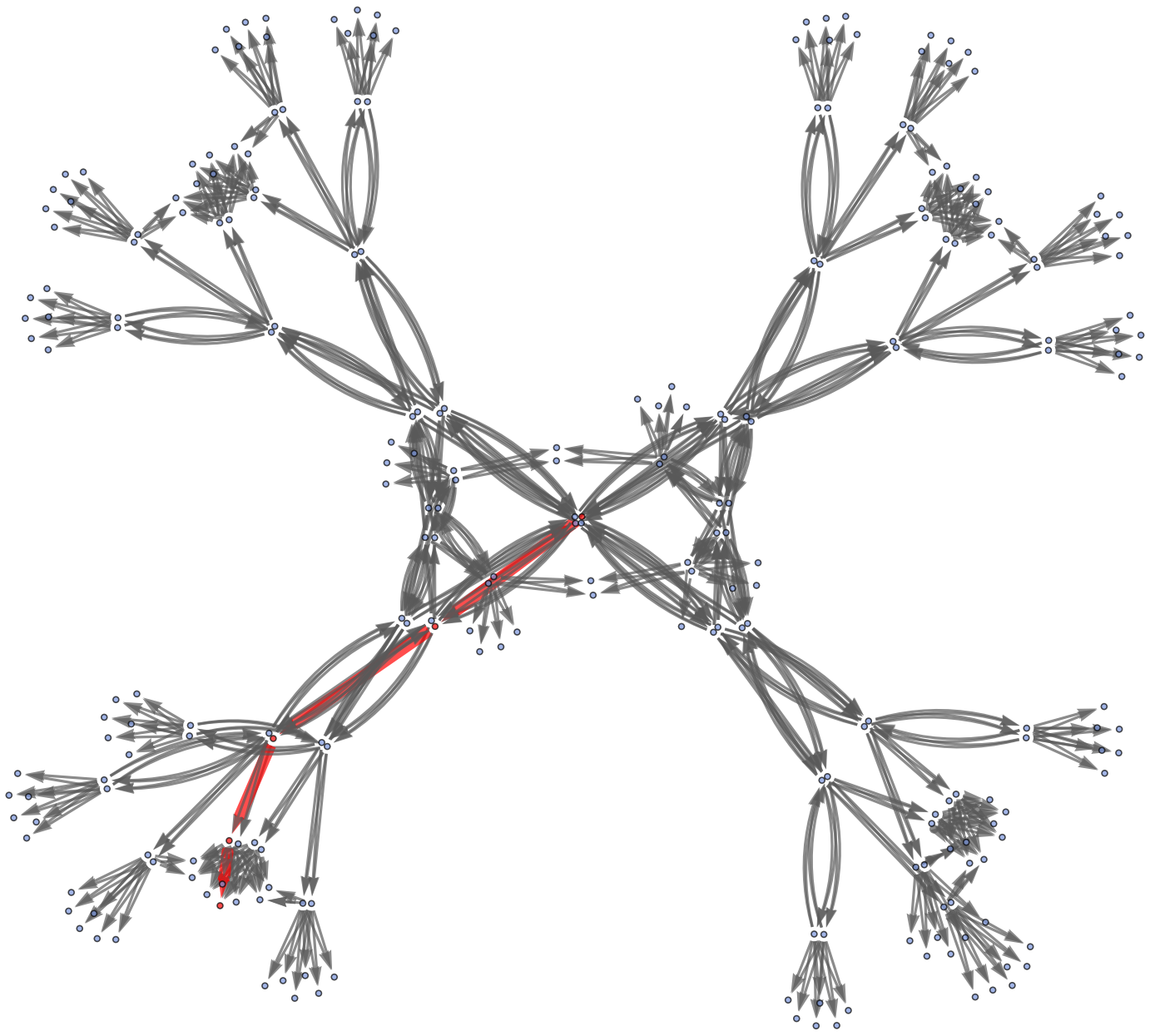

This pictures might make it seem that the paths corresponding to the evolution of deterministic Turing machines are geodesics in the rulial multiway graph. But in general they are definitely not. For example, starting from a blank tape, the rule

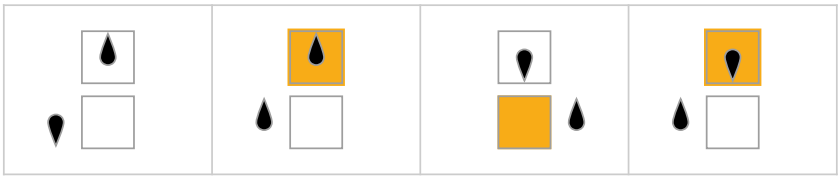

follows this path on the rulial multiway graph:

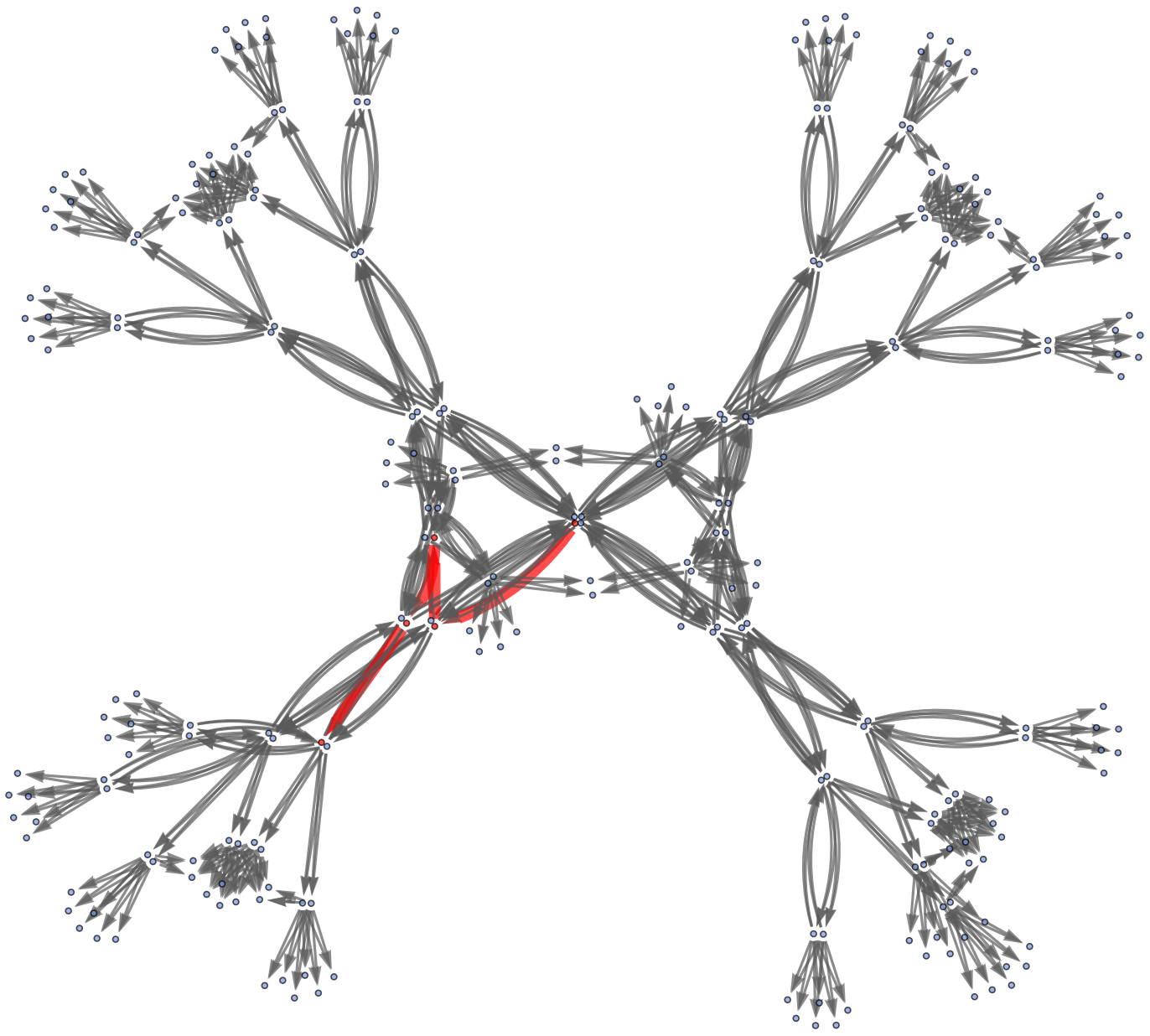

If one allows any possible Turing machine transition at every step, then one can follow a geodesic path from a node corresponding to an initial condition to a node corresponding to any other configuration. But if one restricts oneself to the transitions in a particular (deterministic) Turing machine, then there will in general be many configurations one will never reach, and even those that one can reach, one may reach by a circuitous route in the rulial multiway graph.

Given a particular configuration (corresponding to a node in the rulial multiway graph), there may be many deterministic Turing machines that can reach it from a given initial state. One can consider each of these Turing machines to be implementing a certain algorithm. So then the "optimal algorithm" will be the one which is shortest among deterministic Turing machine paths. As I just mentioned, this won't typically be the shortest possible path: that will usually be achieved by a non-deterministic Turing machine with a particular sequence of transitions. But there is still an optimal case (i.e. an "optimal algorithm") among deterministic Turing machines.

But now we can ask across all possible deterministic Turing machines where they can reach in the rulial multiway graph. Here is the result for 4 steps for the $s = 2$, $k = 2$ Turing machines we have been considering:

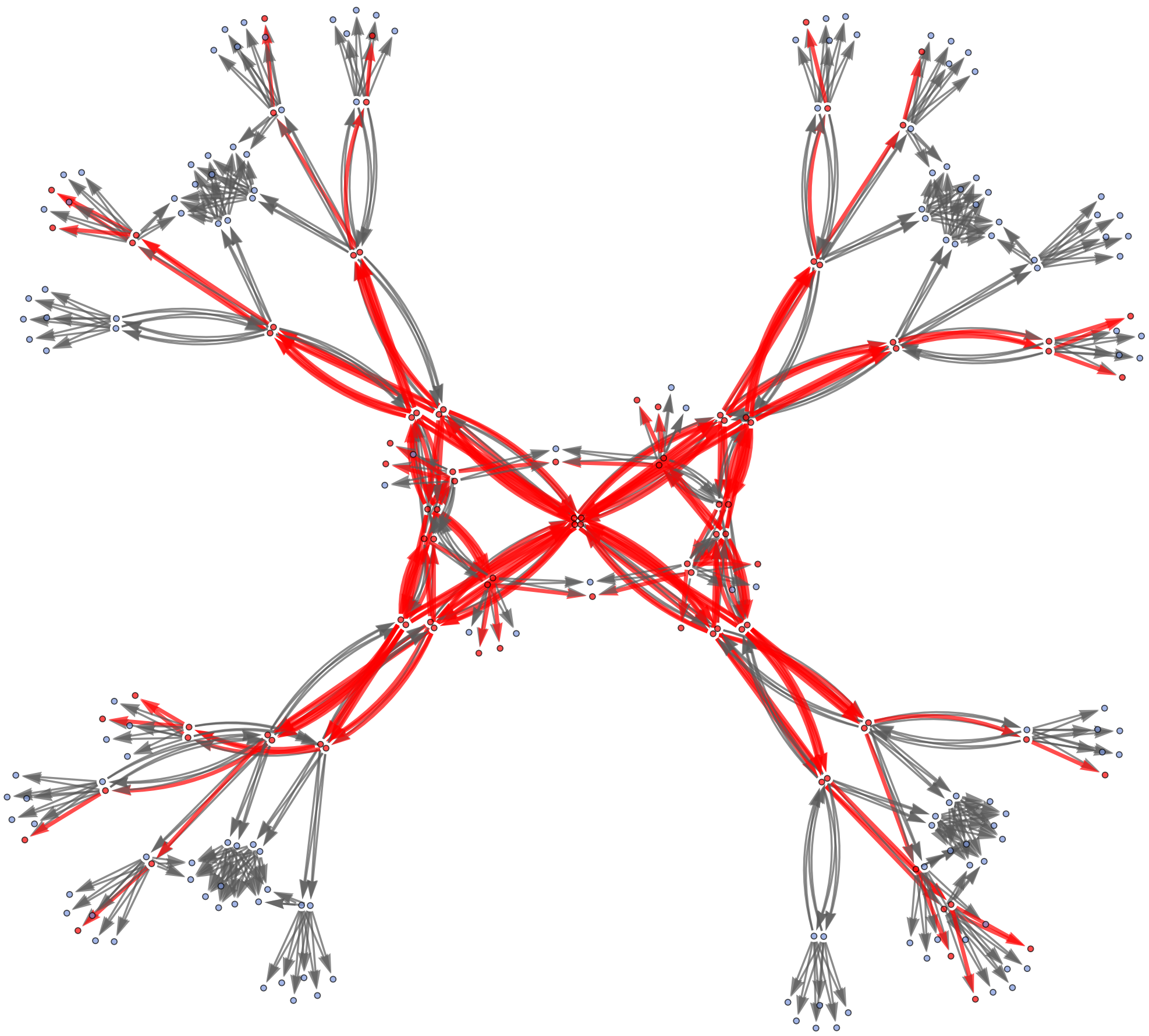

These are the results for 1, 2 and 3 steps:

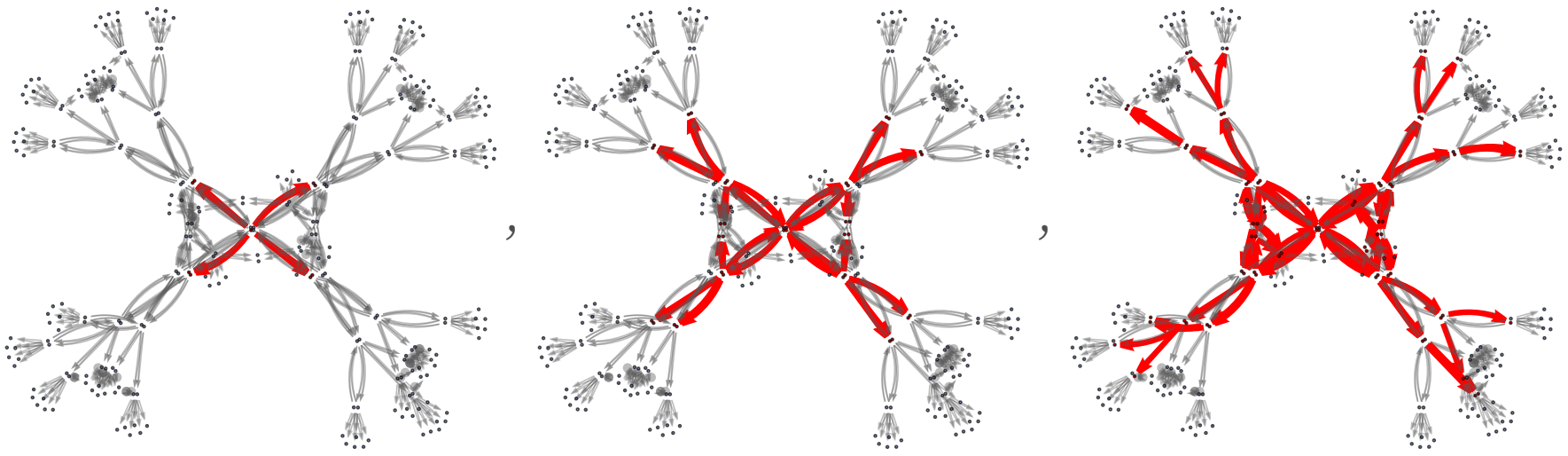

What is the significance of this? Essentially what we're seeing is a comparison of what can be achieved with deterministic computation versus non-deterministic. Taking the 4-step case as an example, the "background" gray rulial multiway graph shows what can be achieved with arbitrary non-deterministic computation in 4 steps. The red region is what deterministic computation can achieve in the same number of steps.

In a sense this is a very simple empirical analog of the P vs. NP problem. Unlike the real P vs. NP case, we're not allowing arbitrary polynomial-time algorithms here; we're just looking at possible $s = 2$, $k = 2$ Turing machine algorithms running specifically for 4 steps. But if we were to generalize this appropriately, P = NP would imply that the "red region" must in some limit in effect "reach anywhere in the graph".

A little more precisely, the official definition of P and NP is for decision problems: you start from some initial condition which defines an instance of a problem ("Is this Boolean formula satisfiable?" or whatever), and the system must eventually evolve to a state representing either "yes" or "no". We can imagine setting things up so that the outcomes correspond to particular configurations of the Turing machine. The inputs are then also encoded as initial configurations of the Turing machine, and we want to know what happens as we consider inputs of progressively larger sizes. We can imagine drawing the rulial multiway graph so that progressively larger inputs are shown "progressively further from the center". If we consider non-deterministic Turing machines (associated with the class of NP computations), then the shortest computation will be a geodesic path in the rulial multiway graph from the input configuration to the outcome configuration. But for deterministic Turing machines (associated with the class P) it will in general be some much more circuitous path.

The standard P vs. NP problem asks about limiting behavior as one increases the size of the input computation—and one might imagine that the question could be "geometrized" in terms of some continuum limit of the rulial multiway graph. Of course, there is no guarantee that any reasonable limit exists. And it could perfectly well be that the question of whether $P \subset NP$ is actually undecidable, or in other words, that no finite proof of it can be given within a standard axiom system (such as Peano arithmetic or ZFC set theory).

One could imagine empirically testing more and more Turing machines, and seeing how they perform on an NP-complete problem. One might think of plotting their running times as a function of $n$ for increasingly large $n$. For a while a particular Turing machine might be the winner. But then another one might take over. And there might be no end to how many "surprises" would occur as one increases $n$. (Somehow this is reminiscent of the story of the Skewes number and whether **LogIntegral**[n] > **Prime**[n].)

In computational complexity theory one usually thinks about explicitly constructing optimal algorithms by standard "engineering-like" methods. But I think there's a lot to be learned from a more empirical approach—in which one doesn't try to construct optimal algorithms, but just finds them "in the wild" by searching all possible programs. In the past, it might not have seemed that "just searching for programs" would ever produce anything terribly interesting. But one of the big consequences of what I discussed in *A New Kind of Science* is that even among tiny programs—small enough that one can, for example, enumerate all of them—there's often very complex and potentially "useful" behavior. And that makes it seem much more reasonable to try to do "empirical computational complexity theory"—enumerating possible programs to find optimal ones, and so on.

Small programs can be thought of as ones with low algorithmic complexity. So searching for "fast programs" among these can be thought of as searching for programs with low time complexity, and low algorithmic complexity. We don't know exactly how strong the constraint of low algorithmic complexity is, but from what I've seen in the computational universe (and what's embodied in things like the Principle of Computational Equivalence), it seems as if it's not such a big constraint.

I studied "empirical computational complexity theory" a bit in *A New Kind of Science*, notably for Turing machines. And one of the interesting observations was that the optimal algorithm for things was often not something that could readily be constructed in a step-by-step engineering way. Instead, it was something that one basically could only find pretty much by doing a search of possible programs—and where there usually didn't seem to be a "general form" that would "apply for all $n$", and let one readily deduce the properties of the $n \to \infty$ limit. In other words, it didn't seem like what we'd now think of as the sequence of paths in rulial space would show any sign of converging to a "continuum limit".

A decent example of all this occurs in sorting networks. Imagine that you are given a collection of $n$ numbers to sort. You can straightforwardly do this by making about $n^2$ pairwise comparisons, and it's pretty easy to optimize this a bit.

But explicit searches have revealed that the actual optimal networks for successive $n$ are:

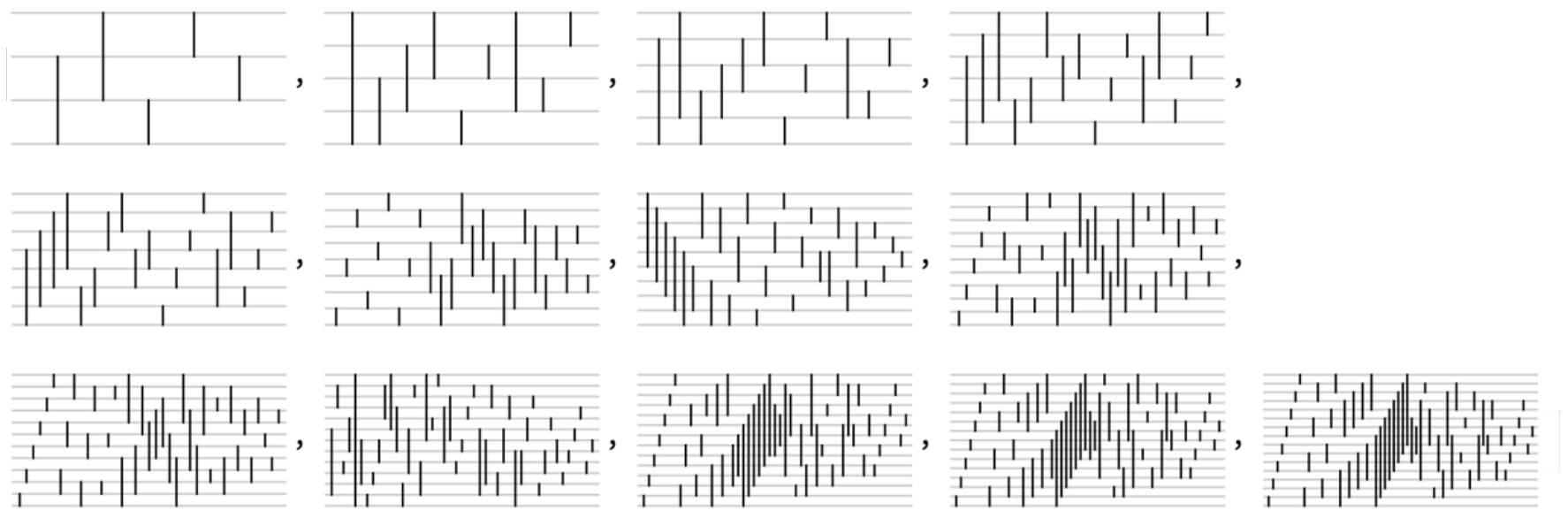

What's notable is how complicated and "random" they look; there doesn't seem to be any obvious pattern to them (and my guess is that there fundamentally isn't). Here's a plot of the sizes of these networks (divided by $n^2$):

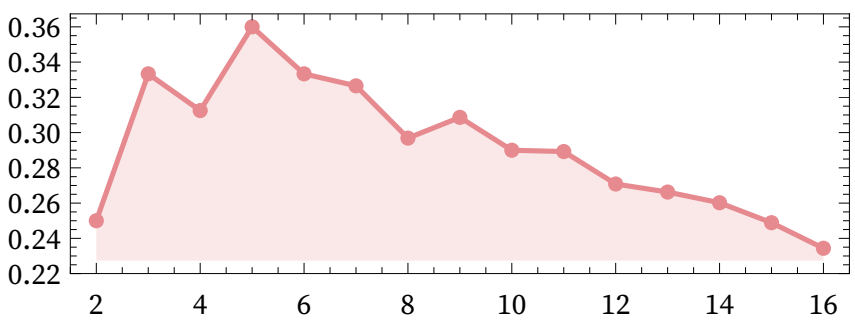

It's worth noting that these are deterministic networks. One could also imagine non-deterministic networks (and indeed one could construct a rulial multiway graph by considering all possible successive placements of pairwise comparisons)—and in a non-deterministic network it's always possible to sort $n$ numbers in at most $n - 1$ steps.

# The Space of Deterministic Turing Machine Computations

We've just seen how the results of deterministic Turing machine computations lay out in the rulial multiway space of all possible non-deterministic Turing machine computations. But what happens if we just look at the graph of deterministic Turing machine computations on their own?

Here are the full rulial multiway graphs for 2 and 3 steps with the graphs of deterministic Turing machine computations superimposed, as before:

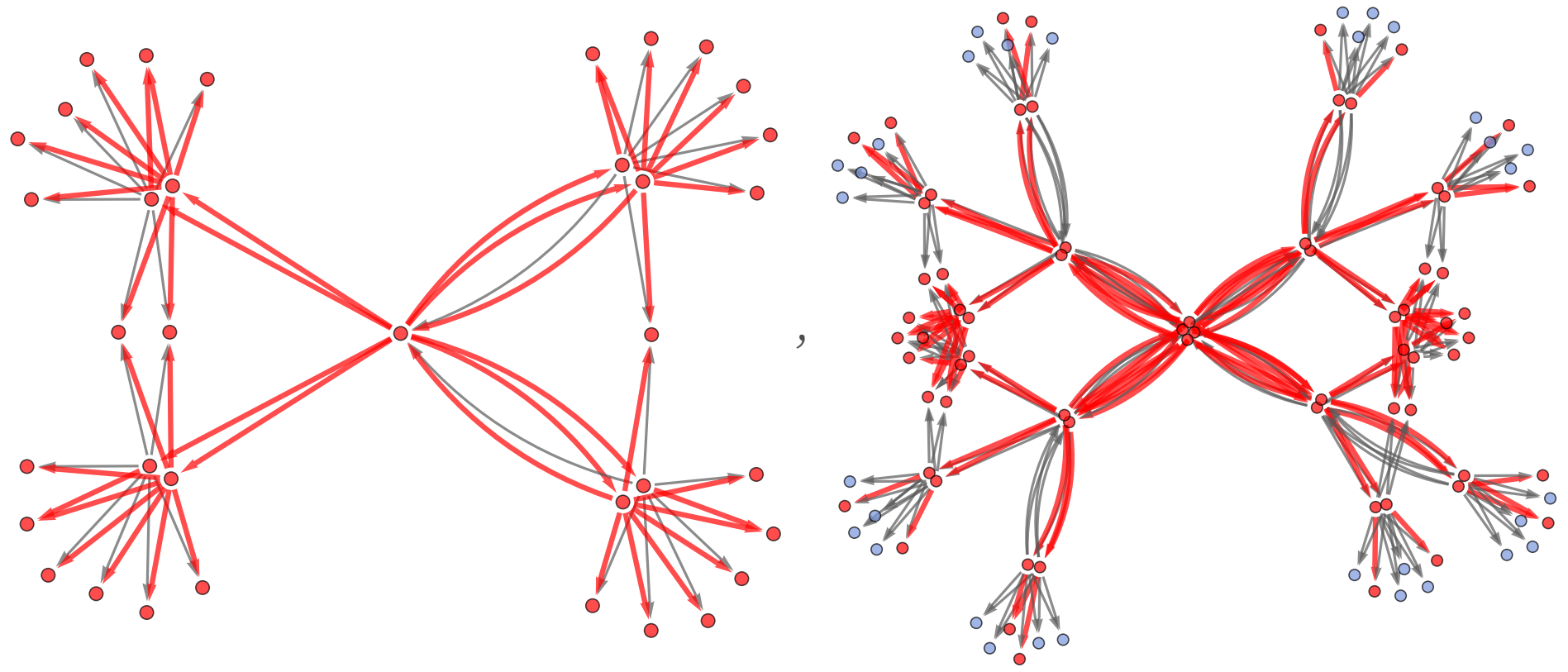

But now let's "pull out just the red subgraphs"—in other words, include as nodes only those configurations that at least one of the 4096 $s = 2$, $k = 2$ Turing machines can reach after 2 or 3 steps:

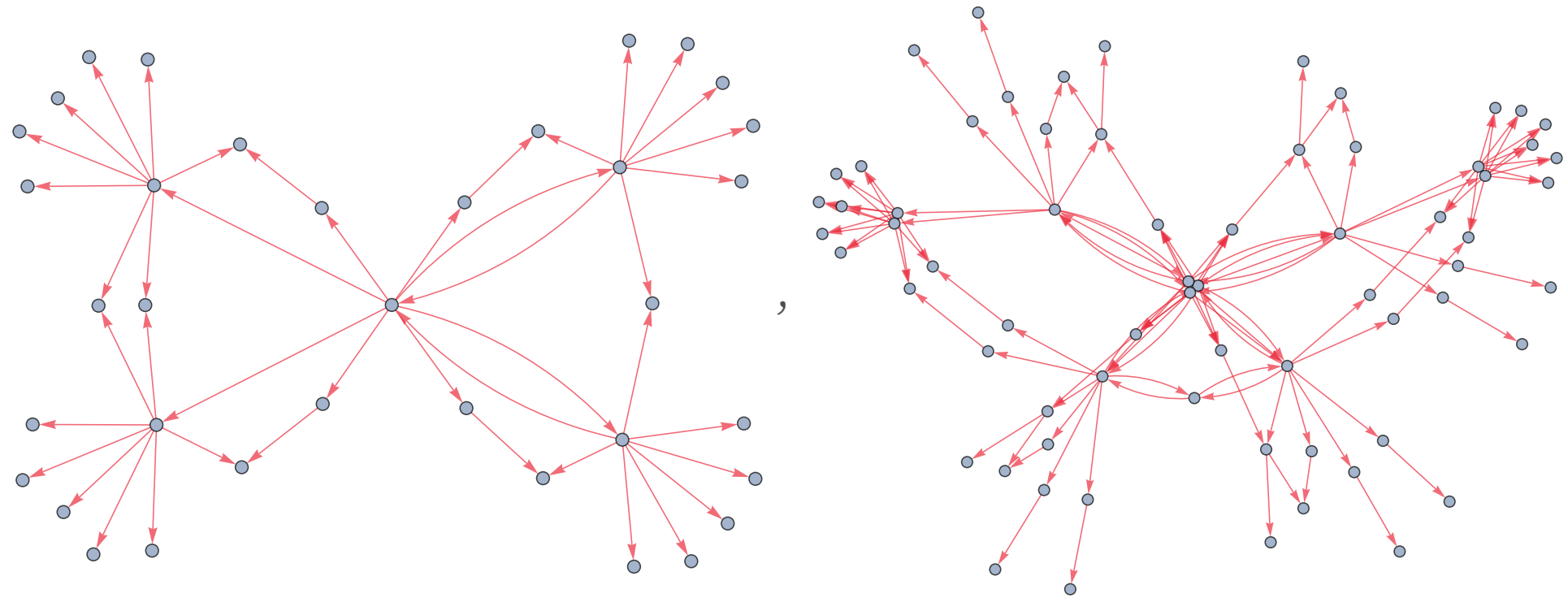

Notice that after 2 steps, deterministic Turing machines can still reach all 36 configurations that non-deterministic ones can reach (though not through quite as many paths). But after 3 steps, among all the deterministic Turing machines, they can only reach 68 possible configurations, while non-deterministic ones can reach 100 configurations.

For all possible non-deterministic Turing machines with $s = 2$, $k = 2$ the total number of configurations that can be reached eventually roughly doubles on successive steps:

{1, 9, 36, 100, 248, 576, 1280, 2768, 5856, 12 224, 25 216}

For deterministic Turing machines, however, the number of possible configurations that can be reached soon increases much more slowly:

{1, 9, 36, 68, 94, 144, 248, 322, 382, 458, 559, 659, 737, 823, 921, 1007}

And in fact there's an obvious bound here: at any given step, the most that can happen is that each of the 4096 $s = 2$, $k = 2$ Turing machines leads to a new configuration—or, in other words, the maximum number of configurations reached increases by 4096.

Looking at the differences on successive steps, we find:

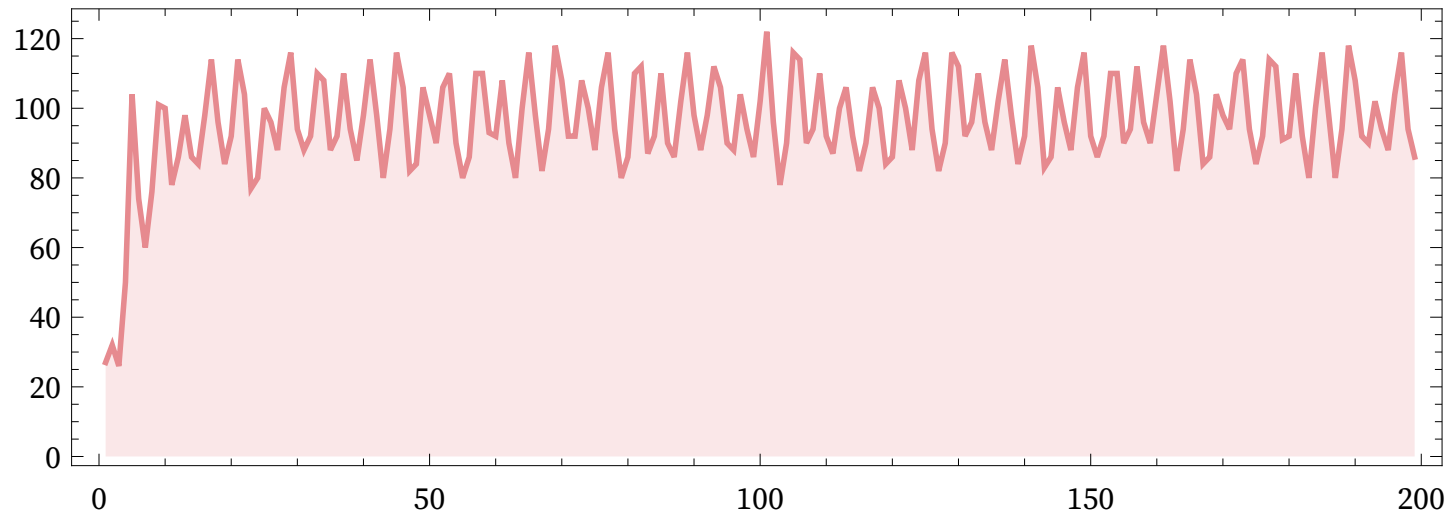

In other words, among all 4096 Turing machines, about 100 "novel configurations" are reached at each successive step. (The actual sequence here looks surprisingly random; it's not clear whether there's any particular regularity.)

Now let's look at the actual graphs formed. With 4 through 7 steps we get:

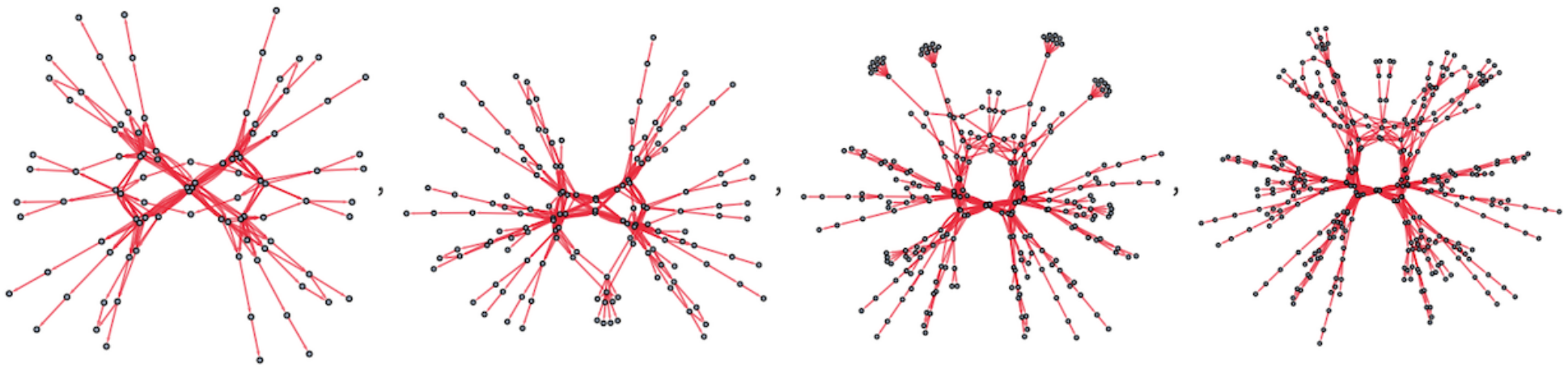

After 10 and 20 steps the results are:

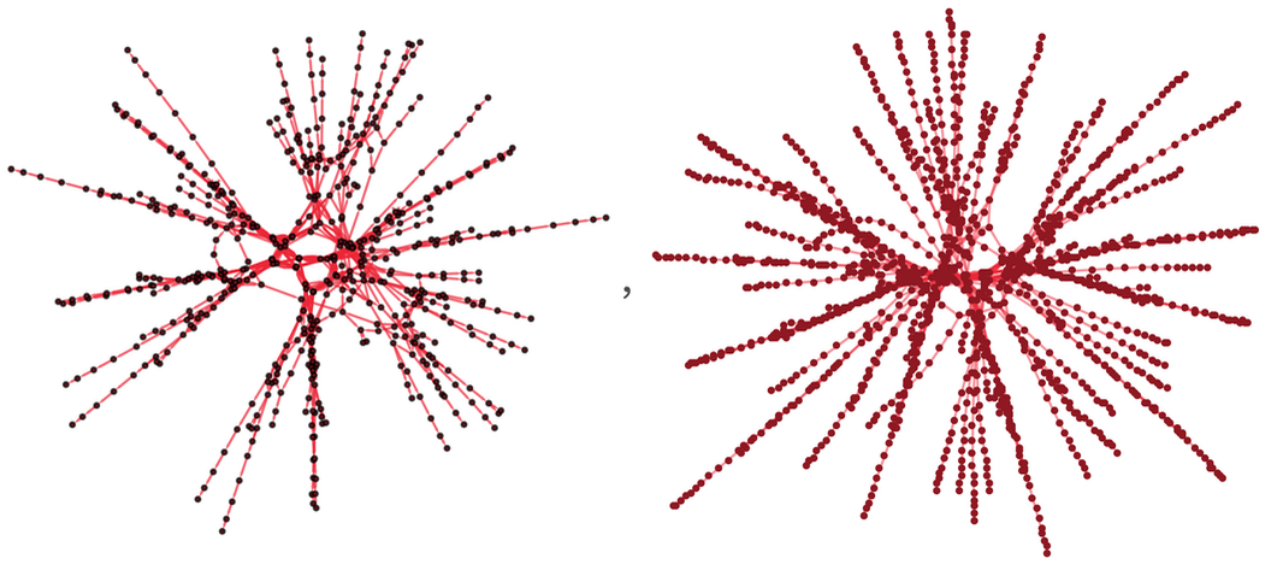

Here is the result after 50 steps:

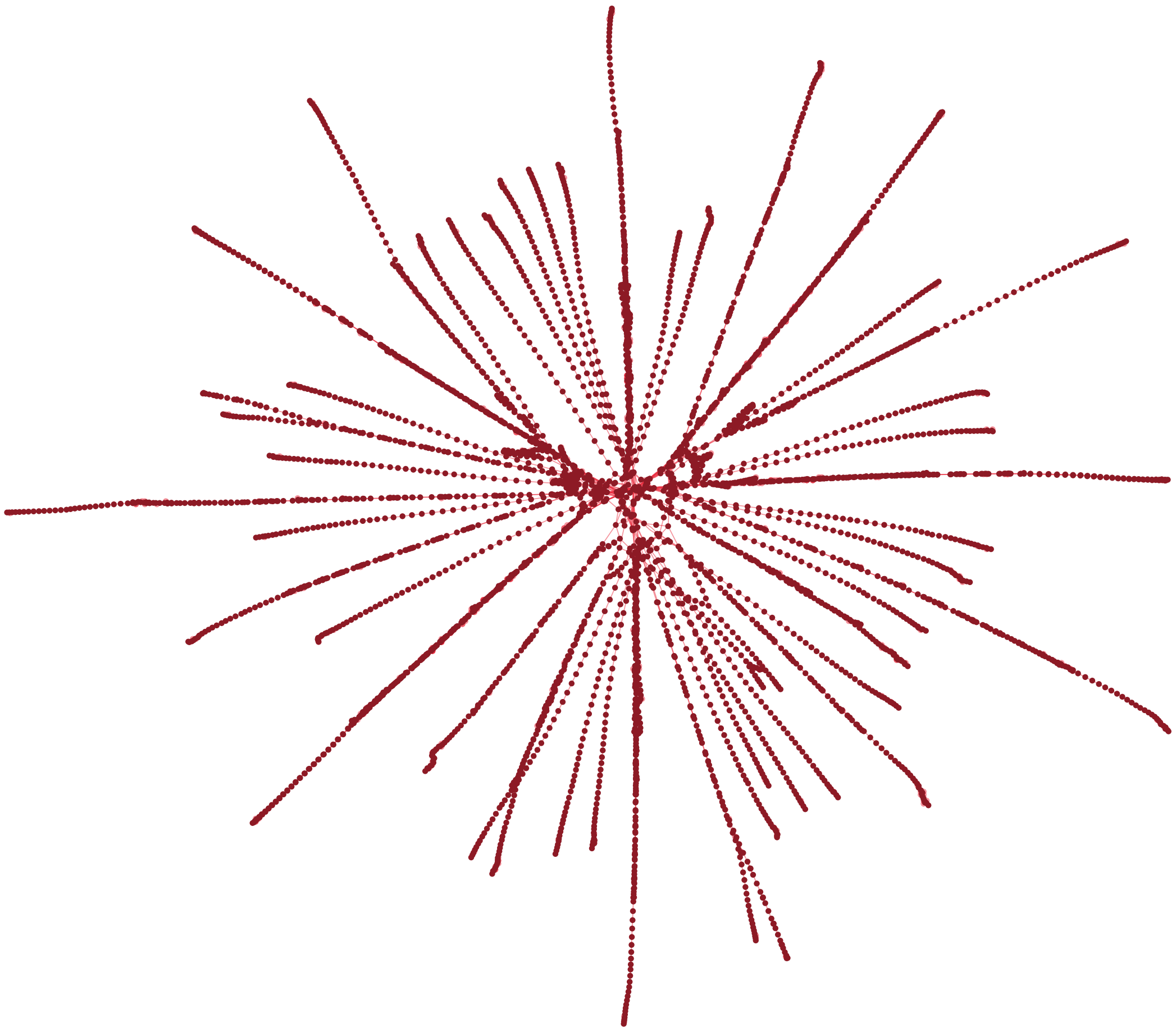

There's a surprising amount of structure in these graphs. There's a "central region" near the initial blank-tape configuration (shown highlighted below) in which many different Turing machines end up visiting the same configurations:

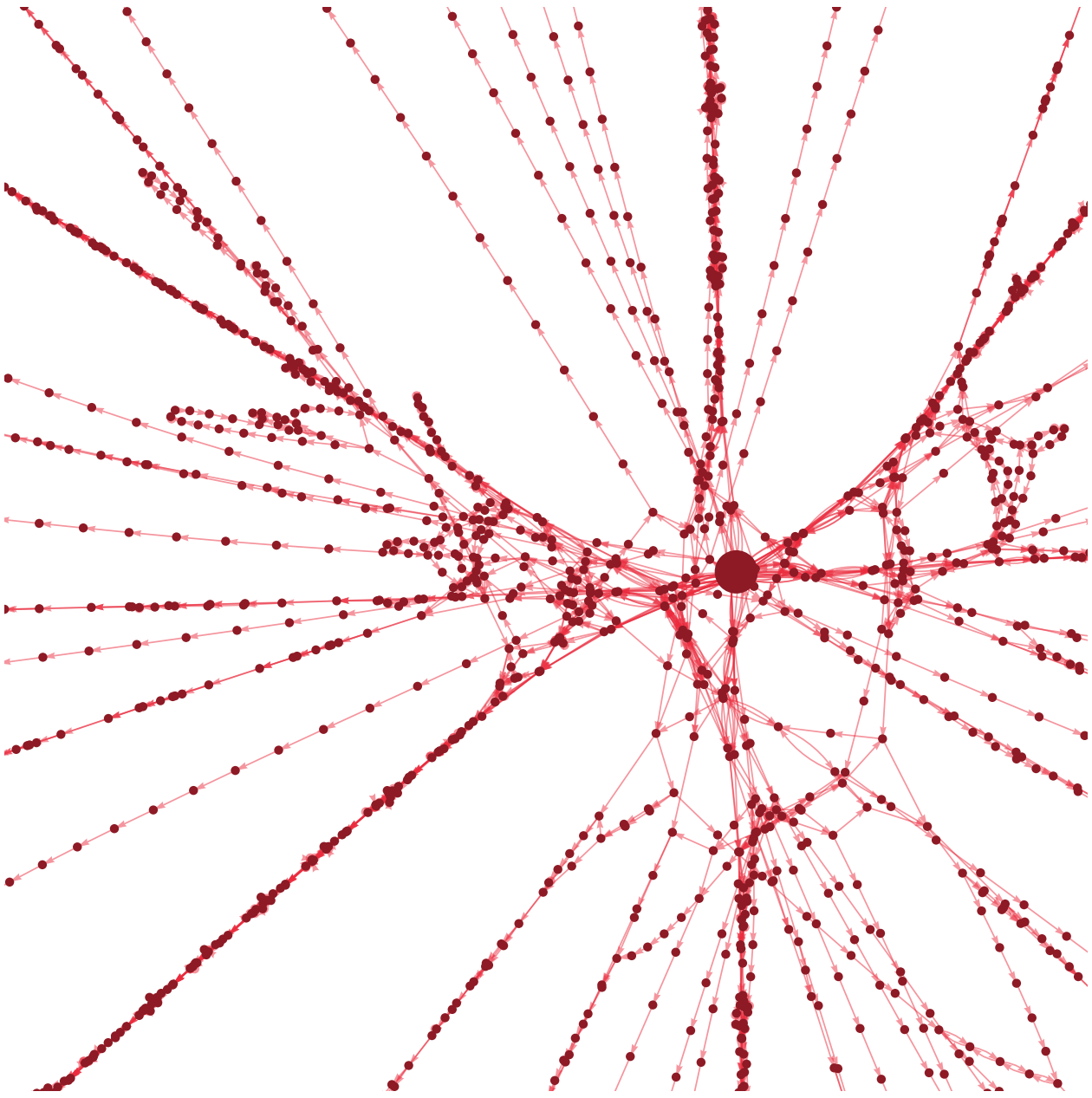

Here's a 3D rendering of this region:

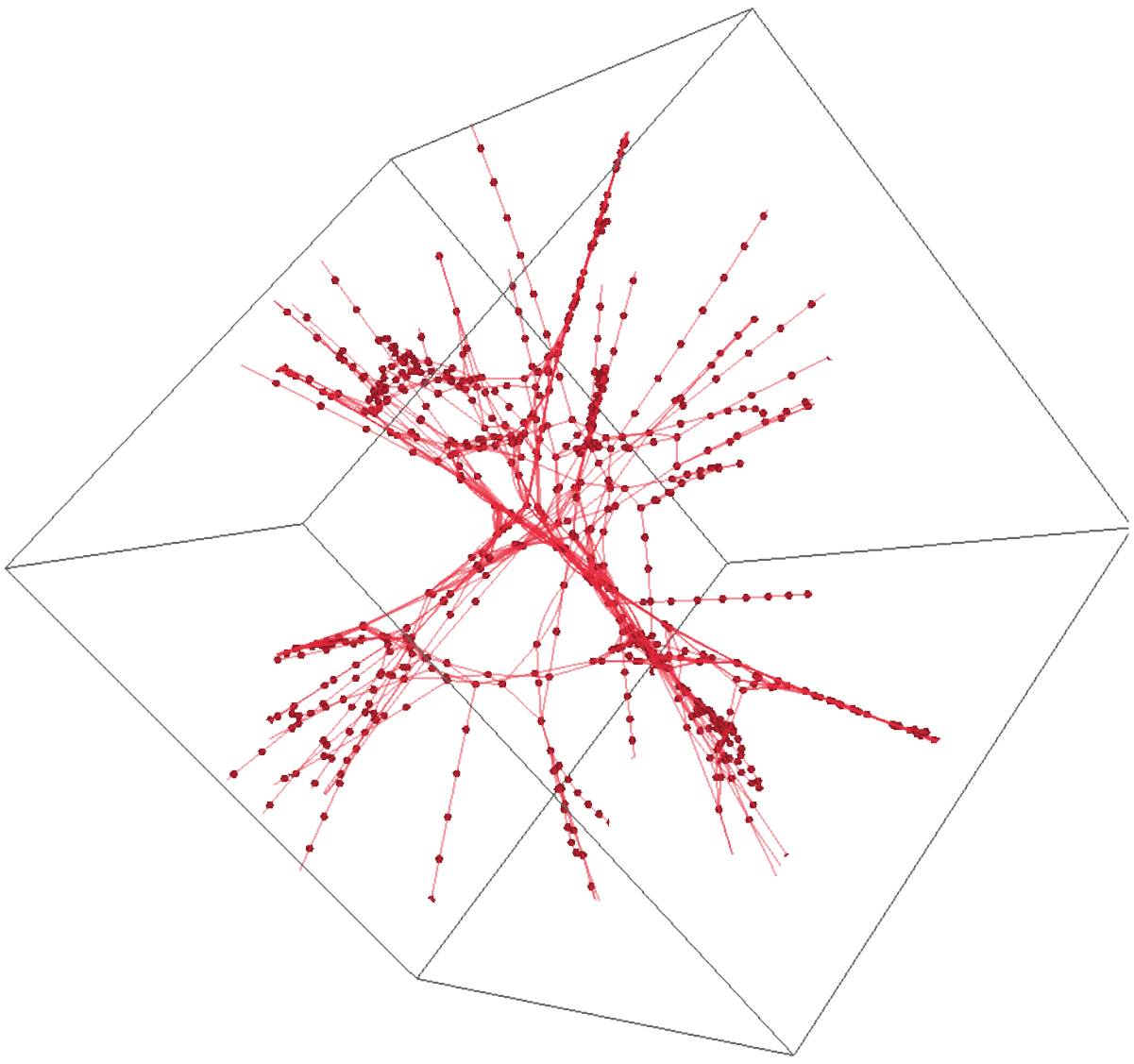

But away from this region there end up being "spokes" (about 100 of them) corresponding to Turing machines that "independently explore new territory" in the space of configurations.

What are those “configurations on the edge” like? Here are sorted collections of them for the first few steps:

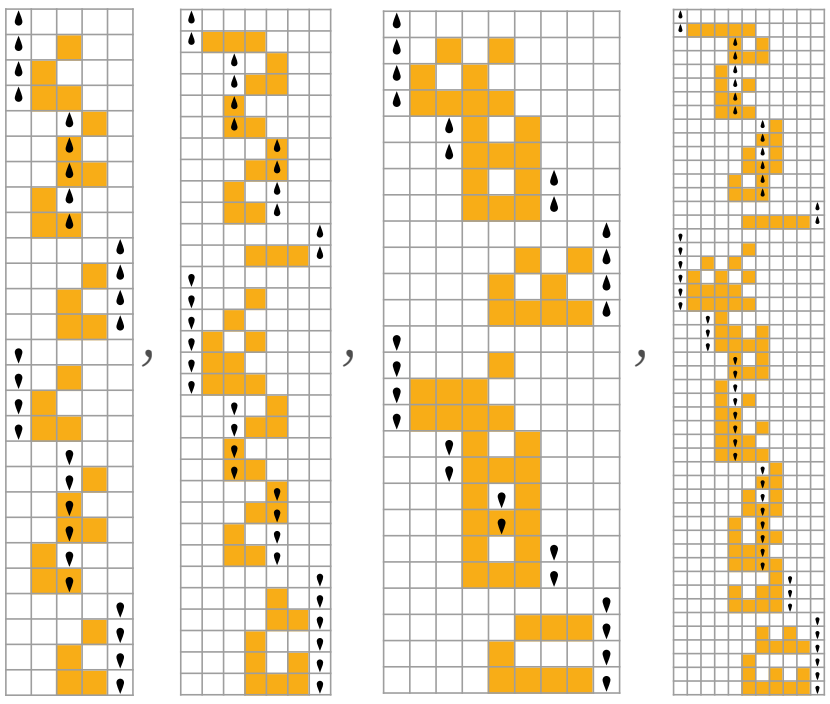

2 3 4 5

For comparison, here is the result for all configurations that can be reached by non-deterministic Turing machines after just 2 steps:

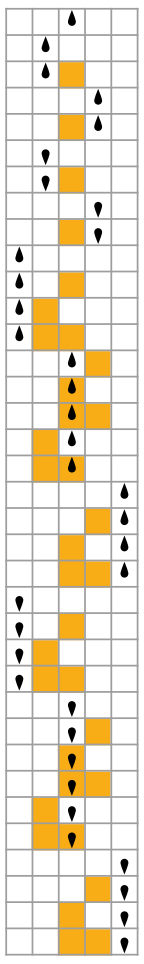

Here are the results for deterministic Turing machines after more steps:

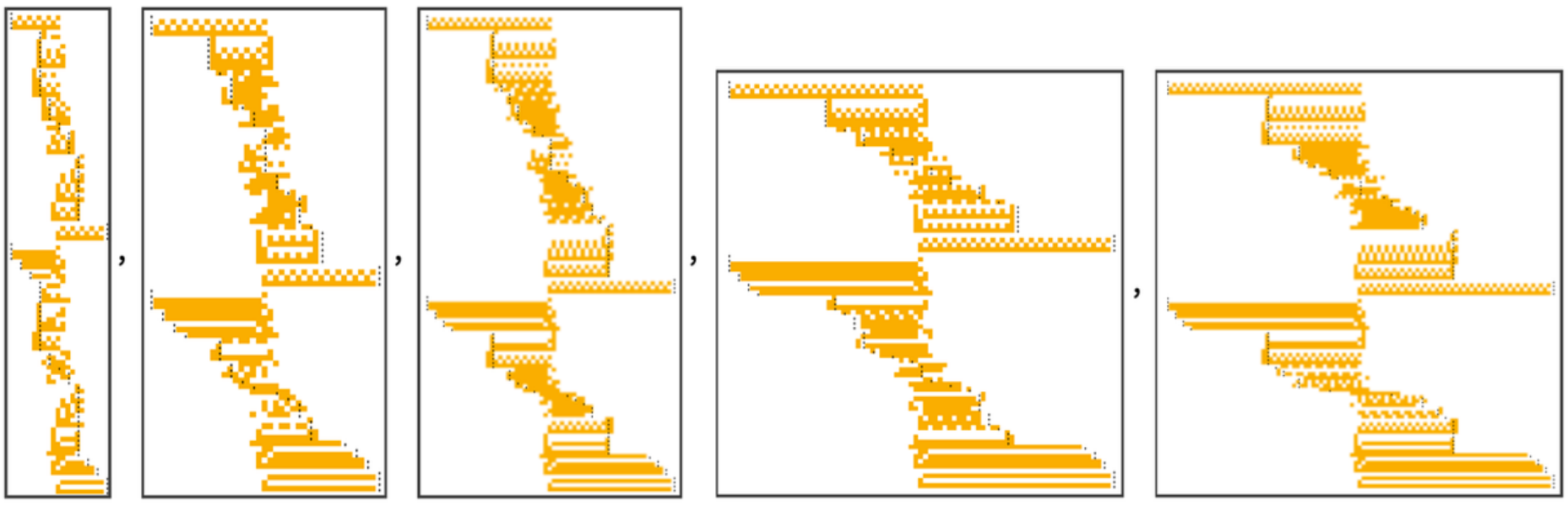

10 20 30 40 50

We can also ask which machines are the ones that typically "explore new territory". Here's the result for 30 steps:

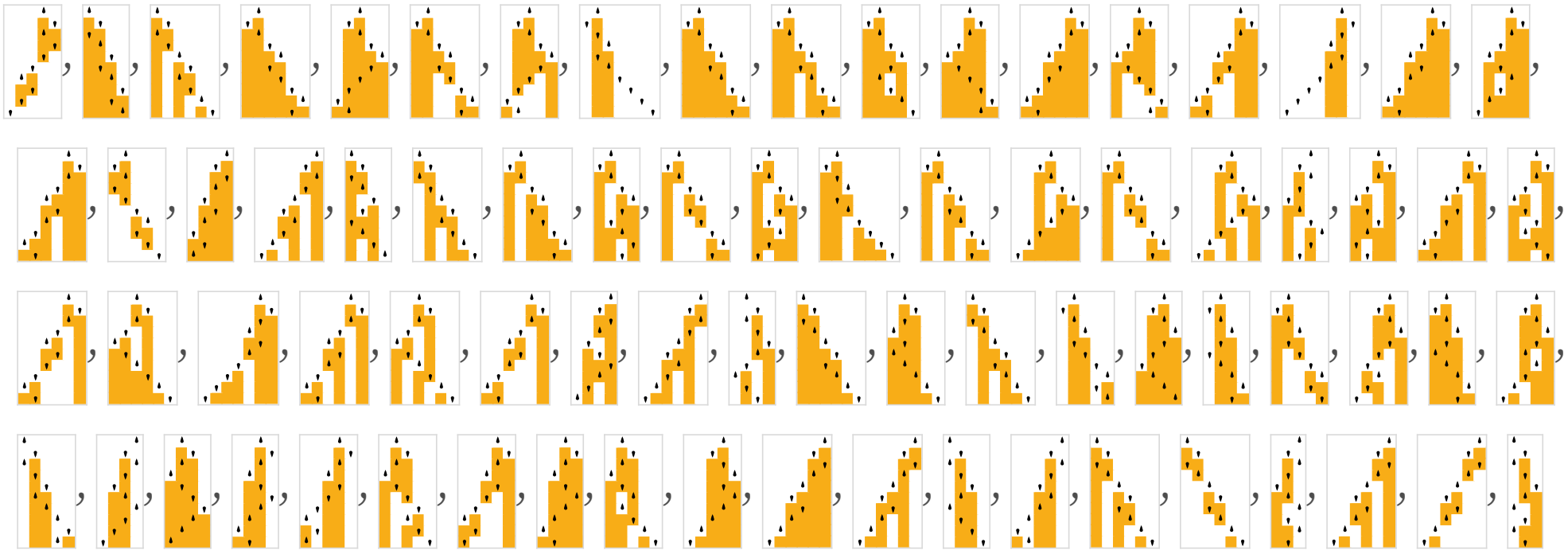

As we go to more steps, the graph of configurations that can be reached by deterministic Turing machines grows. But does at least the core of it reach some kind of limit after sufficiently many steps? We can get a sense of this by looking—as we have done so many times before—at the growth rate of the geodesic ball in the graph starting from the initial blank-tape configuration. The total number of new configurations that can be reached on each new layer of the geodesic ball is at most 4096—and in reality it's much smaller. Here are the numbers of new configurations added on successive layers at steps 100, 200, ..., 500 in the overall evolution:

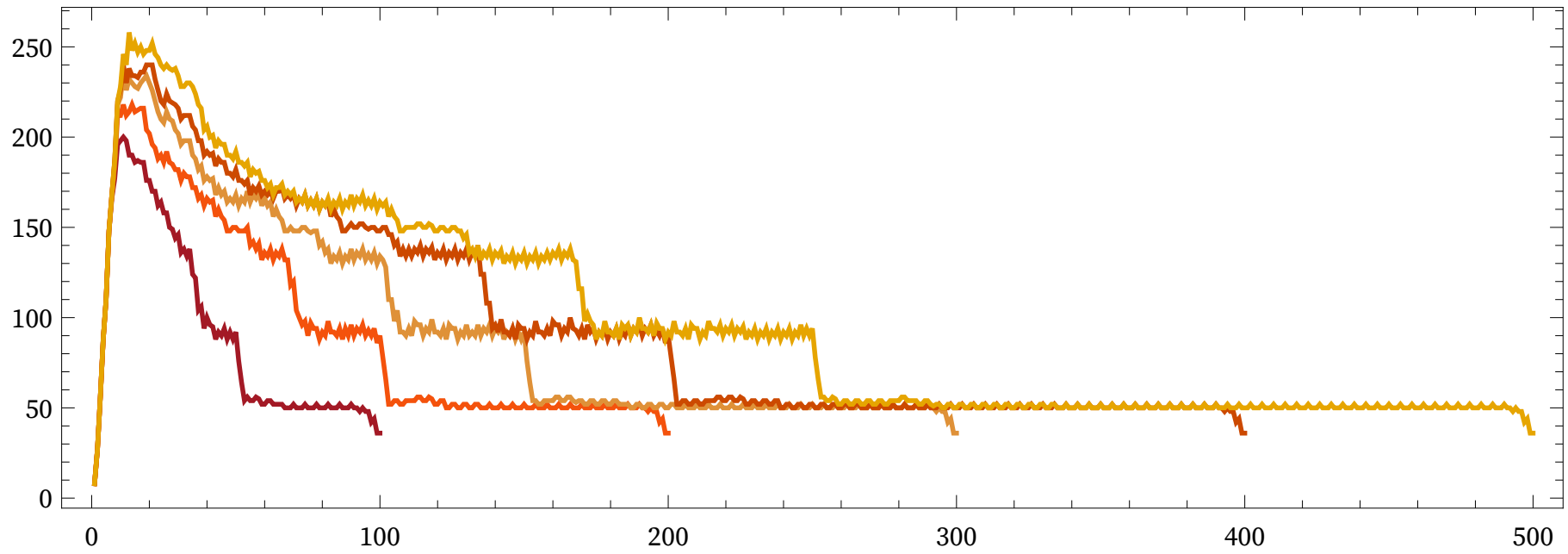

The cause of the "steps down" becomes clearer if one examines more closely the "spokes" in the graph above. Here's one example:

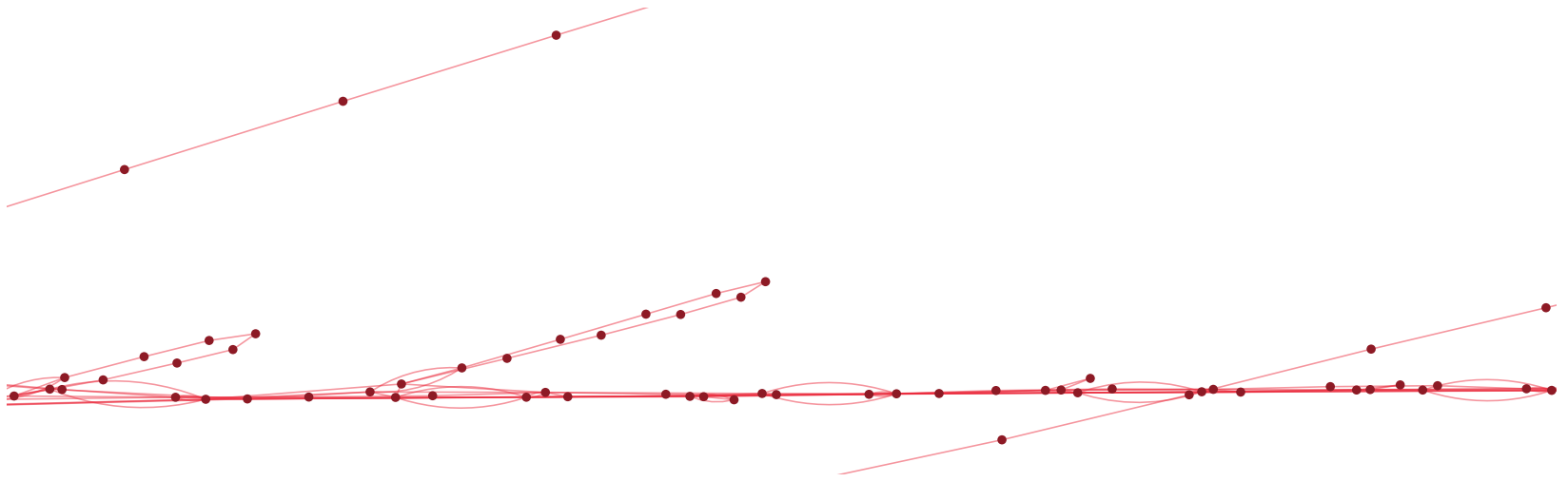

And basically what's happening is that multiple Turing machines are tracing out roughly the same sequences of configurations, but some do it "efficiently", while others "waste time", for example having the head flip around on alternate steps. The "inner parts" of the spokes—that are closer to the initial node—involve both "inefficient" and more efficient Turing machines. But the "inefficient" Turing machines will simply not get as far, so they do not contribute to the outer layers of the geodesic ball. The final "step down" in the plot above—at basically half the total number of steps used for the Turing machines—involves the "petering out" of the roughly half the Turing machines that effectively "waste half their steps".

For the "spoke" shown, here are the actual Turing machine histories involved (there are 8 machines that made identical copies of the first history):

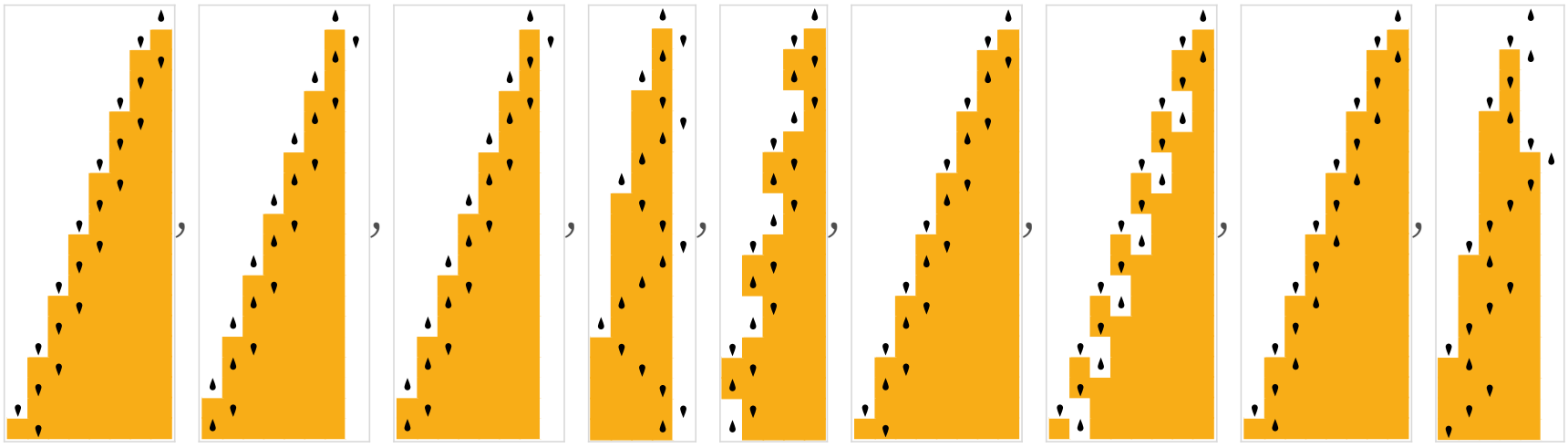

## The Cellular Automaton Analog

The kind of graphs we've just made for deterministic Turing machines can be made for any family of deterministic computational systems. And in particular they can be made for the (at least for me, much more familiar) case of the 256 $k = 2$, $r = 1$ cellular automata. (And, yes, it's somewhat amazing that in all these years I've never made such pictures before—though there's a note on page 956 of *A New Kind of Science* that gets close.)

Here are the results for 5 and 10 steps, starting from an initial condition containing a single black cell:

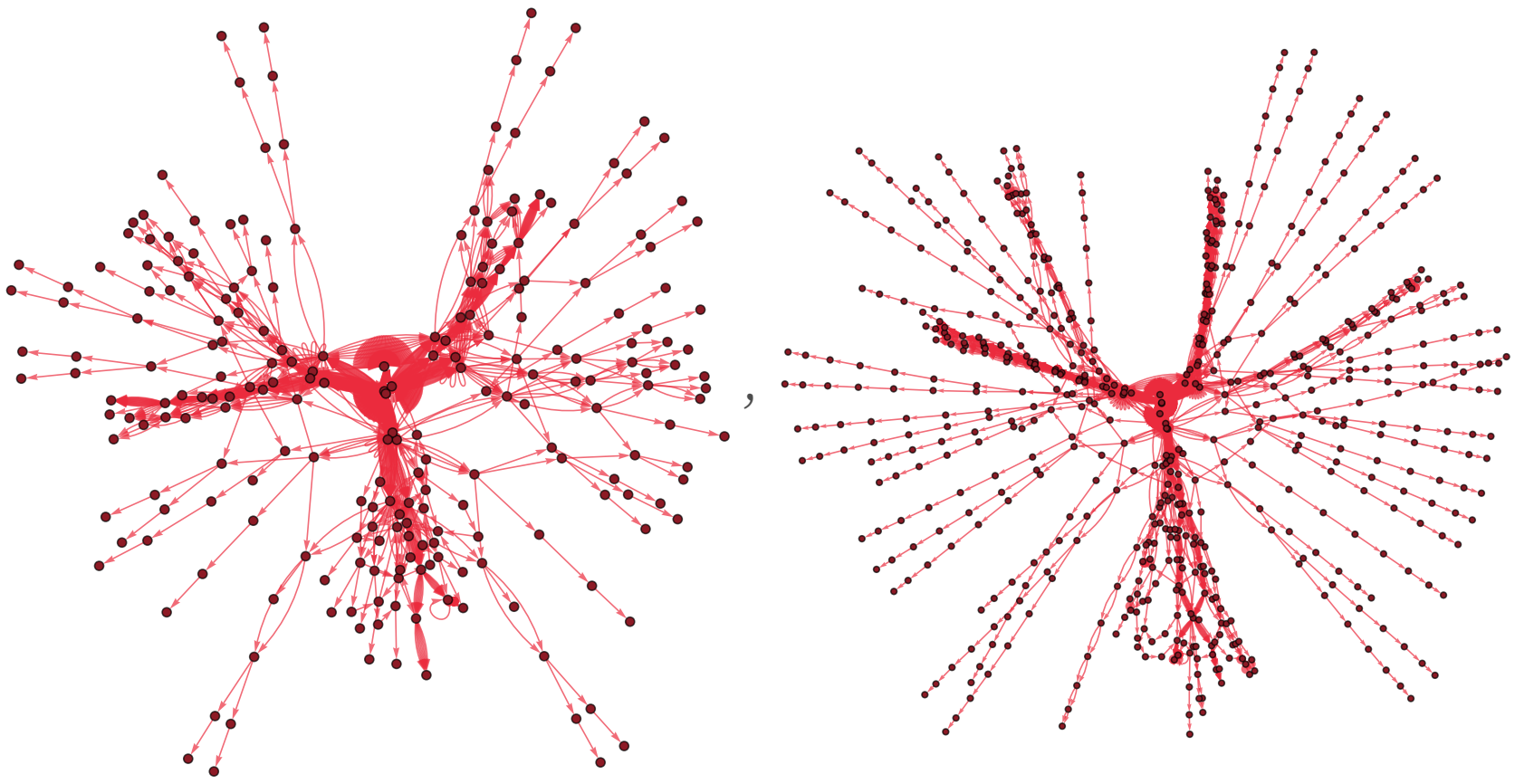

And here is the result for 50 steps:

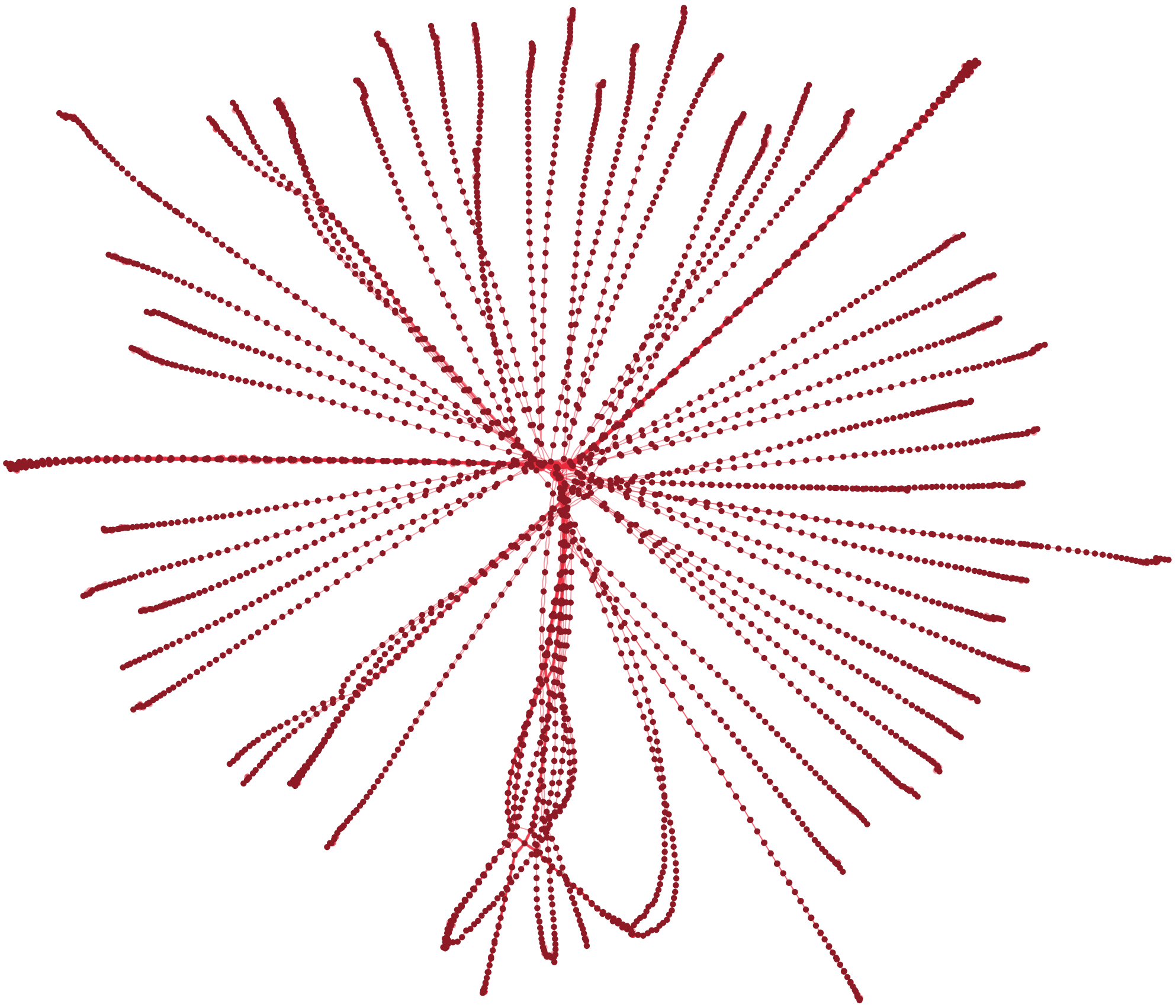

Here is the result for 10 steps, annotated with the actual cellular automaton evolution for the rules that “reach furthest”:

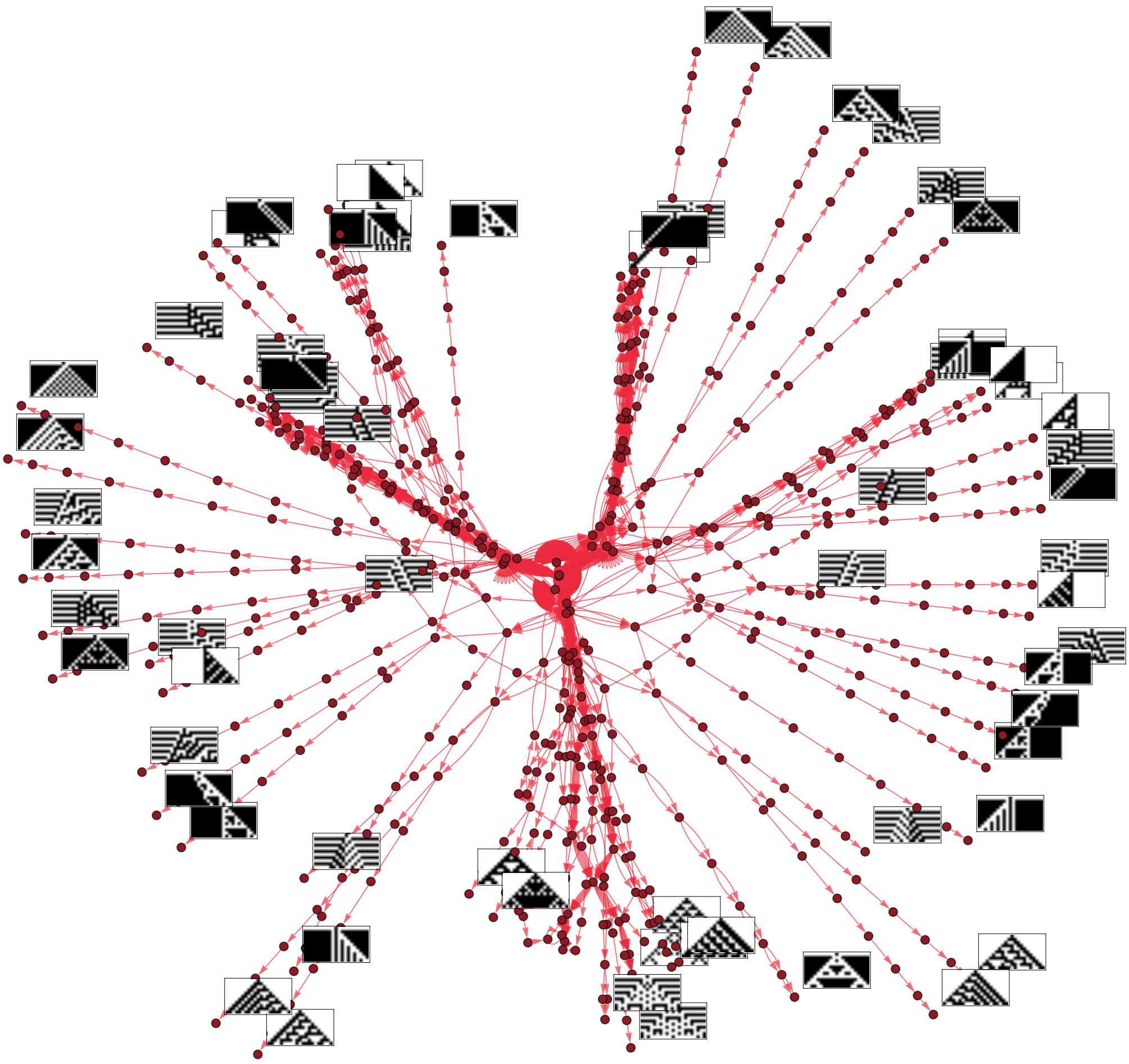

It’s slightly easier to see what’s going on if we include only even-numbered rules (which leave a blank state blank):

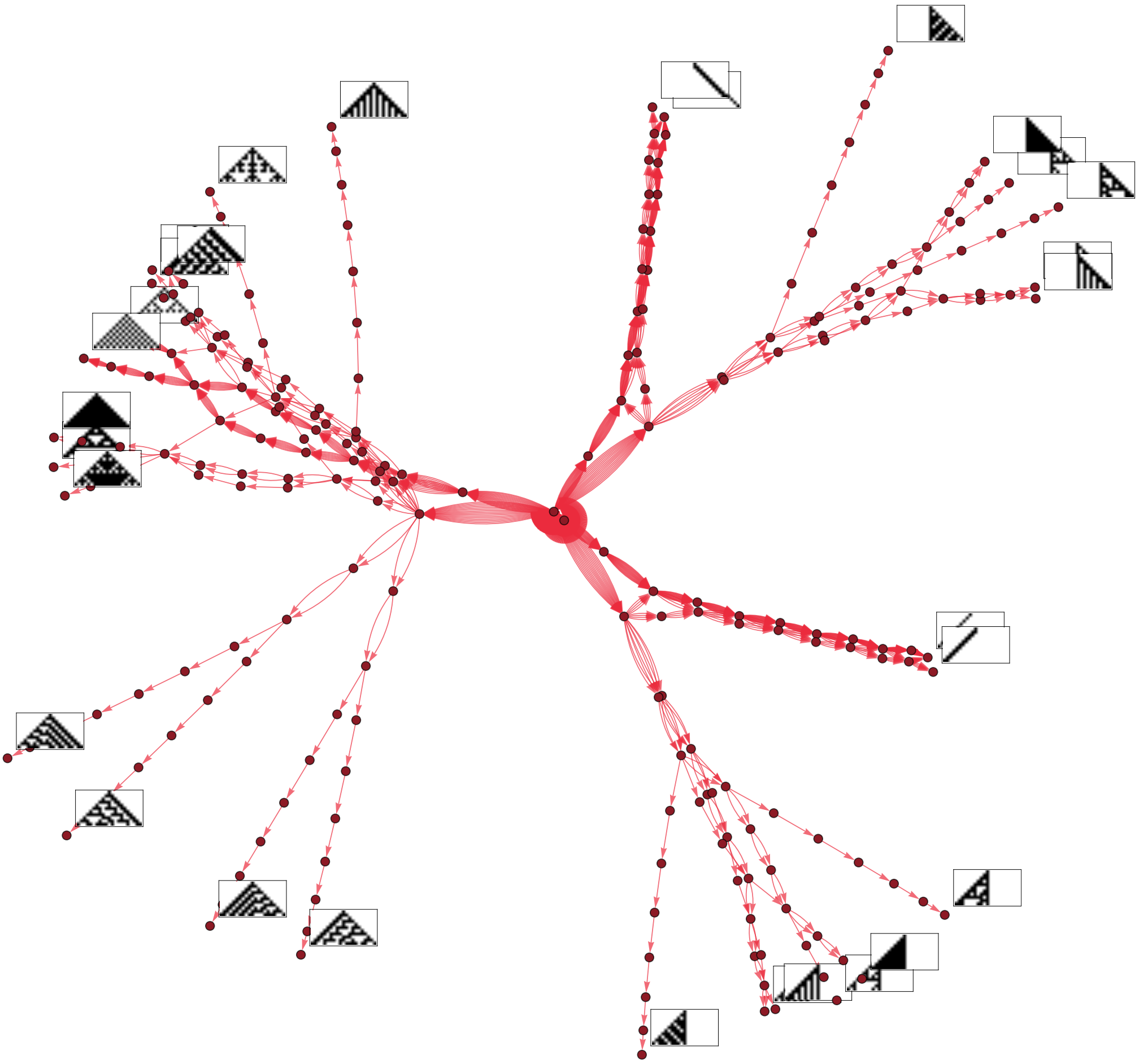

It's an interesting map of "cellular automaton space". (Note the presence of rule 30 on the lower left, and rule 110 on the lower right.)

The total number of new configurations explored by all rules on successive steps has a more regular form than for Turing machines:

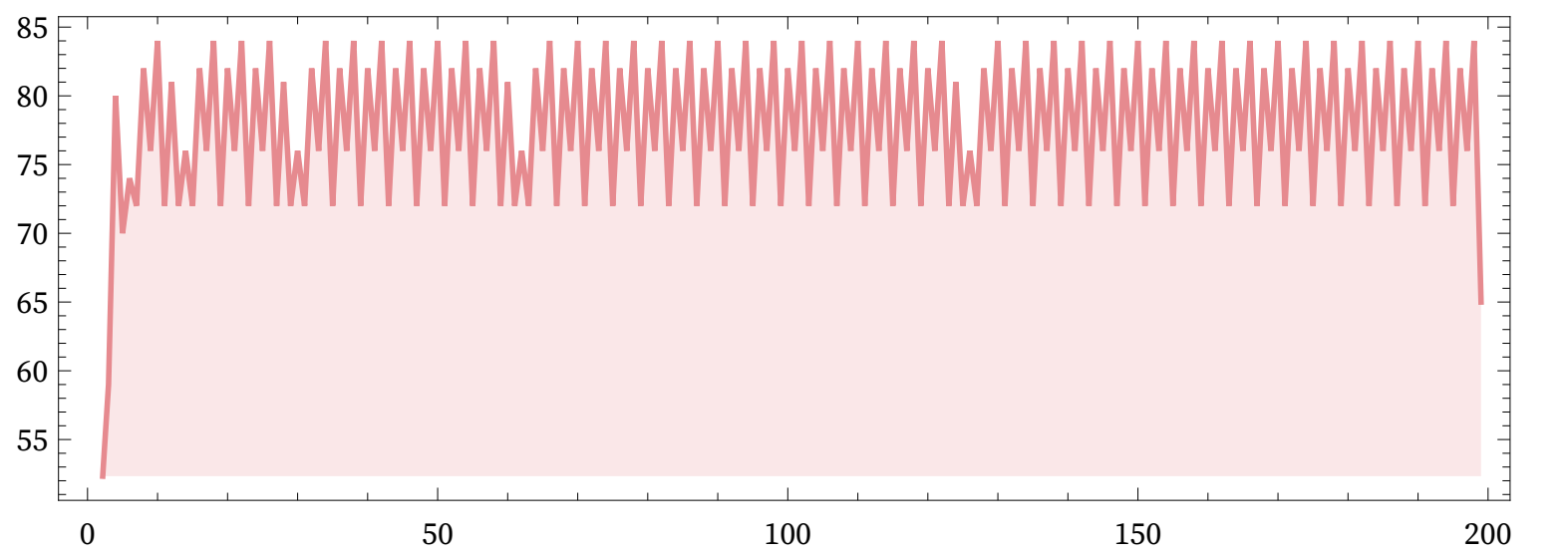

The result mostly alternates with period 4 between 72 and 84, though with dips at steps of the form $2^m$.

If we go for a certain number of steps (say 200), and then look at the geodesic ball centered on the initial condition, the number of configurations in successive layers is just:

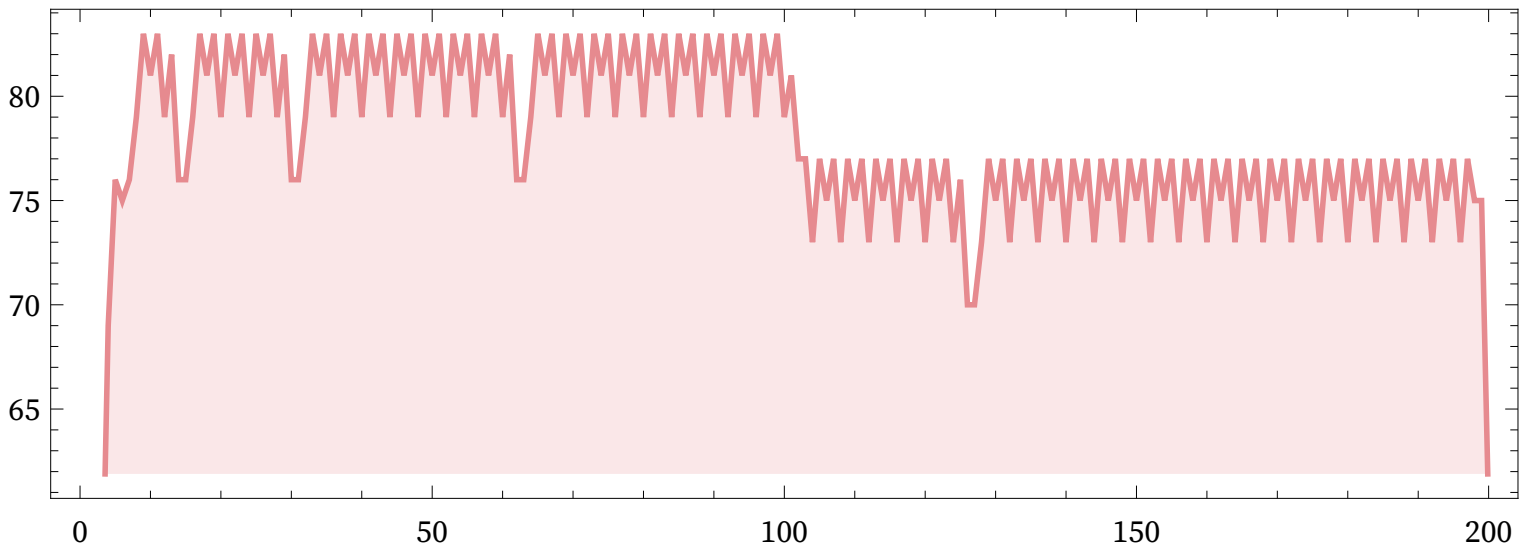

These results are all for ordinary, deterministic cellular automata. So are there non-deterministic cellular automata? Typically, cellular automata are defined to consistently update every cell at every step. But one can also consider sequential cellular automata, where specific cells are updated at each step—and in this case it is straightforward to define a non-deterministic version, for which things like multiway systems can be constructed. (Note that for a full rulial multiway system, such non-deterministic cellular automata are basically the same as non-deterministic mobile automata, which are close to Turing machines.)

## Computation Capabilities and the Structure of Rulial Space

What do the computational capabilities of Turing machines mean for their rulial space? Let's start with computation universality.

An important fact about deterministic Turing machines is that there are universal ones. Among the 4096 $s = 2$, $k = 2$ rules there aren't any. But as soon as one goes to $s = 2$, $k = 3$ rules (of which there are 2,985,984) it's known (thanks to my results in *A New Kind of Science*, and Alex Smith's 2007 proof) that there is a universal machine. The rule is:

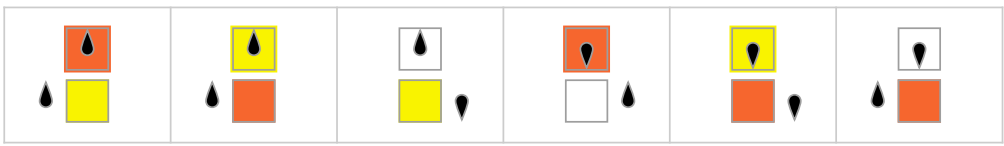

Starting from a blank tape, this machine gives

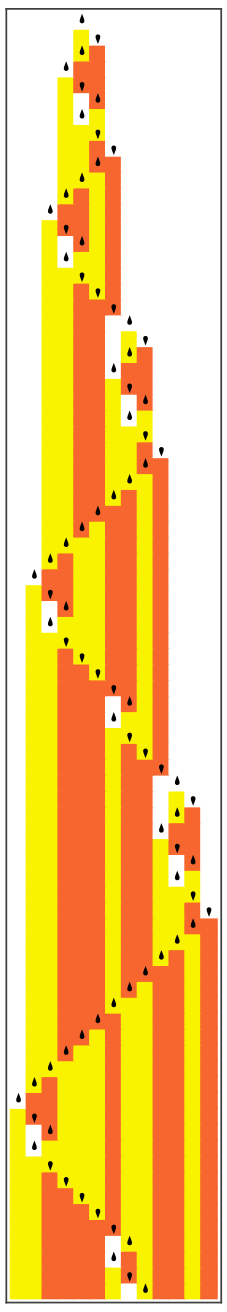

and the corresponding causal graph is:

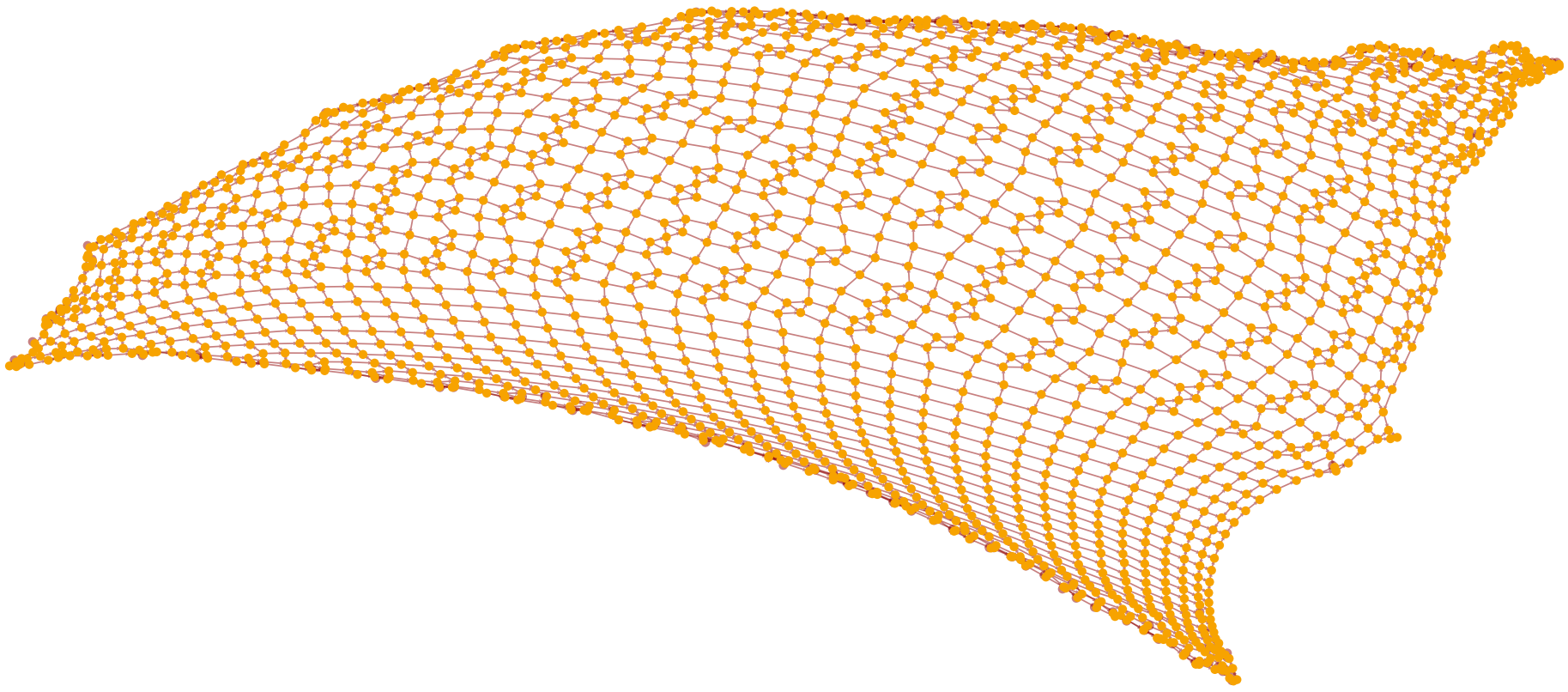

But what does the existence of a universal machine mean in the rulial multiway system? From any given initial condition, any deterministic Turing machine will trace out some trajectory in the rulial multiway graph. And if a machine is universal it means that by appropriately picking its initial conditions it can be "programmed" to "emulate" any other Turing machine, in the sense that its trajectory will track the trajectory of whatever Turing machine it's emulating. What does "track" mean? Basically, that there's some fixed scheme that allows one to go from the states of the universal Turing machine to the states of the machine it's emulating. The "scheme" will correspond to some translation in the rulial multiway graph, and the requirement is that this translation is somehow always limited. In other words, the trajectory of a universal machine will "flail around" as its starting point (i.e. initial condition) moves in the rulial multiway graph, and this "flailing" will be diverse enough that the trajectory can get close to the trajectory of any given other machine.

If, on the other hand, the machine one's dealing with isn't universal, then its trajectory won't "flail around" enough for this work; the trajectory will in a sense be too constrained to successfully "track" all possible other deterministic Turing machine trajectories.

What about non-deterministic Turing machines? What universality can be interpreted to mean in this case is that given a particular initial condition, the output one wants occurs somewhere on the different paths followed by the non-deterministic Turing machine. (If one's trying to do a decision problem—as in the computational complexity class NP—then one can arrange to "signal that one's got an answer" through some feature of the Turing machine state.) In the case of "extreme non-determinism"—as used to construct the rulial multiway graph—computation universality in a sense becomes a statement purely about the structure of the rulial multiway graph. And basically it just requires that the rulial multiway graph is sufficiently connected—which is guaranteed if there's causal invariance (so there's nothing like an "event horizon" anywhere).

But does one have to allow "extreme non-determinism" to get universality? With $s = 2$, $k = 3$ we know that there's a purely deterministic Turing machine that achieves it. And my guess is that among $s = 2$, $k = 2$ there are "slightly multiway" rules that also do. In a standard deterministic $s = 2$, $k = 2$ Turing machine, there are 4 cases of the rule that are each specified to have a unique outcome. But what if even just one of those cases in the rule has two outcomes? The rule is non-deterministic, but it can be thought of as just being the result of specifying 5 defining cases for the rule. And it's my guess that even this will be sufficient to get universality in the system.

A non-deterministic rule like this will not trace out a single path in the rulial multiway graph. Instead, it'll give a bundle of paths, with different paths corresponding to different non-deterministic choices. But the story of universality is very similar to the deterministic case: one simply has to ask whether anything in the bundle successfully manages to track the trajectory of the machine one's emulating.

It's worth remembering that any given rule won't typically be following geodesics in rulial space. It'll be following some more circuitous path (or bundle of paths). But let's say one has some rule that traces out some trajectory—corresponding to performing some computation. The Principle of Computational Equivalence implies that across different possible rules, there's a standard "maximum computational sophistication" for these computations, and many rules achieve it. But then the principle also implies that there's equivalence between these maximally sophisticated computations, in the sense that there's always a limited computation that translates between them.

Let's think about this in rulial space. Let's say we have two rules, starting from the same initial condition. Well, then, at the beginning it's trivial to translate between the rules. But after $t$ steps, the states reached by these rules can have diverged—making it potentially progressively more difficult to translate between them. A key idea of the rulial multiway graph—and rulial space—is that it lets one talk about both computations and translations between them in uniform ways. Let's say that the trajectories of two rules have gone a certain distance in rulial space. Then one can look at their divergence, and see how long a "translation computation" one has to do to get from one to the other.

In ordinary spacetime, let's say a certain time $t$ has elapsed. Then we know that the maximum spatial distance that can have been traversed is $c\,t$, where $c$ is the speed of light. In rulial space, there's something directly analogous: in time $t$, there's a maximum rulial distance that can be traversed, which we can call $\rho\,t$. But here the Principle of Computational Equivalence makes it a crucial contribution: it implies that throughout rulial space, and in all situations, $\rho$ is fixed. It can take an irreducible amount of computational work to successfully translate from the outcome of one rule to another. But this always scales the same way. There's in effect one scale of computational irreducibility, and it's characterized by $\rho$. Like the constancy of the speed of light uniformly limits physical motion, the constancy of $\rho$ uniformly limits rulial motion.

But let's say you're trying to get from one point in rulial space to another. If from your starting point you can follow the path of an irreducible—and effectively universal—computation, then you'll successfully be able to reach the other point. But if from your starting point you can only follow a reducible computation this won't generally be true. And what this means is that "pockets of computational reducibility" in rulial space act a bit like black holes in physical space. You can get into them from regions of irreducibility, but you can't get out of them.

There are probably signs of phenomena like this even in the rulial space for simple Turing machines that we've explored here. But there's considerably more that needs to be worked out to be able to make all the necessary connections.

## The Emerging Picture of Rulial Space

There are lots of analogies between physical space, the branchial space, and rulial space. For example, in physical space, there are light cones that govern the maximum rate at which effects can propagate between different parts of physical space. In branchial space, there are entanglement cones that govern the maximum rate of quantum entanglement. And in rulial space, one can think of "emulation cones", which govern the maximum rate at which one description of behavior can be translated into another.

And when it comes to applying these things to modeling the physical universe, a crucial point is that the observer is necessarily part of the system, governed by the same rules as everything else. And that means that the observer can only be sensitive to certain "appropriately modded-out" aspects of the system. But in actually imagining how an observer "perceives" a system it's almost always convenient to think about coordinatizing the system—by defining some appropriate foliation. In physical space, this involves foliating the causal graph using reference frames like in relativity. In branchial space, it involves our concept of quantum observation frames. And in rulial space, we can invent another such concept: a rulial description frame, or just a rulial frame.

Different rulial frames in effect correspond to describing the evolution of the universe as operating according to different rules. Causal invariance implies that in the end different rulial frames must give equivalent results. But the specific way one describes the time evolution of the universe will depend on what rulial frame one's using. In one frame one would be describing the universe in one way; in another frame, another way. And the "story one tells" about how the universe evolves will be different in the different frames.

Much like with superposition in quantum mechanics, there's probably some notion of regions in rulial space, in which one's somehow viewing the universe as operating according to "rulially entangled" collections of rules.

But while our original motivation was understanding physics, a lot of what we're studying about rulial space also applies to purely computational systems. For example, we can think of rulial space even without having any notion of physical space. And we can in effect imagine that rulial space is some kind of map of a space of possible rules for computational systems. (Notice that because of computation universality and the Principle of Computational Equivalence it ultimately doesn't matter what particular type of rule—Turing machine, cellular automaton, combinator, whatever—is used to "parametrize" rulial space.)

Different places in rulial space in some sense correspond to different rules. Paths at different places in rulial space correspond to evolution according to different rules. So what is the analog of motion in rulial space? In effect it's having a frame which progressively changes the rule one's using. If one's trying to find out what happens in one's system, it's fundamentally most efficient to "stick with one rule". If one progressively changes the rule, one's going to have to keep "translating back" to the original rule, by somehow emulating this rule with whatever rule one's reached. And the result of this is there'll be exactly the analog of relativistic time dilation. Faster motion in rulial space leads to slower progression in time. Of course, to discuss this properly, we really have to talk about the rulial multiway causal graph, etc.

But one thing is clear: motion faster than some maximum speed $\rho$ is impossible. From within the system, you simply can't keep the correct rulial causal connections if you're changing your rulial location faster than $\rho$.

In an abstract study of Turing machines, $\rho$ is just an arbitrary parameter. But in the actual physical universe, it must have a definite value, like the speed of light $c$, or our maximum entanglement speed $\zeta$. It's difficult even to estimate what $\zeta$ might be. And presumably estimating $\rho$ will be even harder. But it's interesting to discuss at least how we might start to think about estimating $\rho$.

The first key observation about rulial space is that in our models, it's discrete. In other words, there's a discrete space of possible rules. Or, put another way, theories are quantized, and there's somehow an elementary distance between theories—or a minimum distance in "theory space" between neighboring theories.

But what units is this distance in? Basically it's in units of rule—or program—size. Given any program—or rule—we can imagine writing that program out in some language (say in Wolfram Language, or as a program for a particular universal Turing machine, or whatever) And now we can characterize the size of the program by just looking at how many tokens it takes to write the program out.

Of course, with different languages, that number will be different—at the simplest level just like the number of decimal digits necessary to represent a number is different from the number of binary digits, or the length of its representation in terms of primes. But it's just like measuring a length in feet or meters: even though the numerical value is different, we're still describing the same length.

It's important to point out that it's not enough to just measure things in terms of "raw information content", or ordinary bits, as discussed in information theory. Rather, we want some kind of measure of "semantic information content": information content that directly tells us what computation to do.

It's also important to point out that what we need is different from what's discussed in algorithmic information theory. Once one has a computation universal system, one can always use it to translate from any one language to any other. And in algorithmic information theory the concept is that one can measure the length of a program up to an additive constant by just expecting to include an "emulation program" that adapts to whatever language one's measuring the length in. But in the usual formalism of algorithmic information theory one doesn't worry about how long it's going to take for the emulation to be done; it's just a question of whether there's ultimately enough information to do it.

In our setup, however, it does matter how long the emulation takes, because that process of emulation is actually part of our system. And basically we need the number of steps needed for the emulation to be in some sense bounded by a constant.

So, OK, what does this mean for the value of $\rho$? Its units are presumably program size per unit time. And so to define its value, we'll have to say how we're measuring program size. Perhaps we could imagine we write our rules in the Wolfram Language. Then there should be a definite value of $\rho$ for our universe, measured in Wolfram-Language-tokens per second.

If we chose to use (2,3)-Turing-machine-tape-values per year then we'd get a different numerical value. But assuming we used the correct conversions, the value would be the same. (And, yes, there's all sorts of subtlety about constant-time or not emulation, etc.)

In some sense, we may be able to think of $\rho$ as the ultimate "processor speed for the universe": how fast tokens in whatever language we're using are being "interpreted" and actually "executed" to determine the behavior of the universe.

Can we estimate the value of $\rho$? If our units are Wolfram-Language-tokens per second we could start by imagining just computing in the Wolfram Language some piece of the rulial multiway graph for our models and seeing how many operations it takes. To allow "all possible rules" we'd have to increase the possible left- (and right-) hand sides of our rules to reflect the size of the hypergraph representing the universe at each step. But now we'd need to divide by the "number of parallel threads" in the rulial multiway graph. So we can argue that all we'd be left with is something like (size of spatial hypergraph represented in Wolfram Language) / (elementary time).

So, based on our previous estimates (which I don't consider anything more than vaguely indicative yet) we might conclude that perhaps:

$\rho \sim 10^{450}$ Wolfram-Language-tokens/second

The number of "parallel threads" in the rulial multiway graph (the rulial analog of Ξ) might then be related to the number of possible hypergraphs that contain about the number of nodes in the universe, or very roughly $(10^{350})$^$(10^{350}) \approx 10^{10^{353}}$. If we ask the total number of Wolfram Language tokens processed by the universe, there'll be another factor ~$10^{467}$, but this "parallelism" will completely dominate, and the result will be about:

$10^{10^{356}}$ Wolfram-Language-tokens

OK, so given a value of $\rho$, how might we conceivably observe it? Presumably there's an analog of quantum uncertainty in rulial space, that's somehow proportional to the value of $\rho$. It's not completely clear how this would show up, but one possibility is that it would lead to intrinsic limits on inductive inference. For example, given only a limited observation time, it might be fundamentally impossible to determine beyond a certain ("rulial") accuracy what rule the universe is following in your description language. There'd be a minimum rate of divergence of behaviors from different rules, associated with the minimum distance between theories—and it would take a certain time to distinguish theories at this rate of divergence.

In our models, just like every causal edge in physical and branchial space is associated with energy, so should every causal edge in rulial space be. In other words, the more processing that happens in a particular part of rulial space, the more physical energy one should consider exists there. And just as with the Einstein equations in physical space, or the Feynman path integral in branchial space, we should expect that the presence of energy in a particular region of rulial space should reflect in a deflection of geodesics there.

Geodesics in rulial space are the shortest paths from one configuration of the universe to another, using whatever sequences of rules are needed. But although that's somewhat like what's considered at a theoretical level in non-deterministic computation, it's not something we're usually familiar with: we're used to picking a particular description language and sticking with it. So exactly what the interpretation of deflections of geodesics in rulial space should be isn't clear.

But there are a few other things we can consider. For example, presumably the universe is expanding in rulial space, perhaps implying that in some sense more descriptions of it are becoming possible over time. What about rulial black holes? As mentioned above, parts of rulial space that correspond to computational reducibility should behave like black holes, where in effect "time stops". Or, in other words, while in most of the universe time is progressing, and irreducible computation is going on, computational reducibility will cause that process to stop in a rulial black hole.

Presumably geodesics near the rulial black hole will be pulled towards it. Somehow when there's a description language that leads to computational reducibility, languages near it will tend to also stop being able to successfully describe computationally irreducible processes.

Can we estimate the density of rulial black holes? Let's consider the Turing machine case. In effect we're going to want to know what the density of non-universality is among all possible non-deterministic Turing machines. Imagine we emulate all Turing machines using a single universal machine. Then this is effectively equivalent to asking what fraction of initial conditions for that machine lead to reducible behaviors—or, in essence, in the traditional characterization of Turing machines, halt. But the probability across all possible inputs that a universal Turing machine will halt is exactly Greg Chaitin's $\Omega$. In other words, the density of rulial black holes in rulial space is governed by $\Omega$.

But $\Omega$ isn't just a number like $\pi$ that we can compute as accurately as we want; it's noncomputable, in the sense that it can't be computed to arbitrary accuracy in any finite time by a Turing machine. Now, in a sense it's not too surprising that the density of rulial black holes is noncomputable—because, given computational irreducibility, to determine for certain whether something is truly a rulial black hole from which nothing can escape one might have to watch it for an unbounded amount of time.

But for me there's something personally interesting about Greg Chaitin's $\Omega$ showing up in any way in a potential description of anything to do with the universe. You see, I've had a nearly 40-year-long debate with Greg about whether the universe is "like $\pi$" or "like $\Omega$". In other words, is it possible to have a rule that will let us compute what the universe does just like we can compute (say, in principle, with a Turing machine) the digits of $\pi$? Or will we have to go beyond a Turing machine to describe our universe? I've always thought that our universe is "like $\pi$"; Greg has thought that it might be "like $\Omega$". But now it looks as if we might both be right!

In our models, we're saying that we can compute what the universe does, in principle with a Turing machine. But what we're now finding out is that in the full rulial space, general limiting statements pull in Ω. In a particular rulial observation frame, we're able to analyze the universe "like $\pi$". But if we want to know about all possible rulial observation frames—or in a sense the space of all possible descriptions of the universe—we'll be confronted with Ω.

In our models, the actual operation of the universe, traced in a particular rulial observation frame, is assumed never to correspond to anything computationally more powerful than a Turing machine. But let's say there was a hypercomputer in our universe. What would it look like? It'd be a place where the effective $\rho$ is infinite—a place where rulial geodesics infinitely diverge—a kind of white hole in rulial space (or perhaps a cosmic event horizon).(We can also think about the hypercomputer as introducing infinite-shortcut paths in the rulial multiway graph which ordinary Turing-machine paths can never "catch up to" and therefore causally affect.)

But given a universe that does hypercomputation, we can then imagine defining a rulial multiway graph for it. And then our universe will show up as a black hole in this higher-level rulial space.

But OK, so if there's one level of hypercomputer, why not consider all possible levels? In other words, why not define a hyperrulial multiway graph in which the possible rules that are used include both ordinary computational ones, but also hypercomputational ones?

And once we're dealing with hypercomputational systems, we can just keep going, adding in effect more and more levels of oracles—and progressively ascending the arithmetic hierarchy. (The concept of "intermediate degrees" might be thought to lead to something that isn't a perfect hierarchy—but I suspect that it's not robust enough to apply to complete rulial spaces.) Within the hyperrulial multiway graph at a particular level, levels below it will be presumably appear as rulial black holes, while ones above will appear as rulial white holes.

And there's nothing to keep us only considering finite levels of the arithmetic hierarchy; we can also imagine ascending to transfinite levels, and then just keeping going to higher and higher levels of infinity. Of course, according to our models, none of this is relevant to our particular physical universe. But at a theoretical level, we can still at least to some extent "symbolically describe it", even in our universe.

## Thanks

*Thanks to Jonathan Gorard for a variety of discussions, as well as to Tali Beynon, Todd Rowland, Ed Pegg, Jose Martin-Garcia and Christopher Wolfram.*

## References

*Links to references are included within the body of this document.*